\documentclass[12pt]{article}
\usepackage{amsmath,amssymb}
\usepackage{bm}
\usepackage{xcolor}
\usepackage{color}
\usepackage{graphicx}
\usepackage{cite}
\usepackage{mathtools}
\usepackage{geometry}  
\usepackage{comment}
\usepackage{hyperref}
\usepackage{slashed}
\usepackage{multirow}
\usepackage{dcolumn}
\usepackage{longtable}
\usepackage{here}
\usepackage{tabularx}
\usepackage{dynkin-diagrams}
\usepackage{url}
\usepackage{amsthm}
\usepackage{amssymb}
\usepackage{amsmath}
\usepackage{mathtools}
\usepackage{tabularx}
\usepackage{dynkin-diagrams}
\usepackage{titlesec}
\usepackage{slashed}
\usepackage{cite}
\usepackage{url}
\usepackage{empheq}
\usepackage{diagbox}
\usepackage{caption}
\usepackage{geometry}
\usepackage{subcaption}
\usepackage{color}

\newcommand{\tr}[1]{\hspace{.1em}\mathrm{tr\hspace{.1em}#1}}

\begin{document}

\newcommand{\specialcell}[2][c]{%
\begin{tabular}[#1]{@{}c@{}}#2\end{tabular}} 

\title{
\begin{flushright}
\ \\*[-80pt]
\begin{minipage}{0.2\linewidth}
\normalsize
EPHOU-25-010 \\*[50pt]
\end{minipage}
\end{flushright}
{\Large \bf
Generation structures and Yukawa couplings \\ in magnetized $T^{2g}/\mathbb{Z}_N$ models
\\*[20pt]}}

\author{Tim Jeric, Tatsuo Kobayashi, Kaito Nasu, and
Shohei Takada
\\*[20pt]
\centerline{
\begin{minipage}{\linewidth}
\begin{center}
{\it \normalsize
Department of Physics, Hokkaido University, Sapporo 060-0810, Japan} \\*[5pt]
\end{center}
\end{minipage}}
\\*[50pt]}
\date{
\centerline{\small \bf Abstract}
\begin{minipage}{0.9\linewidth}
\medskip
\medskip
\small
We study fermion zero-mode wave functions with various chiralities
in magnetized $T^{2g}$, $(g=2,3)$ torus.
First, we consider the wave functions 
satisfying the Dirac equation and the boundary conditions on the magnetized torus. 
Second, we introduce the $SO(3)$ (or parity) transformations and derive the wave functions under the modular transformation. 
Additionally, we calculate the Yukawa couplings with consideration for the chirality.
Lastly, we briefly review how to construct $T^{4}/\mathbb{Z}_N$ ($N=2,3,4,6$) and $T^6/\mathbb{Z}_{12}$ twisted orbifold. Also, we explicitly analyze the number of the wave functions in $\mathbb{Z}_N$ sectors. 
\end{minipage}
}
\begin{titlepage}
\maketitle
\thispagestyle{empty}
\end{titlepage}

\newpage
\tableofcontents
\clearpage

\section{Introduction}
Superstring theory is a promising candidate for unified theory of particle physics. When we start with superstring theory, we need to compactify extra six-dimensional ($6$D) space to realize four-dimensional ($4$D) low energy effective field theory. Specifically, we should consider $4$D chiral theory derived from superstring theory because the standard model is chiral theory.
We expect that string compactification can  solve mysteries in particle physics, e.g. origins of three-generation structure and the mass hierarchy of quarks and leptons.

Many compactifications such as Calabi-Yau manifolds have been proposed. In particular, toroidal compactification $T^6$ is one of the simplest examples of $6$D compactifications. However, 4D $\mathcal{N}=4$ supersymmetry (SUSY) remains in the simple $T^6$ compactification.
It has the $SU(4)_R$ symmetry and leads to 4D non-chiral theory.
Orbifolding $T^6/\mathbb{Z}_N$ is a way to realize $4$D chiral theory, 
where 4D SUSY can be broken to $\mathcal{N} =  1$ or $\mathcal{N} = 0$ and 
the $SU(4)_R$ symmetry is also broken 
\cite{Dixon:1985jw,Dixon:1986jc}, although 
orbifolding can also lead to 4D $\mathcal{N}=2$ SUSY and it is not a chiral theory.

Introduction of background magnetic fluxes is another way to realize 4D chiral theory\cite{Witten:1984dg,Bachas:1995ik,Berkooz:1996km,Blumenhagen:2000wh,Angelantonj:2000hi,Blumenhagen:2000ea}.
The number of zero modes is determined by amount of magnetic fluxes.
Their profiles in the compact space can be written explicitly and they are non-trivial \cite{Cremades:2004wa,Antoniadis:2009bg}.
Yukawa couplings and higher dimensional couplings in 4D low energy effective field theory can be computed by overlap integral of wave functions \cite{Cremades:2004wa,Antoniadis:2009bg,DeAngelis:2013iwa, Abe:2009dr}.
Furthermore, the modular transformation behavior of zero modes was also studied~\cite{
Kobayashi:2017dyu,Kobayashi:2018rad,Kikuchi:2020frp,Kikuchi:2021ogn,Kobayashi:2018bff,Ohki:2020bpo,Kikuchi:2020nxn,Almumin:2021fbk,Kikuchi:2023awe,Kobayashi:2024ysa}. 
The modular symmetry appears in 4D low energy effective field theory and controls allowed couplings.
In these analyses, it is essential that we can write explicitly the wave functions.

One can combine the above orbifold geoemetry and introduction of background magnetic fluxes.
That is, toroidal orbifold compactification with magetic fluxes is also interesting.
$T^2/\mathbb{Z}_N$ compactifications with magnetic fluxes \cite{Abe:2008fi,Abe:2013bca} were studied intensively.
Their extensions to $(T^2\times T^2 \times T^2)/\mathbb{Z}_N $ and $(T^2\times T^2 \times T^2)/(\mathbb{Z}_M\times \mathbb{Z}_N)$ are straightforward, when  
$T^6$ is factorized as $T^6=T^2\times T^2 \times T^2$.
Three-generation models were classified \cite{Abe:2008sx,Abe:2015yva} and realization of quark and lepton masses and their mixing angles was studied 
\cite{Abe:2012fj,Abe:2014vza,Fujimoto:2016zjs,Buchmuller:2017vho,Buchmuller:2017vut,Hoshiya:2022qvr}.

Such a study was also extended to non-factorizable $T^4$ and $T^6$
and their orbifolds with generic magnetic flux background \cite{Cremades:2004wa,Antoniadis:2009bg,Kikuchi:2022lfv,Kikuchi:2022psj,Kikuchi:2023awm}.
These compactifications lead to more rich structure and are important.
For example, these compactifications include more degree of freedom of moduli.
As mentioned above, it is essentially necessary to find explicitly the wave functions in order to write down 4D low energy effective field theory, although only the number of zero modes could be computed by some index theorem \cite{Sakamoto:2020vdy,Kobayashi:2022tti,Aoki:2024rmf}.
Detailed studied of the zero mode wave functions have been done for the zero modes, whose $SO(6)$ chiralities on $T^6$ are $(+,+,+)$ and $(-,-,-)$ including the $SO(4)$ chiralities $(+,+)$ and $(-,-)$ on $T^4$, although the wave functions with general chirality were studied in 
Refs.~\cite{Cremades:2004wa,Antoniadis:2009bg}.
Note that the invariant combinations of $SO(6)$ scalar, spinor and vector representations appear in allowed couplings of 4D low energy effective field theory.
For example, in 4D $\mathcal{N}=1$ SUSY theory, 
the invariant combinations of $SU(3)$ triplets appear in 4D Yukawa couplings, where 
$SU(3) \subset SU(4)_R$.
Therefore, we have to write explicitly the wave functions of zero modes with various chiralities in order to obtain 4D low energy effective field theory.
That is our purpose of this paper.
When $T^6$ is factorized as $T^6=T^2_1 \times T^2_2 \times T_3^2$ and 
magnetic flux backgrounds are non-vanishing on only each $T^2_i$, 
the wave functions with generic chiralities are obtained as products of the wave functions $\psi^{(i)}_{\pm}$ on $T^2_i$ as   $\psi^{(1)}_{\pm}\psi^{(2)}_{\pm}\psi^{(3)}_{\pm}$.
That is straightforward, because $\psi^{(i)}_{\pm}$ are known already.
We study the wave functions of zero modes with various chiralities on 
generic $T^6$ with magnetic flux backgrounds, which can not be realized 
as $T^6=T^2_1 \times T^2_2 \times T_3^2$ with non-vanishing magnetic flux backgrounds on only each $T^2_i$.

This paper is organized as follows.
We first introduce background magnetic fluxes and boundary conditions on $T^6$ in section \ref{section2}.
In section \ref{section3}, we obtain wave functions with various chiralities that satisfy the Dirac equations and the boundary conditions. We also obtain the wave functions through $SO(3)$ rotations in $T^6$. 
In section \ref{section4}, we show how to introduce the modular transformation behavior of zero-mode wave functions. This is crucial to analyze the number of zero-modes in $T^{2g}/\mathbb{Z}_N$, ($g=2,3$).
In section \ref{section5}, we analyze the Yukawa couplings in magnetized $T^6$ taking both the chirality and the $SO(6)$ symmetry into account. 
In section \ref{section6}, we find the numbers of zero-modes in magnetized $T^{2g}/\mathbb{Z}_N$ to verify whether three-generation models can be realized or not.
Section \ref{conclusion} is devoted to conclusion.
We analyze magnetized $T^6$ model with the non-trivial scale factors in Appendix A.
We explain whether the wave functions are the solutions of the Dirac equation and satisfy the boundary condition in Appendices B and C. In Appendices D and E, we give the details of the $Sp(2g,\mathbb{Z})$ symplectic modular symmetry and the calculation for the Yukawa couplings. 
Lastly, we summarize the details of the calculations for zero-modes and spinor transformation in magnetized $T^4/\mathbb{Z}_N$ in Appendices F, G, and H.

\section{Magnetized $T^6$ model}\label{section2}

Here we study $U(1)$ gauge theory on $T^6$ with magnetic flux background, which is originated from ten-dimensional (10D) supersymmetric 
Yang-Mills theory.
We give a review on the magnetic flux background and the Dirac equation.
We can extend this model to 
 $U(N)$ gauge theory with magnetic flux backgrounds 
along the diagonal direction of $U(N)$.

\subsection{Dirac operator}
First, we review how to construct magnetized $T^{6}$ models \cite{Kikuchi:2023awm}.
The $6$D torus $T^6 \simeq \mathbb{C}^3/\Lambda$ is constructed by the 6D lattice $\Lambda$. 
(See Refs.~\cite{Markushevich:1986za,Ibanez:1987pj,Katsuki:1989bf,Kobayashi:1991rp,Lust:2005dy,Lust:2006zg} for various 6D Lie lattices to construct $T^6$ and its orbifolds.)
The 6D lattice $\Lambda$ is spanned by six basis vectors $e^{\prime}_{i}$, ($i=1,\cdots,6$).
We write them as  
\begin{align}
\begin{aligned}
\label{basis}
&e^{\prime}_{1} 
= 2\pi r_1 {e}_1
= 2\pi r_1 
\begin{bmatrix}
1 \\
0 \\
0 \\
\end{bmatrix}
, 
e^{\prime}_{2} 
= 2\pi r_2 {e}_2
= 2\pi r_2 
\begin{bmatrix}
0 \\
1 \\
0 \\
\end{bmatrix}
,
e^{\prime}_{3} 
= 2\pi r_3 {e}_3
= 2\pi r_3 
\begin{bmatrix}
0 \\
0 \\
1 \\
\end{bmatrix}
, \\ 
&e^{\prime}_{4} 
= 2\pi r_i \Omega_{i1} {e}_1
= 2\pi 
\begin{bmatrix}
r_1 \omega_{11} \\
r_2 \omega_{21} \\
r_3 \omega_{31} \\
\end{bmatrix}
,
e^{\prime}_{5} 
= 2\pi r_i \Omega_{i2} {e}_2
= 2\pi 
\begin{bmatrix}
r_1 \omega_{12} \\
r_2 \omega_{22} \\
r_3 \omega_{32} \\
\end{bmatrix}
, \\ \notag
&e^{\prime}_{6} 
= 2\pi r_i \Omega_{i3} {e}_3
= 2\pi 
\begin{bmatrix}
r_1 \omega_{13} \\
r_2 \omega_{23} \\
r_3 \omega_{33} \\
\end{bmatrix}
,
\end{aligned}
\end{align}
where $2\pi r_i$, ($i=1,2,3$), are the scale factors and $e_i$ denote the Euclidean standard basis. Then we define  the complex structure moduli $\Omega =  \Omega_{ij}$ on $T^6$ as 
\begin{align}
\Omega
=
\begin{bmatrix}
\omega_{11} & \omega_{12} & \omega_{13} \\
\omega_{21} & \omega_{22} & \omega_{23} \\
\omega_{31} & \omega_{32} & \omega_{33} \\
\end{bmatrix}
.
\end{align}
We require $\Omega^T = \Omega$ to construct lattice of $T^{2g}/\mathbb{Z}_N$, ($g = 2, 3$).
When we define complex coordinates $\vec{Z}$ on $\mathbb{C}^3$ with real vectors $\vec{x}, \vec{y}$, we also define complex coordinates on $T^6$ written by $z^i = x^i + \Omega_{ij} y^j$:
\begin{align}
{Z}^i \equiv 2\pi r_i ( {x}^i + \Omega_{ij} {y}^j
) =  2\pi r_i {z}^i,
\end{align}
where we impose the identifications $z^i \sim z^i + e_i \sim z^i + \Omega_{ij} e_j$ on $T^6$, namely $x^i \sim x^i + e_i$ and $y^i \sim y^i + e_i$. 
Here, vector $\vec{v}$ may simply be denoted by $v$.
In this paper, since we mainly consider $T^{2g}/\mathbb{Z}_N$ orbifolds with simply-laced Lie Lattices, we restrict the scale factors $2\pi r_i = 1$.\footnote{Magnetized $T^6$ model with the non-trivial scale factors $2\pi r_i$ is discussed in Appendix A.}
We introduce the Siegel upper-half plane $\mathcal{H}^3$ defined as follows
\cite{Siegel:1943},
\begin{align}
\label{H3}
\mathcal{H}^3 
= \{  \Omega \in GL(3,\mathbb{C}) | \Omega^T = \Omega, \text{Im}\Omega  > 0  \}.
\end{align}
We will utilize this when we consider wave functions derived from the magnetic fluxes.  
In section \ref{section3}, we will discuss how to obtain well-defined wave functions.

We introduce the following metric on $\mathbb{C}^3$:
\begin{align}
\label{metric}
ds^2  = \delta_{i, {j}} dZ^i d\bar{Z}^{\bar{j}} \quad \text{for} \quad i, j = 1,2,3.
\end{align}
We also define the Gamma matrices $\Gamma^{z^i}$, $\Gamma^{\bar{z}^i}$ to introduce the Dirac equations on $T^6$:
\begin{align}
\begin{aligned}
\label{Gammamatrix}
&\Gamma^{z^1} \equiv \frac{1}{2\pi r_1} \Gamma^{Z^1} = \frac{1}{2\pi r_1} (\sigma^z \otimes \sigma^3 \otimes \sigma^3), \quad
\Gamma^{\bar{z}^{\bar{1}}} \equiv \frac{1}{2\pi r_1} \Gamma^{\bar{Z}^{\bar{1}}} = \frac{1}{2\pi r_1}
(\sigma^{\bar{z}} \otimes \sigma^3 \otimes \sigma^3), \\
&\Gamma^{z^2} \equiv \frac{1}{2\pi r_2} \Gamma^{Z^2}  
=  \frac{1}{2\pi r_2} ({\bf{1}}_2 \otimes \sigma^z \otimes \sigma^3) , \quad
\Gamma^{\bar{z}^{\bar{2}}} \equiv \frac{1}{2\pi r_2} \Gamma^{\bar{Z}^{\bar{2}}}  
= \frac{1}{2\pi r_2} ({\bf{1}}_2 \otimes \sigma^{\bar{z}} \otimes \sigma^3), \\
&\Gamma^{z^3} \equiv \frac{1}{2\pi r_3} \Gamma^{{Z}^3} 
=  \frac{1}{2\pi r_3} ( {\bf{1}}_2 \otimes {\bf{1}}_2 \otimes \sigma^z) , \quad
\Gamma^{\bar{z}^{\bar{3}}} \equiv \frac{1}{2\pi r_3} \Gamma^{\bar{Z}^{\bar{3}}} 
 =  \frac{1}{2\pi r_3}  ({\bf{1}}_2 \otimes {\bf{1}}_2 \otimes \sigma^{\bar{z}}) ,
\end{aligned}
\end{align}
where ${\bf{1}}_n$ represents $n\times n$ unit matrix and we define the Pauli matrices $\sigma^{i}$ as follows
\begin{align}
\sigma^{1} 
=
\begin{bmatrix}
0 & 1 \\
1 & 0 \\
\end{bmatrix}
,
\sigma^{2} 
=
\begin{bmatrix}
0 & -i \\
i & 0 \\
\end{bmatrix}
,
\sigma^{3} 
=
\begin{bmatrix}
1 & 0 \\
0 & -1 \\
\end{bmatrix}
.
\end{align}
We also define $\sigma^{z} = \sigma^{1} +i \sigma^2$, $\sigma^{\bar{z}} = \sigma^{1} - i \sigma^2$ using the Pauli matrices.
We find that the Gamma matrices satisfy the Clifford algebra:
\begin{align}
\begin{aligned}
&\{ \Gamma^{Z^i} , \Gamma^{Z^j}  \} = 
\{ \Gamma^{\bar{Z}^{\bar{i}}} , \Gamma^{\bar{Z}^{\bar{j}}}  \} = 0, \\
&\{ \Gamma^{Z^i} , \Gamma^{\bar{Z}^{\bar{j}}}  \} = 4 \delta^{i, {j}} .
\end{aligned}
\end{align}
In particular, we can determine the chirality of wave functions on magnetized $T^6$ by the following matrix $\Gamma^{7}$:
\begin{align}
\Gamma^{7} = \sigma^3 \otimes \sigma^3 \otimes \sigma^3 
= \text{diag} [+, -, -, +, -, +, +, -].
\end{align} 
Then, the Dirac operator $\slashed{D}$ defined on $T^6$ is given by
\begin{align}
i\slashed{D} 
\equiv
i (\Gamma^{z^j} D_{z^j} + \Gamma^{\bar{z}^{\bar{j}}} \bar{D}_{\bar{z}^{\bar{j}}}),
\end{align}
where we define following covariant derivatives $D_{z^j}$, $\bar{D}_{\bar{z}^{\bar{j}}}$.
Here, we consider $U(1)$ gauge field theory, where the gauge fields $A_{z}$, $A_{\bar{z}}$ are coupled to fermions with $U(1)$ unit charge $q=+1$:
\begin{align}
\begin{aligned}
\label{covariant}
&D_{z^j} \equiv D_j = \partial_{z^j} - i A_{z^j}, \\
&\bar{D}_{\bar{z}^{\bar{j}}} \equiv \bar{D}_j = \partial_{\bar{z}^{\bar{j}}} - i A_{\bar{z}^{\bar{j}}}.
\end{aligned}
\end{align}
In order to show both the gauge fields written in Eq. \eqref{covariant} and boundary conditions of spinor wave functions on $T^6$ explicitly, we will introduce background magnetic fluxes in the next subsection 2.2.

\subsection{Magnetic fluxes and boundary condition on $T^6$}

Here, $U(1)$ background magnetic fluxes $F$ on $T^6$ are written as follows:
\begin{align}
\begin{aligned}
\label{F1}
F
&\equiv  \frac{1}{2} (p_{xx})_{ij} dx^i \wedge dx^j + \frac{1}{2} (p_{yy})_{ij} d{y}^i \wedge d{y}^j + (p_{xy})_{ij} dx^i \wedge dy^j \\ 
&=\frac{1}{2} (F_{zz})_{ij} dz^i \wedge dz^j + \frac{1}{2} (F_{\bar{z}\bar{z}})_{ij} d\bar{z}^i \wedge d\bar{z}^j + (F_{z\bar{z}})_{ij} (i dz^i \wedge d\bar{z}^j),
\end{aligned}
\end{align}
where
\begin{align}
\begin{aligned}
\label{F2}
&(F_{zz})_{ij} = \left(
[(\bar{\Omega} - \Omega)^{-1}]^T (\bar{\Omega}^T p_{xx} \bar{\Omega} + p_{yy} + p^{T}_{xy} \bar{\Omega} - \bar{\Omega}^T p_{xy}) (\bar{\Omega} - \Omega)^{-1} \right)_{ij}, \\
&(F_{\bar{z}\bar{z}})_{ij} = \left(
[(\bar{\Omega} - \Omega)^{-1}]^T ({\Omega}^T p_{xx} {\Omega} + p_{yy} + p^{T}_{xy} {\Omega} - {\Omega}^T p_{xy}) (\bar{\Omega} - \Omega)^{-1}
\right)_{ij}, \\
&(F_{{z}\bar{z}})_{ij} = i \left( [(\bar{\Omega} - \Omega)^{-1}]^T (\bar{\Omega}^T p_{xx} {\Omega} + p_{yy} + p^{T}_{xy} {\Omega} - \bar{\Omega}^T p_{xy}) (\bar{\Omega} - \Omega)^{-1} \right)_{ij}. 
\end{aligned}
\end{align}
In what follows, we suppose $p_{xx} = p_{yy} = 0$ for simplicity. We then consider the Hermitian Yang-Mills equation to keep SUSY in $10$D $\mathcal{N} = 1$ super Yang-Mills theory (see Refs. \cite{Cremades:2004wa,Abe:2012ya}). That is, since $F$ has to be $(1,1)$-form,
we then find 
\begin{align}
p^{T}_{xy} {\Omega} = {\Omega}^T p_{xy}.
\end{align}
Also, since we can define the integer fluxes $N$ as $p_{xy} = 2\pi N^T$ through the Dirac quantization condition, 
we find
\begin{align}
\label{SUSY}
(N\Omega)^T = N\Omega.
\end{align}
This is called the $F$-term SUSY condition which realizes $F$-flat vacua satisfying the Hermitian Yang-Mills equation. 
In particular, when both $N^T = N$ and $\Omega^T = \Omega$ hold under Eq. \eqref{SUSY}, we find $[N, \Omega] = 0$. That is, $N$ and $\Omega$ can be diagonalized simultaneously. 
We will restrict ourselves to the symmetric matrices, 
$N^T = N$ and
$\Omega^T = \Omega$. \footnote{The following discussion holds even if we introduce the asymmetric $N$ and $\Omega$ under the $(1, 1)$-form background magnetic fluxes and the $F$-term SUSY condition $(N\Omega)^T = N\Omega$.}
We use the following 
magnetic flux background:
\begin{align}
    F=\pi[ N^T (\text{Im}\Omega)^{-1} ]_{ij}(idz^i \wedge d\bar z^j) \equiv F_{i\bar{j}} (idz^i \wedge d\bar z^j).
\end{align}
We find the corresponding gauge potential $A$ defined as $F=dA$:
\begin{align}
\begin{aligned}
A(z, \bar{z})
&= \pi \text{Im}[ (N(\bar{z} + \bar{\eta}))^T (\text{Im}\Omega)^{-1} dz ] \\ 
&= -\frac{i\pi}{2} [(N(\bar{z} + \bar{\eta}))^T (\text{Im}\Omega)^{-1}]_i dz^i 
+ \frac{i\pi}{2} [(N({z} + {\eta}))^T (\text{Im}\Omega)^{-1}]_i d\bar{z}^i \\
&\equiv A_{z^i} dz^{i} + A_{\bar{z}^{\bar{i}}} d\bar{z}^i,
\end{aligned}
\end{align}
where $\eta$ denotes the Wilson line. For simplicity, we use the vanishing Wilson line, namely $\eta = \vec{0}$.
Then the values of the gauge fields $A_{z^i}$, $A_{\bar{z}^{\bar{i}}}$ lead to the following commutation relations with Eq. \eqref{covariant}
\begin{align}
\begin{aligned}
\label{commutation1}
&[ D_{i}, \bar{D}_{j}   ] = F_{i\bar{j}}, \\
&[ D_i, D_j ] = F_{ij} = 0, \\
&[ \bar{D}_{{i}}, \bar{D}_{{j}}  ] = F_{\bar{i}\bar{j}} = 0.
\end{aligned}
\end{align}
Also, the $U(1)$ gauge background $A(z,\bar{z}) = A(z)$ satisfies the following boundary conditions:
\begin{align}
\begin{aligned}
\label{gauge}
&A(z + e_k) = A (z) + d\zeta_{e_k} (z), \\
&A(z + \Omega e_k) = A (z) + d\zeta_{\Omega e_k} (z),
\end{aligned}
\end{align}
where $k = 1,2,3$. 
Thus we obtain 
\begin{align}
\begin{aligned}
\zeta_{e_k} (z) &= \pi [N^T (\text{Im}\Omega)^{-1} \text{Im} z]_k, \\
\zeta_{\Omega e_k} (z) &= \pi \text{Im}[ (N \bar{\Omega})^T (\text{Im}\Omega)^{-1} z]_k \\
&= \pi[N (\text{Re}\Omega (\text{Im}\Omega)^{-1}\text{Im}z -\text{Re} z) ]_k ,
\end{aligned}
\end{align}
and impose boundary conditions of spinor wave functions in magnetized $T^6$
\begin{align}
\begin{aligned}
\label{boundary}
\psi (z+e_k) &=  e^{i\zeta_{e_k} (z)} \psi (z) = e^{i\pi[N^T (\text{Im}\Omega)^{-1} \text{Im}z]_k} \psi (z), \\
\psi (z+\Omega e_k) &=  e^{i\zeta_{\Omega e_k} (z)} \psi (z) = e^{i\pi[N  (\text{Re}\Omega (\text{Im}\Omega)^{-1}\text{Im}z -\text{Re} z) ]_k} \psi (z),
\end{aligned}
\end{align} 
under $(N\Omega)^T = N\Omega$.

\section{Wave functions with various chiralities in magnetized $T^6$ model}\label{section3}

Here, we study wave functions of zero modes with all chiralities on $T^6$ with magnetic flux background.

\subsection{Direct way to introduce wave functions with various chiralities}

We denote components of the spinor field by 
\begin{align}
\label{Diracspinor}
\begin{aligned}
\Psi \equiv 
\begin{bmatrix}
+ \\
- \\
\end{bmatrix} 
\otimes 
\begin{bmatrix}
+ \\
- \\
\end{bmatrix} 
\otimes 
\begin{bmatrix}
+ \\
- \\
\end{bmatrix} 
&=[\psi_{+++}, \psi_{++-}, \psi_{+-+}, \psi_{+--}, \psi_{-++}, \psi_{-+-}, \psi_{--+}, \psi_{---}]^T \\
&\equiv
[\psi^{\prime}_{1+}, \psi_{3-}, \psi_{2-}, \psi_{1+}, \psi_{1-}, \psi_{2+}, \psi_{3+}, \psi^{\prime}_{1-}]^T.
\end{aligned}
\end{align}
We define $\psi_{j+}$ and $\psi_{j-}$, $(j=1,2,3)$, as the positive and negative chirality components, respectively.
Also, we define $\psi_{j+}' = \psi_{+++}$ and  $\psi_{j-}' = \psi_{---}$. 
The Dirac equations, $i\slashed{D}_6 \Psi = 0$, are written explicitly by components as 
\begin{align}
\left\{ \,
\begin{aligned}
\label{dirac}
 D_{3} \psi_{3-} +  D_{2} \psi_{2-} +  D_{1} \psi_{1-} &= 0, \\
 \bar{D}_{3} \psi^{\prime}_{1+} -  D_{2} \psi_{1+} -  D_{1} \psi_{2+} &= 0, \\
 \bar{D}_{2} \psi^{\prime}_{1+} +  D_{3} \psi_{1+} -  D_{1} \psi_{3+} &= 0, \\
 -  \bar{D}_{2} \psi_{3-} +  \bar{D}_{3} \psi_{2-} +  D_{1} \psi^{\prime}_{1-} &= 0, \\
\bar{D}_{1} \psi^{\prime}_{1+} +  D_{3} \psi_{2+} +  D_{2} \psi_{3+} &= 0, \\
- \bar{D}_{1} \psi_{3-} +  \bar{D}_{3} \psi_{1-} -   D_{2} \psi^{\prime}_{1-} &= 0, \\
-  \bar{D}_{1} \psi_{2-} +  \bar{D}_{2} \psi_{1-} +   D_{3} \psi^{\prime}_{1-} &= 0, \\
 \bar{D}_{1} \psi_{1+} -  \bar{D}_{2} \psi_{2+} +  \bar{D}_{3} \psi_{3+} &= 0.
\end{aligned}
\right.
\end{align}
First, we briefly review the solution that only $\psi_{+++} = \psi^{\prime}_{1+}$ is non-vanishing and the others are vanishing.
For such a solution, the Dirac equations can be written by 
\begin{align}
\label{eq:DiracEq+++}
    \bar{D}_1\psi^{\prime}_{1+} =\bar{D}_2\psi^{\prime}_{1+} =\bar{D}_3\psi^{\prime}_{1+} = 0.
\end{align}
Its solution is written by \cite{Cremades:2004wa,Antoniadis:2009bg,Kikuchi:2023awm}
\begin{align}
\label{po}
\psi^{J}_{+++}  = \mathcal{N} \cdot e^{i\pi [ N z ]^T (\text{Im}\Omega)^{-1} \text{Im} z } \cdot 
\theta
\begin{bmatrix}
J^TN^{-1} \\
0 \\
\end{bmatrix}
(Nz, N\Omega),
\end{align}
where $J \in \Lambda_{N^T}$ denotes lattice vectors in lattice $\Lambda_{N^T}$ spanned by $N^T e_i$, $(i=1,2,3)$. 
Thus we find $|\det{N}|$ of independent degenerated zero-modes.
We have introduced the Riemann-Theta function with characters defined as 
\begin{align}
\label{riemann}
\theta
\begin{bmatrix}
a \\
b \\
\end{bmatrix}
(z, \Omega)
\equiv
\sum_{m \in \mathbb{Z}^3} e^{i\pi (m+a)^T \Omega (m + a)} e^{2\pi i (m+a) (z + b)}, \quad  a, b \in \mathbb{R}^3, \quad \Omega \in \mathcal{H}^3.
\end{align}
The factor $\mathcal{N}$ denotes normalization constant, 
\begin{align}
    \mathcal{N} = (\det{N})^{\frac{1}{4}} \cdot [\text{Vol} (T^6)]^{-\frac{1}{2}}=  (\det{N})^{\frac{1}{4}} \cdot [\det{(\text{Im}\Omega)}]^{-\frac{1}{2}},
\end{align}
 if $\det{\text{Im}\Omega} > 0$.
 Here, we normalize the wave function as
 \begin{align}
\int_{T^6} d^3{z} d^3{\bar{z}} (\psi^{J})^{*} \psi^{K} = [\det{(2\text{Im}\Omega)}]^{-1/2} \delta_{J,K},
\end{align}
following Ref.~\cite{Kikuchi:2023awe}.  
Note that $\psi_{+++}$ shown in Eq. \eqref{po} are well-defined only if all of three eigenvalues of $F_{i\bar j}=\pi [N^T ({\rm Im}\Omega)^{-1}]_{i j}$ are positive.
For other $F_{i\bar j}=\pi [N^T ({\rm Im}\Omega)^{-1}]_{ij}$, 
the mode $\psi_{+++}$ has no zero mode, but other modes like $\psi_{\underline{+--}}$ have zero modes, where the underline denotes all of the possible permutations.

We can also obtain the solution for only the component $\psi_{---}$ 
by flipping all chiralities in $\psi_{+++}$, we find that the sign of $\det{N}$ has to be flipped. Then $\psi_{---}$ is obtained as
\begin{align}
\psi^J_{---} = \mathcal{N} \cdot e^{i\pi [ N \bar{z} ]^T (\text{Im}\Omega)^{-1} \text{Im} z } \cdot 
\theta
\begin{bmatrix}
J^T N^{-1} \\
0 \\
\end{bmatrix}
(N\bar{z}, N\bar{\Omega}),
\end{align}
under the following replacement:
\begin{align}
\label{eq:+to-}
    z\rightarrow \bar{z}, \quad \Omega \rightarrow \bar{\Omega}, \quad {N}  \rightarrow -{N} \equiv N  .
\end{align}
Next, we study the solutions that other components are non-vanishing.
Hereafter, we focus on wave functions of $\psi_{1+} = \psi_{+--}$, $\psi_{2+} = \psi_{-+-}$, and $\psi_{3+} = \psi_{--+}$.
The others of $\psi_{1-} = \psi_{-++}$, $\psi_{2-} = \psi_{+-+}$, and $\psi_{3-} = \psi_{++-}$ 
are obtained by the above replacement (\ref{eq:+to-}).
Now, let us study the solution that only $\psi_{1+}$ is non-vanishing and the others are vanishing.
The corresponding Dirac equations are written by 
\begin{align}
\label{dirac3}
D_2 \psi_{1+} = D_3 \psi_{1+} = \bar{D}_1 \psi_{1+} = 0 .
\end{align}
Note that both $D_i$ and $\bar{D}_i$ appear in this equation, while 
only $\bar{D}_i$ appear in Eq.~(\ref{eq:DiracEq+++}).
By using Eq. \eqref{commutation1}, we find $F_{1\bar{2}} \psi_{1+} = F_{1\bar{3}} \psi_{1+} = 0$.
We can not obtain non-trivial solution in generic case including $F_{1 \bar 2},F_{1 \bar 3} \neq 0$.
When $F_{1\bar 2},F_{1 \bar 3} = 0$, the first plane $z^1$ and the other are decomposed to the factorizable torus, $T^2\times T^4$.
In order to understand the non-factorizable magnetized $T^6$, 
we study the solution that $\psi_{1+} = \psi_{+--}$, $\psi_{2+} = \psi_{-+-}$, and $\psi_{3+} = \psi_{--+}$ are excited simultaneously.
These states with different chiralities are mixed each other.
Following Ref.~\cite{Antoniadis:2009bg}, we make Ansatz for the solution that 
these components can be written by a single function $\psi$, i.e., $\psi_{i+} = \alpha_i \psi$, ($i=1,2,3$), where $\alpha_i$ are coefficients.
By use of this Ansatz, their Dirac equations can be written as 
\begin{align}
\left\{ \,
\begin{aligned}
\label{dirac4}
(\alpha_1 D_{2} + \alpha_2 D_{1})\psi &= 0, \\
(\alpha_1 D_{3} - \alpha_3 D_{1})\psi &= 0, \\
(\alpha_2 D_{3} + \alpha_3 D_{2})\psi &= 0, \\
(\alpha_1 \bar{D}_{1} - \alpha_2 \bar{D}_{2} + \alpha_3 \bar{D}_{3})\psi &= 0.
\end{aligned}
\right.
\end{align} 
Note that the first three equations are not independent of each other, because 
\begin{align}
\alpha_3(\alpha_1 D_{2} + \alpha_2 D_{1})\psi + \alpha_2(\alpha_1 D_{3} - \alpha_3 D_{1})\psi = \alpha_1 (\alpha_2 D_{3} + \alpha_3 D_{2}) \psi = 0 .
\end{align}

Suppose  $\alpha_1 \neq 0$.
Then, we use the ratios $q_{1} = \frac{\alpha_2}{\alpha_1}$, $q_{2} = -\frac{\alpha_3}{\alpha_1}$ 
and the notation $\vec{q} = (q_{1}, q_{2})^T$.\footnote{If $\alpha_1=0$, we use the ratio $q=\frac{\alpha_3}{\alpha_2}$. Then, we can carry out a similar discussion.}
Eq. \eqref{dirac4} leads to  the following integrable conditions
\begin{align}
\left\{ \,
\begin{aligned}
\label{integrable}
&F_{2\bar{1}} - q^2_{1} F_{1\bar{2}} +q_{1} (F_{1\bar{1}} - F_{2\bar{2}}) -q_2 F_{2\bar{3}} -q_1 q_2 F_{1\bar{3}} = 0, \\ 
&F_{3\bar{1}} - q_{1} F_{3\bar{2}} -q_{2} (F_{3\bar{3}} - F_{1\bar{1}}) -q_1 q_2 F_{1\bar{2}} -q^{2}_2 F_{1\bar{3}} = 0,
\end{aligned}
\right.
\end{align}
so as to obtain a non-trivial solution of $\psi$.
Since $F$ is the $(1,1)$-form, 
$F_{i\bar{j}}=  \pi  [N^T (\text{Im}\Omega)^{-1}]_{ij}$ is a real symmetric matrix.
It has six degrees of freedom, but there are two constraints due to Eq.~(\ref{integrable}).
As a result, we have four degrees of freedom in $F$.
We parametrize them by a real number $\hat{N}^{\prime}$ and $2\times 2$ real symmetric matrix $\tilde{N^{\prime}}$.
By use of them, we can write $F_{i\bar{j}}$ with the coefficients $q_i$ as
\begin{align}
\begin{aligned}
\label{flux}
&F_{i\bar{j}}
\equiv 
\pi [ \hat{N}^T (\text{Im}\Omega)^{-1}]_{ij} 
+ 
\pi [ \tilde{N}^T (\text{Im}\Omega)^{-1}]_{ij} , \\ 
& \pi [ \hat{N}^T (\text{Im}\Omega)^{-1}]_{ij} 
= 
\hat{N}^{\prime}
\begin{bmatrix}
1 & -\vec{q}^T \\
-\vec{q} & \vec{q} (\vec{q})^T \\
\end{bmatrix}
=\hat{N}^{\prime}
\begin{bmatrix}
1 & -q_1 & -q_2 \\
-q_1 & q^2_{1} & q_1 q_2 \\
-q_2 & q_1 q_2 & q^2_2 \\
\end{bmatrix},\\
&\pi [ \tilde{N}^T (\text{Im}\Omega)^{-1}]_{ij} 
=
\begin{bmatrix}
\vec{q}^T \tilde{N}^{\prime} \vec{q} & \vec{q}^T \tilde{N}^{\prime} \\
\tilde{N}^{\prime} \vec{q} & \tilde{N}^{\prime} \\
\end{bmatrix} 
\\
&=
\begin{bmatrix}
\tilde{N}^{\prime}_{11} q^2_1 + 2\tilde{N}^{\prime}_{12} q_1 q_2 + \tilde{N}^{\prime}_{22} q^2_2 &  \tilde{N}^{\prime}_{11} q_1 + \tilde{N}^{\prime}_{12} q_2  &  \tilde{N}^{\prime}_{12} q_1 + \tilde{N}^{\prime}_{22} q_2  \\
\tilde{N}^{\prime}_{11} q_1  + \tilde{N}^{\prime}_{12} q_2 & \tilde{N}^{\prime}_{11} & \tilde{N}^{\prime}_{12} \\
 \tilde{N}^{\prime}_{12} q_1 + \tilde{N}^{\prime}_{22} q_2 & \tilde{N}^{\prime}_{12} & \tilde{N}^{\prime}_{22} \\
\end{bmatrix}
,
\end{aligned}
\end{align}
such that they satisfy the condition (\ref{integrable}), 
where $N\equiv \hat{N} + \tilde{N}$.

One can rotate $F_{i \bar j}$ as 
\begin{align}
\label{part}
&F^{\text{part}}_{i\bar{j}} \equiv P^{-1}_{12} F_{i\bar{j}} P_{12} 
=\begin{bmatrix}
F^{\text{part}}_{1\bar{1}} & 0 & 0 \\
0 & F^{\text{part}}_{2\bar{2}} & F^{\text{part}}_{2\bar{3}}\\
0 & F^{\text{part}}_{2\bar{3}} & F^{\text{part}}_{3\bar{3}}\\    
\end{bmatrix}
,
\end{align}
by the following rotation:
\begin{align}
P_{12}
=
\begin{bmatrix}
\cos{\theta_2} & 0 & \sin{\theta_2} \\
0 & 1 & 0 \\
-\sin{\theta_2} & 0 & \cos{\theta_2} \\
\end{bmatrix}
\begin{bmatrix}
\cos{\theta_1} & -\sin{\theta_1} & 0 \\
\sin{\theta_1} & \cos{\theta_1} & 0 \\
0 & 0 & 1 \\
\end{bmatrix}
,
\end{align}
where 
\begin{align}
q_1 = -\frac{\tan{\theta_1}}{\cos{\theta_2}}, \quad 
q_2 = \tan{\theta_2}.
\end{align}
The explicit entries in $F^{\text{part}}_{i\bar{j}}$ are obtained as 
\begin{align}
F^{\text{part}}_{1\bar{1}}
&= [F_{1\bar{1}} + F_{2\bar{2}}q^2_1 + F_{3\bar{3}} q^2_2 
- 2 (F_{1\bar{2}} q_1 - F_{2\bar{3}} q_1 q_2 + F_{1\bar{3}} q_2) ] \cos^2{\theta_1} \cos^2{\theta_2}, \\ \notag
F^{\text{part}}_{2\bar{2}}
&= [[2(F_{1\bar{2}} - F_{2\bar{3}} q_2)q_1 + (-2F_{1\bar{3}} q_2 + F_{1\bar{1}} + F_{3\bar{3}} q^2_2) q^2_1 \cos^2{\theta_2} ] \cos^2{\theta_2} + F_{2\bar{2}} ] \cos^2{\theta_1}, \\ \notag
F^{\text{part}}_{3\bar{3}}
&= [F_{1\bar{1}} q^2_2 + F_{3\bar{3}} + 2F_{1\bar{3} 
} q_2] \cos^2{\theta_2}, \\ \notag
F^{\text{part}}_{2\bar{3}}
&= [ [2F_{1\bar{3}} + (F_{1\bar{1}} - F_{3\bar{3}}) q_2] q_1 \cos^2{\theta_2} + (F_{1\bar{2}} q_2 + F_{2\bar{3}} - F_{1\bar{3}} q_1) ] \cos{\theta_1} \cos{\theta_2}.
\end{align}
Therefore, the parameters $q_1$ and $q_2$ as well as $\alpha_i$ correspond to 
parameters of $SO(6)$ rotation, in particular the holomorphic rotation among $z^i$ ($\bar z^i$).
In this rotated basis, the solution $\psi$ corresponds to $\psi_{+--}$ if 
$F^{\text{part}}_{1\bar{1}} >0$.
If $F^{\text{part}}_{1\bar{1}} <0$, the solution $\psi$ corresponds 
to the combination of $\psi_{-+-}$ and $\psi_{--+}$.
Furthermore, we can diagonalize the bottom right $2\times 2$ matrix of 
$F^{\text{part}}_{i\bar{j}}$ by the rotation $P_{3}$ of $z^2$ and $z^3$ in the above basis, i.e.,
\begin{align}
P_{3} 
=
\begin{bmatrix}
1 & 0 & 0 \\
0 & \cos{\theta_3} & -\sin{\theta_3} \\
0 & \sin{\theta_3} & \cos{\theta_3} \\
\end{bmatrix}.
\end{align}
Thus, the flux background is written by the diagonal matrix $
(F^{\text{diag}}_{1\bar{1}}, F^{\text{diag}}_{2\bar{2}}, F^{\text{diag}}_{3\bar{3}})$,
where 
\begin{align}
F^{\text{diag}}_{1\bar{1}} &= F^{\text{part}}_{1\bar{1}}, \\ \notag
F^{\text{diag}}_{2\bar{2}} &= [[\sin{\theta_3} \tan{2\theta_3} + \cos{\theta_3}]\cos{\theta_3}]F^{\text{part}}_{2\bar{2}}
+
[[-\cos{\theta_3} \tan{2\theta_3} + \sin{\theta_3}]\sin{\theta_3}]F^{\text{part}}_{3\bar{3}},
\\ \notag
F^{\text{diag}}_{3\bar{3}} &= [[-\cos{\theta_3} \tan{2\theta_3} + \sin{\theta_3}]\sin{\theta_3}]F^{\text{part}}_{2\bar{2}}
+
[[\sin{\theta_3} \tan{2\theta_3} + \cos{\theta_3}]\cos{\theta_3}]
F^{\text{part}}_{3\bar{3}}, \\ \notag
\end{align}
and
\begin{align}
\tan{2\theta_3} &= \frac{2F^{\text{part}}_{2\bar{3}}}{F^{\text{part}}_{2\bar{2}} - F^{\text{part}}_{3\bar{3}}}, 
\end{align}
if 
$F^{\text{part}}_{2\bar{2}} \neq F^{\text{part}}_{3\bar{3}}$.
If $F^{\text{part}}_{2\bar{2}} = F^{\text{part}}_{3\bar{3}}$, we find 
\begin{align}
F^{\text{part}}_{2\bar{3}} (\cos^{2} \theta_3 - \sin^{2} \theta_3) = 0.
\end{align}
Since we consider the most general case where $F^{\text{part}}_{2\bar{3}} \neq 0$, we find $\tan^{2}{\theta_3} = 1$. 
The diagonal basis corresponds to 
\begin{align}
    w^i = (P^{-1})^{ij}z^j, 
\end{align}
where $P = P_{12}P_3$.
In this basis, the solution $\psi$ corresponds to a single component 
of $\psi_{+--}, \psi_{-+-}, \psi_{--+}$.
Note that $P_{3}$ depends on values of the flux.  Furthermore, $\theta_1$ and $\theta_2$ depend on $q_1$ and $q_2$.

By use of the above background fluxes, we can write the solutions of the Dirac equation consistent with the boundary conditions Eq. \eqref{boundary} by\footnote{The solution with these chiralities was studied for $\Omega=i {\bf 1}_n$ in Ref.~\cite{Antoniadis:2009bg}.}
\begin{align}
\label{psi}
\psi^{J}_{M,N} = \mathcal{N} \cdot f(z, \bar{z}) \cdot \hat{\Theta}(z,\bar{z}),
\end{align} 
where 
\begin{align}
\begin{aligned}
f(z, \bar{z})
&\equiv \exp{[i\pi ( (\hat{N}z)^T (\text{Im}\Omega)^{-1} \text{Im}z  - (\tilde{N}\bar{z})^{T} (\text{Im}\Omega)^{-1} \text{Im}\bar{z})]} \\ 
&= \exp{[i\pi (\hat{N}z + \tilde{N} \bar{z})^T (\text{Im}\Omega)^{-1} \text{Im}z]} \\
&=  \exp{[i\pi (N\text{Re}z +i M \text{Im}{z})^T (\text{Im}\Omega)^{-1} \text{Im}z]}, \\
\end{aligned}
\end{align}
\begin{align}
\begin{aligned}
\hat{\Theta} (z, \bar{z})
&\equiv \sum_{m\in \mathbb{Z}^3} e^{\pi i (m +  J^T N^{-1})^T (\hat{N}\Omega + \tilde{N}\bar{\Omega}) (m+ J^T N^{-1})} e^{2\pi i (m+ J^T N^{-1})^T (\hat{N} z+\tilde{N} \bar{z})} \\ 
&= \sum_{m\in \mathbb{Z}^3} e^{\pi i (m +  J^T N^{-1})^T (N \text{Re}\Omega +i M\text{Im}\Omega) (m+ J^T N^{-1})} e^{2\pi i (m+ J^T N^{-1})^T (N \text{Re}z +i M\text{Im}z)} \\ 
&=
{\theta} 
\begin{bmatrix}
J^T N^{-1} \\
0 \\
\end{bmatrix}
(N \text{Re}z +i M\text{Im}z, N \text{Re}\Omega +i M\text{Im}\Omega).
\end{aligned}
\end{align}
We have defined 
\begin{align}
\label{apparent}
    M \equiv \hat{N} - \tilde{N}.
\end{align}
Similar to $\psi_{+++}$, the normalization constant $\mathcal{N}$ in Eq.~(\ref{psi}) is written by 
\begin{align}
&\mathcal{N}
= |\det{N}|^{\frac{1}{4}} \cdot [\text{Vol} (T^6)]^{-\frac{1}{2}} =  |\det{N}|^{\frac{1}{4}} \cdot [|\det{(\text{Im}\Omega)}|]^{-\frac{1}{2}}. 
\end{align}
In addition, we can replace $|\det{N}|$ by $|\det{M}|$ in the above equation of $\mathcal{N}$, because of $|\det{N}|=|\det{M}|$.

In this paper, we impose the $F$-term SUSY condition $(N\Omega)^T = N\Omega$ in Eq. \eqref{SUSY} and the Riemann condition. From Eq. \eqref{H3}, when we consider the Riemann condition $(N \text{Re}\Omega +i M\text{Im}\Omega)^T = N \text{Re}\Omega +i M\text{Im}\Omega$ and $M\text{Im}\Omega > 0$, namely $N \text{Re}\Omega +i M\text{Im}\Omega \in \mathcal{H}^3$, the spinor wave functions shown in Eq. \eqref{psi} converge and satisfy both Dirac equation and the boundary conditions.  (See Appendices B and C.) Thus, we can obtain the wave functions with chiralities $(\underline{+--})$.

In addition, when $N=M$, the wave functions in Eq. $\eqref{psi}$ correspond to $\psi_{+++}$ in Eq. \eqref{po}. Also wave functions flipped chirality are derived from appropriate transformations such as $z\rightarrow \bar{z}$, $\Omega \rightarrow \bar{\Omega}$. 
We therefore find all spinor wave functions in $T^6$ with non-factorizable background magnetic fluxes $F$ in Eq. \eqref{flux}.

We can relate the above wave function solution with chiralities $(\underline{+--})$ 
to the wave function solution $\psi_{+++}$ by six dimensional rotation.
Suppose that 
$F^{\text{diag}}_{1\bar{1}} >0$, $F^{\text{diag}}_{2\bar{2}} <0$, and $F^{\text{diag}}_{3\bar{3}}<0$.
Then, the solution corresponds to $\psi_{+--}$ in the basis of $w^i$.
We can rotate further $w^{\prime i} = \text{Re}w^{i} + i(U\text{Im}w)^i$, where 
\begin{align}
U \equiv
\begin{bmatrix}
1 & 0 & 0 \\
0 & -1 & 0 \\
0 & 0 & -1 \\
\end{bmatrix}.
\end{align}
In the basis of $w^{\prime i}$, each of fluxes is positive, and the solution corresponds to $\psi_{+++}$.
Thus, we can relate the solution $\psi_{+--}$ for $F^{\text{diag}}_{1\bar{1}} >0$, $F^{\text{diag}}_{2\bar{2}} <0$, and $F^{\text{diag}}_{3\bar{3}}<0$ to the solution 
$\psi_{+++}$ for 
$F^{\text{diag}}_{1\bar{1}} >0$, $F^{\text{diag}}_{2\bar{2}} >0$, and $F^{\text{diag}}_{3\bar{3}}>0$.
In generic signs of the fluxes, we use the following $U$ matrix:
\begin{align}
U \equiv
\begin{bmatrix}
{\rm sgn}(F^{\text{diag}}_{1\bar{1}}) & 0 & 0 \\
0 & {\rm sgn}(F^{\text{diag}}_{2\bar{2}}) & 0 \\
0 & 0 & {\rm sgn}(F^{\text{diag}}_{3\bar{3}}) \\
\end{bmatrix},
\end{align}
to relate them, where ${\rm sgn}(x)=1$ and $-1$ for $x>0$ and $x<0$, respectively.
By use of the original coordinate $z^i$, we denote 
\begin{align}
\label{zprime}
z^{\prime} \equiv  \text{Re} z + i R \text{Im}z,
\end{align}
where $R \equiv P U P^{-1}$ with $R^2 = {\bf{1}}_3$.
In addition, we define 
\begin{align}
\begin{aligned}
\label{parity}
&{\Omega^{\prime}} \equiv \text{Re} \Omega + i R \text{Im} \Omega,
\end{aligned}
\end{align}
where $z^{\prime} = x + \Omega^{\prime} y$.
By use of them, we can relate the solution of 
$\psi^{\prime J}_{+++}(z',\Omega')$ with the chirality $(+++)$ in $z^{\prime}$ system  
to the solution $\psi^J_{M,N}(z,\Omega)$ with 
one of chiralities $(\underline{+--})$ as 
\begin{align}
\begin{aligned}
\label{positivechirality}
&\psi^{\prime J}_{+++} ({z^{\prime}}, \Omega^{\prime}) \\
&= \mathcal{N}\cdot e^{\pi i [N{z^{\prime}}]^T (\text{Im}\Omega^{\prime})^{-1} \text{Im} {z^{\prime}}} 
\cdot
\theta
\begin{bmatrix}
{J}^T N^{-1} \\
0 \\
\end{bmatrix}
(N{z^{\prime}} , N\Omega^{\prime}) \\ 
&=\mathcal{N}\cdot e^{\pi i [N \text{Re}{z} + i M \text{Im}{z}]^T (\text{Im}\Omega)^{-1} \text{Im} {z} }
\cdot
\theta
\begin{bmatrix}
J^T N^{-1} \\
0 \\
\end{bmatrix}
(N \text{Re}{z} + i M \text{Im}{z} , N\text{Re}\Omega + i M\text{Im}\Omega) \\
&= \psi^{J}_{M,N} (z,\Omega).
\end{aligned}
\end{align} 
Note that  $R = N^{-1} M$ and this denotes the $SO(3)$ rotation for $\text{Im}z$ direction.

Here we give a comment about the metric in $z^{\prime}$ system.
Specifically, we have introduced the metric Eq. \eqref{metric} in $z$ system. In real basis, we can write the metric in $z$ system, $g$, as 
\begin{align}
g 
= 
dz^i d\bar{z}^{\bar{i}}
= 
\begin{bmatrix}
dx & dy
\end{bmatrix}
\begin{bmatrix}
{\bf{1}}_3 & \text{Re} \Omega \\
\text{Re} \Omega & \Omega \bar{\Omega} \\
\end{bmatrix}
\begin{bmatrix}
dx \\ 
dy \\
\end{bmatrix}
.
\end{align}
Meanwhile, a metric in $z^{\prime}$ system, $g^{\prime}$, can be described as 
\begin{align}
g^{\prime} 
=
dz^{\prime i} d\bar{z}^{\prime \bar{i}}
=
\begin{bmatrix}
dx & dy
\end{bmatrix}
\begin{bmatrix}
{\bf{1}}_3 & \text{Re} \Omega^{\prime} \\
\text{Re} \Omega^{\prime} & \Omega^{\prime} \bar{\Omega}^{\prime} \\
\end{bmatrix}
\begin{bmatrix}
dx \\ 
dy \\
\end{bmatrix}
.
\end{align}Since we can see $\text{Re}\Omega^{\prime} = \text{Re}\Omega$, and the $SO(3)$ rotation keeps the metric invariant, we can get $\Omega\bar{\Omega} =  \Omega^{\prime} \bar{\Omega}^{\prime}$.
In what follows, we suppose $[\text{Re}\Omega, \text{Im}\Omega] = 0$, or $[\Omega, \bar{\Omega}] = 0$.
Then, we can find $[R,\Omega] = 0$ because of $[\Omega, \Omega^{\prime}] = [\Omega, \bar{\Omega}] = 0$, which leads to $\Omega\bar{\Omega} =  \Omega^{\prime} \bar{\Omega}^{\prime}$:
\begin{align}
&\Omega^{\prime} \bar{\Omega}^{\prime} = (\text{Re}\Omega)^2 + R^2(\text{Im}\Omega)^2
= \Omega\bar{\Omega}.
\end{align}
In $z^{\prime}$ system, the chirality operator $\Gamma^{7\prime}$ is defined as 
\begin{align}
\Gamma^{7\prime} \equiv \sigma^3 \otimes (-\sigma^3) \otimes (-\sigma^3) = \text{diag} [+,-,-,+,-,+,+,-]
\end{align}
by using $U = \text{diag}(+, -, -)$. This $\Gamma^{7\prime}$ is the same as $\Gamma^{7}$ in the six-dimensional theory way. 
Similar to Eq. \eqref{Gammamatrix}, we can define the Gamma matrices in $z^{\prime}$ system satisfying the Clifford algebra under the same metric $g = g^{\prime}$ and the vielbein.
Then, we will consider the wave functions with positive chirality $(+++)$ in $z^{\prime}$ system, $\psi^{\prime}_{+++}$, as solutions of the Dirac Equation described in $z^{\prime}$ system. Also, by making use of the $\psi^{\prime}_{+++}$, we will find the eigen equation of the Laplacian and the $D$-term SUSY condition in $z^{\prime}$ system.

{\subsection{Laplacian and the $D$-term SUSY condition}
We can define the Laplacian $\Delta$ in $z$ system as follows
\begin{align}
\Delta \equiv -2\sum_{j=1}^{3} \{ D_{z^j}, \bar{D}_{\bar{z}^{\bar{j}}}  \} . 
\end{align}
If all the three eigenvalues of $F_{i\bar j}= \pi [N^T ({\rm Im}\Omega)^{-1}]_{i j}$ are positive, 
the solution $\psi_{+++}$ satisfies the Dirac equation.
One can check that it also satisfies the Laplace equation:
\begin{align}
\Delta \psi_{+++} = 2 \left[ \sum_{j=1}^{3} F_{z^j \bar{z}^{\bar{j}}}  \right] \psi_{+++} = 2 \cdot (\text{tr} F_{z^i\bar{z}^{\bar{j}}} ) \psi_{+++}.
\end{align}
where the eigenvalue is non-negative, i.e., $\text{tr} F_{z^i \bar{z}^{\bar{j}}} \geq 0$. This is consistent with the positive semi-definiteness of $\Delta$ which is generally true on a compact manifold.
We can make use of $\Delta$ when we analyze mass spectrum of bosonic modes. The eigenvalue corresponds to the mass squared in scalar field on $T^6$. 
The vector modes $(A_{z^i},A_{\bar z^i})$ on $T^6$ have additional effects, 
and we can obtain their squared mass matrix $\mathcal{M}^2$ using $F_{z^i\bar{z}^{\bar{j}}} = F_{i\bar{j}}$ and $\Delta$
\begin{align}
\mathcal{M}^2 = \Delta + 4
\begin{bmatrix}
-F_{i\bar{j}} & O \\
O & F_{i\bar{j}} \\
\end{bmatrix}
.
\end{align}
Since we have found the eigenvalue of $\Delta$, we can rewrite $\mathcal{M}^2$ in the diagonal basis
\begin{align}
\text{diag}\mathcal{M}^2 
=
2 (\lambda_1 + \lambda_2 + \lambda_3) 
+
4\cdot 
\text{diag}[-\lambda_1, -\lambda_2, -\lambda_3, +\lambda_1, +\lambda_2, +\lambda_3],
\end{align}
where $\lambda_i$ denote the eigenvalues of $F_{i\bar j}=  \pi [N^T ({\rm Im}\Omega)^{-1}]_{i j}$, and note that ${\rm tr}F_{i \bar{j}}$ is the lowest eigenvalue of $\Delta$.
From the above formula, the largest eigenvalue $\lambda_I$ must be equal to summation of the remaining two eigenvalues $\lambda_J$, $\lambda_K$ to realize the massless scalar in $4$D effective field theory, namely 
\begin{align}
\lambda_I = \lambda_J + \lambda_K.
\end{align}
This corresponds to the $D$-term SUSY condition, that is, the condition to realize the D-flat solution \cite{Cremades:2004wa,Abe:2012ya}.
On the other hand, the eigen equation of $\psi_{---}$ for $\Delta$ is given by
\begin{align}
\Delta \psi_{---} = -2 \left[ \sum_{j=1}^{3}  F_{z^j \bar{z}^{\bar{j}}}  \right] \psi_{---}.
\end{align}

Next, we study the model, where two eigenvalues of $F_{i\bar j}$ are negative and the other one is positive.
The eigenvalues of the squared mass matrix for $4$D scalar $\mathcal{M}^2$ correspond to the squared $4$D scalar masses, which are independent of the complex basis given by the $SO(3)$ rotation. Then, the squared $4$D scalar masses $m^2_{4D, I}$, ($I=1,2,3$), are written as
\begin{align}
\text{diag}(m^2_{4D, 1}, m^2_{4D. 2}, m^2_{4D, 3}) =
D^{-1} \mathcal{M}^2_{z} D = D^{\prime -1} \mathcal{M}^{\prime 2}_{z^{\prime}} D^{\prime},
\end{align}
where $\mathcal{M}^2_z$, $\mathcal{M}^{\prime 2}_{z^{\prime}}$ denote the squared mass matrices written in $z$, $z^{\prime}$ basis, respectively. Also, $D$, $D^{\prime}$ denote the digonalization matrices of $\mathcal{M}^2_z$, $\mathcal{M}^{\prime 2}_{z^{\prime}}$, respectively.
We will describe $\mathcal{M}^{\prime 2}_{z^\prime}$ using only $\psi^{\prime}_{+++}$. 
From the previous review and the definition of $z^{\prime}$, the spinor in $z^{\prime}$, $\Psi^{\prime}$, can be written as
\begin{align}
\label{positivechiralityprime}
\Psi^{\prime} = [0, 0, 0, \psi^{\prime}_{+++}, 0, 0, 0, 0]^T.
\end{align} 
Then, the Dirac Equation can be described as
$D_{{\bar{z}^{\prime \bar{j}}}} \psi^{\prime}_{+++} = 0$, $(j = 1, 2, 3)$ and Eq. \eqref{positivechiralityprime} is its solution.
We describe the Laplacian $\Delta^{\prime}$ in $z^{\prime}$ system
\begin{align}
\Delta^{\prime} \equiv 
-2\sum_{j=1}^{3} \{ D_{z^{\prime j}}, \bar{D}_{\bar{z}^{\prime \bar{j}}}  \} 
\end{align}
and introduce the eigen equation of $\Delta^{\prime}$
\begin{align}
\Delta^{\prime} \psi^{\prime}_{+++} 
= 
2 \left[ \sum_{j=1}^{3} F^{\prime}_{z^j \bar{z}^{\bar{j}}}  \right] \psi^{\prime}_{+++} = 2 \cdot (\text{tr} F^{\prime}_{z^i\bar{z}^{\bar{j}}} ) \psi^{\prime}_{+++}.
\end{align}
Therefore, we can write the squared mass matrix $\mathcal{M}^{\prime 2}_{z^{\prime}}$ as
\begin{align}
\mathcal{M}^{\prime 2}_{z^{\prime}} 
=
\Delta^{\prime} + 4 
\begin{bmatrix}
-F^{\prime}_{i\bar{j}} & O \\
O & F^{\prime}_{i\bar{j}} \\
\end{bmatrix}
,
\end{align}
where $F^{\prime}_{i\bar{j}}  = F^{\prime}_{z^{\prime i} \bar{z}^{\prime \bar{j}}}$.
We finally obtain the following formula through the diagonalization
\begin{align}
D^{\prime -1} \mathcal{M}^{\prime 2}_{z^{\prime}} D^{\prime}
=
2 (\lambda^{\prime}_1 + \lambda^{\prime}_2 + \lambda^{\prime}_3) 
+
4\cdot
\text{diag}[-\lambda^{\prime}_1, -\lambda^{\prime}_2, -\lambda^{\prime}_3, +\lambda^{\prime}_1, +\lambda^{\prime}_2, +\lambda^{\prime}_3],
\end{align}
where $\lambda^{\prime}_i$ denote the eigenvalues of $F^{\prime}_{i\bar{j}}$. We therefore conclude that the $D$-term SUSY condition can be briefly described as follows
\begin{align}
\lambda^{\prime}_{I} = \lambda^{\prime}_{J} + \lambda^{\prime}_{K}, 
\end{align}
where $\lambda^{\prime}_{I}$ denotes the largest eigenvalue of $F^{\prime}_{i\bar{j}}$ and $\lambda^{\prime}_{J}, \lambda^{\prime}_{K}$ denote the other two eigenvalues. In other words, we can write the $D$-term SUSY condition using the eigenvalues of $F_{z^i\bar{z}^{\bar{j}}}$, that is,  ones with the largest absolute value $\lambda_I$, the others $\lambda_J$, $\lambda_K$
\begin{align}
|\lambda_I| = |\lambda_J| + |\lambda_K|.    
\end{align}
}

\section{Wave functions under the $Sp(6,\mathbb{Z})$ modular transformation}\label{section4}

The $T^6$ compactification has the $Sp(6,\mathbb{Z})$ modular symmetry\cite{Ding:2020zxw,Siegel:1943,Igusa:1972,Freitag:1983,Freitag:1991,Klingen:1990,Geer:2008,Fay:1973,Mumford:1984,Kikuchi:2023awe}.
The modular transformations have been used for realizing $\mathbb{Z}_N$-twist and counting zero mode numbers generation structure in magnetized orbifold models \cite{Kobayashi:2017dyu,Kikuchi:2022psj,Kikuchi:2023awm}.
Also, the modular symmetry is important to control 4D low energy effective field theory.\footnote{Recently, modular flavor symmetries have been studied intensively \cite{Feruglio:2017spp}.
(See for early works Refs.~~\cite{Kobayashi:2018vbk,Penedo:2018nmg,Criado:2018thu,Kobayashi:2018scp,Novichkov:2018ovf,Novichkov:2018nkm,deAnda:2018ecu,Okada:2018yrn,Kobayashi:2018wkl,Novichkov:2018yse} and for reviews Refs.~~\cite{Kobayashi:2023zzc,Ding:2023htn}.)} 
In what follows, we will consider behavior of wave functions
under the $Sp(6,\mathbb{Z})$ modular transformation.

According to Ref. \cite{Kikuchi:2023awm}, the $Sp(6,\mathbb{Z})$ modular symmetry comes from $6$D lattice spanned by the basis vectors $e_i$, $\Omega_{ij} e_j$, $(i,j=1,2,3)$. \footnote{The definition of the $Sp(2g, \mathbb{Z})$ symplectic modular group is described in Appendix D.} 
The generators are represented by the matrices $S$ and $T$ 
\begin{align}
S 
=
\begin{bmatrix}
O_3 & {\bf 1}_3 \\
-{\bf 1}_3 & O_3 \\
\end{bmatrix} 
,\quad
T = 
\begin{bmatrix}
{\bf 1}_3 & B \\
0_3 & {\bf 1}_3 \\
\end{bmatrix}
,
\end{align}
where the $B$ matrix corresponds to $3\times 3$ integer matrix. Then, $z$ and $\Omega$ are transformed as
\begin{align}
S: (z,\Omega) \Rightarrow (-\Omega^{-1}z, -\Omega^{-1}), 
\quad T: (z,\Omega) \Rightarrow (z, \Omega + B),
\end{align}
under the $S$ and $T$ transformation.
In what follows, we will describe how the wave functions with various chiralities in magnetized $T^6$ are transformed under the modular $S$ and $T$ transformation.

\subsection{Modular $T$ transformation}
First, let us consider the $T$ transformation under $Sp(6,\mathbb{Z})$ symmetry.
When we consider the background magnetic fluxes $F$ shown in Eqs. \eqref{F1} and \eqref{flux}, we require the following condition 
\begin{align}
(NB)^T = NB \quad \text{where} \quad (NB)_{ii} \in 2\mathbb{Z}
\end{align}
to be consistent with the relation between the wave functions and the $T$ transformation.

The wave functions are transformed as
\begin{align}
\psi^{J}_{M,N} (z, \Omega + B)
= e^{\pi i J^T N^{-1} B J} \psi^{J}_{M,N} (z, \Omega)
\end{align}
under the $T$ transformation. This can be also described as 
\begin{align}
\psi^{\prime J}_{+++} (z^{\prime}, \Omega^{\prime} + B)
= e^{\pi i J^T N^{-1} B J} \psi^{\prime J}_{+++} (z^{\prime}, \Omega^{\prime})
\end{align}
in $z^{\prime}$ system.

\subsection{Modular $S$ transformation}
Next we consider the wave functions behaviors under the $S$ transformation.
When we consider the background magnetic fluxes, 
the integer fluxes $N$ are transformed as $N^T$ under the $S$ transformation.
In the following, we consider the wave functions behaviors and suppose that both $N$ and $M$ are invariant under the $S$ transformation, that is, the $R=N^{-1}M$ is also $S$-invariant. 
We stress that since we have considered the wave functions with positive chirality $(+++)$ in $z^{\prime}$ under $R^{2} = {\bf{1}}_3$ and $[\Omega, \bar{\Omega}] = 0$, the transformation of $z^{\prime}$ and $\Omega^{\prime}$ corresponds to the original $S$ transformation 
\begin{align}
\begin{aligned}
&\Omega^{\prime} \Rightarrow -\Omega^{\prime -1} \quad \text{under S: $\Omega \rightarrow -\Omega^{-1},$} \\
&z^{\prime} \Rightarrow -\Omega^{\prime -1} z^{\prime} \quad \text{under S: $z \rightarrow -\Omega^{-1} z.$ } 
\end{aligned}
\end{align}
We conclude that the wave functions are transformed as
\begin{align}
\begin{aligned}
\psi^{\prime J}_{+++} (-\Omega^{\prime -1}z^{\prime}, -\Omega^{\prime -1}) 
=  
\frac{\sqrt{\det(-i\Omega^{\prime})}}{\sqrt{|\det{N}|}} \sum_{K \in {\Lambda}_N}
e^{2\pi i J^T N^{-1} K}\cdot
\psi^{\prime K}_{+++} (z^{\prime}, \Omega^{\prime}),
\end{aligned}
\end{align}
under the $S$ transformation in $z^{\prime}$ system.
When we analyze the zero-mode numbers in $T^{2g}/\mathbb{Z}_N$ orbifolds, we consider $(ST\cdots)$-invariant moduli such that $\Omega = -(\Omega + B)^{-1}$. Thus we derive
\begin{align}
\Omega^{\prime} = -(\Omega^{\prime} + B)^{-1}
\end{align}
and
\begin{align}
\psi^{J}_{N,R} (\Omega^{\prime} z^{\prime}, \Omega^{\prime}) = \rho_{JK}(ST\cdots) \psi^{K}_{N,R} (z^{\prime}, \Omega^{\prime})
\end{align}
through $(ST\cdots)$-transformation, where 
$\rho_{JK} (ST\cdots)$ denotes a unitary representation derived from the modular transformations and the magnetic fluxes.
We will make use of the above formula and the spinor representation matrix to count the number of zero-modes in magnetized $T^4/\mathbb{Z}_N$ and $T^6/\mathbb{Z}_{12}$ orbifolds in the Section 6.

\section{Yukawa couplings in magnetized $T^{2g}$ model}
\label{section5}
\subsection{Calculation for Yukawa couplings}
We consider the Yukawa couplings taking the chiralities into account. 
The Yukawa couplings can be described by the spinor wave functions in left-handed, right-handed, and Higgs sectors.
They may have different magnetic fluxes and chiralities. 
We are assuming 4D ${\cal N}=1$ SUSY. 
For the Higgs sector, we denote the chirality of higgsino fields.
We denote magnetic fluxes for left-handed and right-handed sectors by $N_L$ and $N_R$, respectively. 
The magnetic flux appearing in the Higgs sector must satisfy $N_H = N_L + N_R$ because of gauge invariance. 
We also use the notation, $M_L$, $M_R$, $M_H$, where $M = NR$.
According to Ref. \cite{Kikuchi:2023dow}, since we have imposed $[N, \Omega] = 0$ derived from the $F$-term SUSY condition, we can generally describe $N$ as linear combinations of the unit matrix ${\bf{1}}_3$, the moduli $\Omega$, and the inverse $\Omega^{-1}$:
\begin{align}
N = k_1 {\bf{1}}_3 + k_2 \Omega + k_3 \Omega^{-1},
\end{align}
where $k_i$, $(i=1,2,3)$, denotes complex coefficients such that $N$ can be the integer fluxes.
Then, when we consider the integer fluxes in left-handed, right-handed sectors, or $N_L$, $N_R$, we can describe them as follows:
\begin{align}
\begin{aligned}
&N_L =  k_{1L} {\bf{1}}_3 + k_{2L} \Omega + k_{3L} \Omega^{-1}, \\
&N_R =  k_{1R} {\bf{1}}_3 + k_{2R} \Omega + k_{3R} \Omega^{-1}.
\end{aligned}
\end{align}
We therefore obtain $[N_L, N_R] = 0$ under $[N, \Omega] = 0$, which means we can find different chiralities in each sector.
Similarly, we can find $[M_L, M_R] = 0$ and $[N, M] = 0$ under $[N, \Omega] = [R, \Omega] = 0$.

When we consider the wave functions in left-handed, right-handed, and Higgs sectors on the compact space, an overlap integral constructed by products of these wave functions can introduce three-point couplings, namely Yukawa couplings. 
The Yukawa couplings among modes with the same chirality $(+++)$ were studied in Ref.~\cite{Antoniadis:2009bg}.
Here, we study the Yukawa couplings among other combinations of chiralities.
We denote wave functions of the left-handed matter field, the right-handed matter one, and the Higgs field by $\psi^I_L$, ($I\in \Lambda_{N_L}$), $\psi^J_R$, ($J\in \Lambda_{N_R}$), and $\psi^K_H$, ($K\in \Lambda_{N_H}$), where all $N$ are integer matrices,
the Yukawa couplings $Y^{I,J,K}$ are described as
\begin{align}
\begin{aligned}
\label{Y1}
Y^{I, J, K}
&\equiv \sigma_{abc} g^{\prime}  
\int_{T^6} d^3 z d^3 \bar{z} ~ \psi^{I}_{L} \cdot \psi^{J}_{R} \cdot (\psi^{K}_{H})^{*},
\end{aligned}
\end{align}
where $g^{\prime}$ and $\sigma_{abc}$ denote the gauge coupling constant in higher dimensional theory and a sign defined by the product of differences of fluxes between D-branes, respectively \cite{Cremades:2004wa}.
Under Eq. \eqref{Y1}, when we focus on the product of the wave functions with various chiralities, we can analyze the following integral:
\begin{align}
\begin{aligned}
\label{C1}
&C
\equiv 
\int_{T^6} d^3 z d^3 \bar{z} ~\psi^{I}_{N_L, M_L} (z_L, \Omega_L) \cdot \psi^{J}_{N_R, M_R} (z_R, \Omega_R) \cdot (\psi^{K}_{N_H, M_H} (z_H, \Omega_H))^{*} \\
&= |-2i \text{Im}\Omega|  \\ 
&\cdot \int_{0}^{1} d^3x \int_{0}^{1} d^3 y ~\psi^{I}_{N_L, M_L} (z_L, \Omega_L) \cdot \psi^{J}_{N_R, M_R} (z_R, \Omega_R) \cdot (\psi^{K}_{N_L + N_R, M_H} (z_H, \Omega_H))^{*},
\end{aligned}
\end{align}
where we have made use of $N_H = N_L + N_R$ under the gauge symmetry. Also, we have introduced the following measure derived from the Jacobian 
\begin{align}
d^3z d^3 \bar{z} = |-2i\text{Im}\Omega| d^3x d^3y, 
\end{align}
and the identifications, i.e., $x \sim x+e$ and $y\sim y+e$. 
Here, we describe $f_{N_{\ell}}$ and $\Theta^{J}_{N_{\ell}}(z^{\prime}_{\ell},\Omega^{\prime}_{\ell})$ derived from Eq. \eqref{psi}:
\begin{align}
\begin{aligned}
f_{N_{\ell}} \equiv e^{i\pi [N_{\ell}z^{\prime}_{\ell}]^T (\text{Im}\Omega^{\prime}_{\ell})^{-1} 
\text{Im} z^{\prime}_{\ell}},
\quad 
\Theta^{J}_{N_{\ell}}(z^{\prime}_{\ell},\Omega^{\prime}_{\ell})
\equiv \theta
\begin{bmatrix}
J^T N^{-1}_{\ell} \\
0 \\
\end{bmatrix}
(N_{\ell}z^{\prime}_{\ell}, N_{\ell}\Omega^{\prime}_{\ell}),
\end{aligned}
\end{align}
where we have defined $J \in \Lambda_{N_{\ell}}$, $z^{\prime}_\ell = \text{Re}z + i R_\ell \text{Im}z$, $\Omega^{\prime}_\ell = \text{Re}\Omega + i R_\ell \text{Im}\Omega$, and $R_\ell = N^{-1}_{\ell} M_{\ell}$,  $(\ell = L, R, H)$. 
We will use the following relation in the Higgs sector: 
\begin{align}
\label{Higgstheta}
\left( \Theta^{K}_{N_H}(z^{\prime}_H,\Omega^{\prime}_H) 
\right)^{*} = \Theta^{-K}_{-N_H} (\bar{z}^{\prime}_H, \bar{\Omega}^{\prime}_H).
\end{align}
The wave functions above converge since $ - M_H \text{Im}\bar{\Omega} =
M_H \text{Im} \Omega > 0$.
Thus $C$ is proportional to
\begin{align}
\begin{aligned}
\int_{T^6} d^3 z d^3 \bar{z} ~\left[ f_{N_L} \cdot 
\Theta^{I}_{N_{L}}(z^{\prime}_{L},\Omega^{\prime}_{L}) \right]
\cdot \left[  f_{N_R} \cdot 
\Theta^{J}_{N_{R}}(z^{\prime}_{R},\Omega^{\prime}_{R})  \right]
\cdot  
\left[ {f}^{*}_{N_H} \cdot 
\Theta^{-K}_{-N_H} (\bar{z}^{\prime}_H, \bar{\Omega}^{\prime}_H) \right]
.
\end{aligned}
\end{align}

In order to evaluate Eq. \eqref{C1}, we need to calculate the following product
\begin{align}
\psi^{I}_{N_L, M_L} (z^{\prime}_L, \Omega^{\prime}_L)\cdot
\psi^{J}_{N_R, M_R} (z^{\prime}_R, \Omega^{\prime}_R),
\end{align}
where
$I = N_L \vec{i} \in \Lambda_{N_L}$ and $J = N_R \vec{j} \in \Lambda_{N_R}$.
In this subsection, we will use $\vec{i}$ and $\vec{j}$ when calculating and we can generally discuss the product regardless of the dimension of the torus. 
By use of $z= \text{Re}z +i \text{Im}z  =  x+(\text{Re}\Omega)y + i (\text{Im}\Omega)y$, the product up to the normalization constant is calculated as follows:
\begin{align}
\begin{aligned}
&\psi^{I}_{N_L, M_L} (z^{\prime}_L, \Omega^{\prime}_L)\cdot
\psi^{J}_{N_R, M_R} (z^{\prime}_R, \Omega^{\prime}_R) \\ 
&\propto e^{i\pi [N_L z^{\prime}_L]^T (\text{Im}\Omega)^{-1} \text{Im} z} \cdot 
e^{i\pi [N_R z^{\prime}_R]^T (\text{Im}\Omega)^{-1} \text{Im} z}\cdot 
\Theta^{I}_{N_L}(z^{\prime}_L,\Omega^{\prime}_L) \cdot
\Theta^{J}_{N_R}(z^{\prime}_R,\Omega^{\prime}_R) \\
&= e^{i\pi [x^T (N_L + N_R) y + y^T [(N_L + N_R)\text{Re}\Omega + i (M_L + M_R) \text{Im}\Omega]y]} \\ \cdot 
&\sum_{\vec{\ell}_1, \vec{\ell}_2 \in \mathbb{Z}^g} e^{i\pi (i) [\vec{L}^T \hat{Q}_{N, M} \vec{L} ]} \cdot e^{2\pi i [\vec{L}^T Q_{N} \vec{X}]} \cdot 
e^{2 \pi i (i) [\vec{L}^T \hat{Q}_{N, M} \vec{Y} ]}
,
\end{aligned}
\end{align}
where we have defined the $2g$-D real vectors:
\begin{align}
\label{hoge1}
\vec{L} = 
\begin{bmatrix}
 \vec{i} + \vec{\ell}_1 \\
 \vec{j} + \vec{\ell}_2 \\
\end{bmatrix}
,
\vec{X} =
\begin{bmatrix}
\vec{x} \\
\vec{x} \\
\end{bmatrix}
,
\vec{Y} =
\begin{bmatrix}
\vec{y} \\
\vec{y} \\
\end{bmatrix}
,
\end{align}
and the following $2g \times 2g$ matrices:
\begin{align}
\label{hoge2}
\begin{aligned}
&Q_{N} =
\begin{bmatrix}
N_L & O \\
O & N_R \\
\end{bmatrix}
,\\ 
&\hat{Q}_{N, M}
=
\begin{bmatrix}
M_L \text{Im} \Omega - iN_L \text{Re}\Omega & O \\
O & M_R \text{Im} \Omega - iN_R \text{Re}\Omega \\
\end{bmatrix}
\equiv
\begin{bmatrix}
M^{\prime}_L & O \\
O & M^{\prime}_R \\
\end{bmatrix}
.
\end{aligned}
\end{align}
Here, we have defined $M^{\prime} = M_L \text{Im} \Omega - iN_L \text{Re}\Omega$.
In what follows, we will focus on how to take the summation for indices $\vec{i} N_L, \vec{j} N_R, \vec{k} (N_L + N_R)$. 
In that case, note that $M^{\prime}$ are irrelavant to the index summation.
According to Ref. \cite{Antoniadis:2009bg}, we can make use of the following transformation matrix $T^{\prime}$ under $[N_L, N_R] = [M_L, M_R] = 0$, defined as 
\begin{align}
\begin{aligned}
\label{hoge3}
&T^{\prime}\equiv
\begin{bmatrix}
1_g & 1_g \\
\alpha N^{-1}_L & -\alpha N^{-1}_R \\
\end{bmatrix}
, 
(T^{\prime})^T
=
\begin{bmatrix}
1_g & N^{-1}_L \alpha^T \\
1_g & -N^{-1}_R \alpha^T \\
\end{bmatrix}
, \\
&(T^{\prime})^{-1}
=
(N^{-1}_L + N^{-1}_R)^{-1} 
\begin{bmatrix}
N^{-1}_R & \alpha^{-1} \\
N^{-1}_L & -\alpha^{-1} \\
\end{bmatrix}
, \\
&((T^{\prime})^{-1})^T 
=
(N^{-1}_L + N^{-1}_R)^{-1} 
\begin{bmatrix}
N^{-1}_R & N^{-1}_L\\
(\alpha^{-1})^{T}  & -(\alpha^{-1})^{T} \\
\end{bmatrix}
,
\end{aligned}
\end{align}
with $g\times g$ matrix $\alpha$,  and $g\times g$ unit matrix $1_g$.
In the process of calculating series in the product $\psi^{I}_{N_L, M_L} (z^{\prime}_L, \Omega^{\prime}_L)\cdot
\psi^{J}_{N_R, M_R} (z^{\prime}_R, \Omega^{\prime}_R)$, we take $\alpha = |N_L||N_R|1_g$ when $|N_L|$ and $|N_R|$ are coprime. This comes from the condition that $\alpha (N^{-1}_L + N^{-1}_R)$ has integer values since both $N_L$ and $N_R$ are integer matrices. This condition is necessary to maintain consistency with the series. We can then choose $\alpha$ as the least common multiple of $|N_L|$ and $|N_R|$ when they are not coprime. 
In general, we can choose the following $\alpha$
\begin{align}
\alpha
=
\left\{ \,
    \begin{aligned}
    & |N_L||N_R|1_g \quad \text{for $|N_L|$, $|N_R|$ coprime},\\
    & \text{L.C.M.}(|N_L|, |N_R|) 1_g \quad \text{for the others}.
    \end{aligned}
\right.
\end{align}
After the calculation for the Yukawa couplings \footnote{The detail of the calculation for the Yukawa couplings is described in Appendix E.}, when $\alpha
=  |N_L||N_R|1_g$ and $g = 3$, we can find the integral part of the Yukawa couplings in magnetized $T^6$ through the Riemann-theta function with character:
\begin{align}
\begin{aligned}
\label{CYukawa}
C
&=|-2i \text{Im}\Omega|\cdot \mathcal{N}_i \mathcal{N}_j \mathcal{N}_k \cdot \\
&\int_{0}^{1}  d^3 {y} [ e^{-\pi [y^T [M_L + M_R + M_H ] (\text{Im}\Omega) y  ]} \cdot \\ &\sum_{\vec{\ell_3}, \vec{\ell_4} \in \mathbb{Z}^3}  \sum_{\vec{p}, \vec{\tilde{p}}}  e^{\pi i (i) [\vec{\bf{K}} + \vec{\bf{L_1}}]^T \cdot {\bf{\hat{Q^{\prime}  }  } }^{|N_L||N_R|}_{N, M^{\prime}} \cdot    [\vec{\bf{K}} + \vec{\bf{L_1}}] } \cdot e^{2\pi i (i)  [\vec{\bf{K}} + \vec{\bf{L_1}}]^T \Vec{\bf{{Y}^{\prime}}}^{|N_L||N_R|}_{N, M^{\prime}}   }    ]   \\
&= |-2i \text{Im}\Omega|\cdot \mathcal{N}_i \mathcal{N}_j \mathcal{N}_k \cdot \\ 
&\sum_{\vec{p}, \vec{\tilde{p}}} 
\int_{0}^{1} { d^3 {y} \left[  e^{-\pi [y^T [M_L + M_R + M_H ] (\text{Im}\Omega) y ]} \cdot 
\theta
\begin{bmatrix}
\vec{\bf{K}} \\
0 \\
\end{bmatrix}
(i\Vec{\bf{{Y}^{\prime}}}^{|N_L||N_R|}_{N, M^{\prime}}  , i {\bf{\hat{Q^{\prime}  }  } }^{|N_L| |N_R|}_{N, M^{\prime} })
  \right]  },
\end{aligned}
\end{align}
where
\begin{align}
\begin{aligned}
&\vec{\bf{L_1}} = 
\begin{bmatrix}
\vec{\ell_3} \\
\vec{\ell_4} \\
\end{bmatrix}
,
\quad
\vec{\bf{K}}
=
\begin{bmatrix}
\vec{k} \\
\frac{N_R (N_L + N_R)^{-1} N_L}{|N_L||N_R|} (\vec{i} - \vec{j} + \vec{\tilde{m}}) 
\end{bmatrix}
,
\end{aligned}
\end{align}
\begin{align}
\begin{aligned}
\vec{\bf{{Y}^{\prime}}}^{|N_L||N_R|}_{N, M^{\prime}} 
&\equiv
\begin{bmatrix}
[M^{\prime}_L + M^{\prime}_R +(M^{\prime}_H)^{*}] {y} \\
[(|N_L||N_R|) (M^{\prime}_L N^{-1}_{L} - M^{\prime}_R N^{-1}_{R} )] y
\end{bmatrix}
\\ \notag
&=
\begin{bmatrix}
[(M_L + M_R + M_H) \text{Im} \Omega] {y} \\
|N_L||N_R| [(R_L - R_R)  \text{Im} \Omega ] y
\end{bmatrix}
,
\end{aligned}
\end{align}
and
\begin{align}
\begin{aligned}
&{\bf{\hat{Q^{\prime}  }  } }^{|N_L| |N_R|}_{N, M^{\prime}} \\
&\equiv 
\begin{bmatrix}
M^{\prime}_L + M^{\prime}_R +(M^{\prime}_H)^{*} & |{N_L}| |{N_R}| (M^{\prime}_L N^{-1}_L - M^{\prime}_R N_R^{-1})  \\
|{N_L}| |{N_R}|  (N^{-1}_L M^{\prime}_L  - N^{-1}_R M^{\prime}_R ) & (|{N_L}| |{N_R}|)^2 (N^{-1}_L M^{\prime}_L N^{-1}_L  + N^{-1}_R M^{\prime}_R N^{-1}_R )
\end{bmatrix}
,
\end{aligned}
\end{align}
where $\text{Re} {\bf{\hat{Q^{\prime}  }  } }^{|N_L| |N_R|}_{N, M^{\prime}} > 0$ and with the values of $\alpha = |N_L| |N_R| 1_g$, $M^{\prime}_L = M_L \text{Im} \Omega - iN_L \text{Re}\Omega$, $M^{\prime}_R = M_R \text{Im} \Omega - iN_R \text{Re}\Omega$, and $(M^{\prime}_H)^{*} = M_H \text{Im}\Omega + i (N_L + N_R)\text{Re}\Omega$. 
 Note that when we assign the integer fluxes $N$, complex structure moduli $\Omega$, and chirality, we can also apply $M=NR$ in numerical computations.
We have replaced $\vec{m}$ satisfying $\vec{m} N_L (N_L + N_R)^{-1} \in \mathbb{Z}$ with $\vec{\tilde{m}} + \vec{p} |{N_L}| (N_L + N_R) N^{-1}_L + \vec{\tilde{p}} |{N_R}| (N_L + N_R) N^{-1}_R$, where $\vec{\tilde{m}}$ is the solution for $(N_L + N_R) \vec{k} = N_L \vec{i} + N_R \vec{j} + N_L \vec{\tilde{m}}$, and  $\vec{p} \in \Lambda_{|{N_R}| N^{-1}_R }, \vec{\tilde{p}} \in \Lambda_{|{N_L}| N^{-1}_L }$. 
We have also used $[N_L, N_R] = 0$.

On the other hand, when we take $\alpha = \text{L.C.M.} (|N_L|, |N_R|) 1_g \equiv L_R 1_g$ and $g=3$ under the condition that $|N_L|$ and $|N_R|$ are not coprime, we obtain the integral part of the Yukawa couplings as follows
\begin{align}
\label{lcmYukawa}
\begin{aligned}
C
&=|-2i \text{Im}\Omega|\cdot \mathcal{N}_i \mathcal{N}_j \mathcal{N}_k \\
&\cdot \int_{0}^{1}  d^3 {y} [ e^{-\pi [y^T [M_L + M_R + M_H ] (\text{Im}\Omega) y  ]} \cdot \\ &\sum_{\vec{\ell_3}, \vec{\ell_4} \in \mathbb{Z}^3}  \sum_{\vec{p}, \vec{\tilde{p}}}  e^{\pi i (i) [\vec{\bf{K}} + \vec{\bf{L_1}}]^T \cdot {\bf{\hat{Q^{\prime}  }  } }^{L_R}_{N, M^{\prime} } \cdot    [\vec{\bf{K}} + \vec{\bf{L_1}}] } \cdot e^{2\pi i (i)  [\vec{\bf{K}} + \vec{\bf{L_1}}]^T \vec{\bf{{Y}^{\prime}}}^{L_R}_{N, M^{\prime}}   }    ]   \\
&= |-2i \text{Im}\Omega|\cdot \mathcal{N}_i \mathcal{N}_j \mathcal{N}_k \\ 
&\cdot \sum_{\vec{p}, \vec{\tilde{p}}} 
\int_{0}^{1} { d^3 {y} \left[  e^{-\pi [y^T [M_L + M_R + M_H ] (\text{Im}\Omega) y ]} \cdot 
\theta
\begin{bmatrix}
\vec{\bf{K}} \\
0 \\
\end{bmatrix}
(i\vec{\bf{{Y}^{\prime}}}^{L_R}_{N, M^{\prime}}  , i {\bf{\hat{Q^{\prime}  }  } }^{L_R}_{N, M^{\prime} })
  \right]  },
\end{aligned}
\end{align}
where
\begin{align}
\begin{aligned}
&\vec{\bf{L_1}} = 
\begin{bmatrix}
\vec{\ell_3} \\
\vec{\ell_4} \\
\end{bmatrix}
,
\quad
\vec{\bf{K}}
=
\begin{bmatrix}
\vec{k} \\
\frac{N_R (N_L + N_R)^{-1} N_L}{L_R } (\vec{i} - \vec{j} + \vec{\tilde{m}}) 
\end{bmatrix}
,
\end{aligned}
\end{align}
and
\begin{align}
\begin{aligned}
\vec{\bf{{Y}^{\prime }}}^{L_R}_{N, M^{\prime}} 
\equiv
\begin{bmatrix}
[M^{\prime}_L + M^{\prime}_R +(M^{\prime}_H)^{*}] {y} \\
[L_R (M^{\prime}_L N^{-1}_{L} - M^{\prime}_R N^{-1}_{R} )] y
\end{bmatrix}
=
\begin{bmatrix}
[(M_L + M_R + M_H) \text{Im} \Omega] {y} \\
L_R  [(R_L - R_R)  \text{Im} \Omega ] y
\end{bmatrix}
\end{aligned}
,
\end{align}
\begin{align}
\begin{aligned}
{\bf{\hat{Q^{\prime}  }  } }^{L_R}_{N, M^{\prime}} 
\equiv 
\begin{bmatrix}
M^{\prime}_L + M^{\prime}_R +(M^{\prime}_H)^{*} & L_R  (M^{\prime}_L N^{-1}_L - M^{\prime}_R N_R^{-1})  \\
L_R  (N^{-1}_L M^{\prime}_L  - N^{-1}_R M^{\prime}_R ) & (L_R)^2 (N^{-1}_L M^{\prime}_L N^{-1}_L  + N^{-1}_R M^{\prime}_R N^{-1}_R )
\end{bmatrix}
\end{aligned}
,
\end{align}
where $\text{Re} {\bf{\hat{Q^{\prime}  }  } }^{L_R}_{N, M^{\prime}} > 0$,
with the values of $M^{\prime}_L = M_L \text{Im} \Omega - iN_L \text{Re}\Omega$, $M^{\prime}_R = M_R \text{Im} \Omega - iN_R \text{Re}\Omega$, and $(M^{\prime}_H)^{*} = M_H \text{Im}\Omega + i (N_L + N_R)\text{Re}\Omega$, similar to the calculations under $\alpha = |N_L||N_R| 1_g$. Also, we have replaced $\vec{m}$ satisfying $\vec{m} N_L (N_L + N_R)^{-1} \in \mathbb{Z}$ with $\vec{\tilde{m}} + \vec{p} (N_L + N_R) N^{-1}_L + \vec{\tilde{p}} (N_L + N_R) N^{-1}_R$, where $\vec{\tilde{m}}$ is the solution for $(N_L + N_R) \vec{k} = N_L \vec{i} + N_R \vec{j} + N_L \vec{\tilde{m}}$, and  $\vec{p} \in \Lambda_{L_R N^{-1}_R }, \vec{\tilde{p}} \in \Lambda_{L_R N^{-1}_L }$.

In general, when we impose
\begin{align}
\begin{aligned}
\label{condition1}
&N^T = N, M^T = M, \Omega^T = \Omega, \\
&R = N^{-1}M = P U P^{-1} = PU P^T = R^T = R^{-1},\\ &[N,\Omega] = [N_L, N_R] = [R, \Omega] =[N, R] = [\Omega, \bar{\Omega}] = 0, \\
&N_H = N_L + N_R, (M\text{Im}\Omega)^T = M\text{Im}\Omega > 0,\\
&N_L \vec{i} + N_R \vec{j} + N_L \vec{\tilde{m}} = (N_L + N_R) \vec{k}, \\
&\vec{p} \in \Lambda_{L_R N^{-1}_R}, \vec{\tilde{p}} \in \Lambda_{L_R N^{-1}_L}, \\
&L_R = \text{L.C.M.} (|N_L|, |N_R|), \mathcal{N}_i = |N|^{1/4} ||\text{Im}\Omega||^{-1/2},
\end{aligned}
\end{align}
we obtain the $C$
\begin{align}
\label{ExplicitC}
C 
= 
&\frac{|-2i\text{Im} \Omega| \cdot \mathcal{N}_i\mathcal{N}_j \mathcal{N}_k}{\sqrt{|(M_L + M_R + M_H) \text{Im}\Omega|}}  
\cdot \sum_{\vec{p}, \vec{\tilde{p}}} 
\theta
\begin{bmatrix}
 \frac{N_R (N_L + N_R)^{-1} N_L}{L_R} (\vec{i}-\vec{j}+\vec{\tilde{m}}) \\
 \vec{0} \\
\end{bmatrix}
(\vec{0}, \tilde{\Omega}_{Y} ) \\ \notag
&\cdot \delta_{ (N_L + N_R) \vec{k}, N_L \vec{i} + N_R \vec{j} + N_L \vec{\tilde{m}}}
,
\end{align}
where 
\begin{align}
\tilde{\Omega}_Y 
\equiv 
&(L_R)^2 [(N^{-1}_L + N^{-1}_R)\text{Re}\Omega \\ \notag
&+ i[(R_L N^{-1}_L + R_R N^{-1}_R ) - (R_L - R_R) (M_L + M_R + M_H)^{-1} (R_L - R_R)]\text{Im}\Omega].
\end{align}
We prove Eq. \eqref{ExplicitC} in the Appendix E.2.
We can also see the modular $T$ transformation of the C as follows:
\begin{align}
T : C \Rightarrow e^{i\pi (\vec{i}-\vec{j}+\vec{\tilde{m}})^T B N_R (N_L + N_R)^{-1} N_L (\vec{i}-\vec{j}+\vec{\tilde{m}})} \cdot C,   
\end{align}
where $B$ satisfies with $(L_R)^2 [(N^{-1}_L + N^{-1}_R)B]_{ii} \in 2\mathbb{Z}$
under $T: \Omega \Rightarrow \Omega + B$.
We therefore may realize mass hierarchy, mixing angle in both lepton and quark sectors through such Yukawa couplings.
In the future, we would study numerical computations for the Yukawa couplings and verify whether realistic models can be realized or not.

\section{Zero mode numbers in magnetized $T^{2g}/\mathbb{Z}_N$}\label{section6}
Here, we study the number of zero modes in magnetized $T^4/\mathbb{Z}_N$ and $T^6/\mathbb{Z}_N$ orbifold models. (See Refs.~\cite{Markushevich:1986za,Ibanez:1987pj,Katsuki:1989bf,Kobayashi:1991rp,Lust:2005dy,Lust:2006zg} for various 6D Lie lattices to construct $T^6$ and its orbifolds.)

\subsection{Reduction from $T^6$ to $T^4$}
We will analyze the zero-mode number of wave functions with negative chirality in magnetized $T^4/\mathbb{Z}_N$.
We can discuss the number in $T^4$ from the discussion on magnetized $T^6$.
We have represented the Dirac equation Eq. \eqref{dirac4}, the integrable conditions shown in Eq. \eqref{integrable}, and the background magnetic fluxes $F^{(T^6)}_{z\bar{z}}$ in Eq. \eqref{flux}.

As we have discussed in Section 3, we find the Dirac equation, the magnetic fluxes, and the integrable conditions on $T^6$. In the following, we will introduce the magnetic fluxes on $T^4$. 
In order to realize magnetized $T^4 \times T^2$, when we take $\tilde{N}^{\prime}_{12} = 0$ and $q_2 = 0$, or $\theta_2 =0$,  we find the following formulae
\begin{align}
\label{part1}
F^{\text{part}}_{1\bar{1}}
&=[
F_{1\bar{1}} + F_{2\bar{2}}q^2_1 
- 2 F_{1\bar{2}} q_1  ] \cos^2{\theta_1} 
=
\frac{\hat{N}^{\prime}}{\cos^2{\theta_1}},
\\ \notag
F^{\text{part}}_{2\bar{2}}
&= [2F_{1\bar{2}}q_1 +   F_{1\bar{1}}  q^2_1  + F_{2\bar{2}} ] \cos^2{\theta_1} 
=
\frac{\tilde{N}^{\prime}_{11}}{\cos^2{\theta_1}},
\\ \notag
F^{\text{part}}_{3\bar{3}}
&= F_{3\bar{3}} 
= \tilde{N}^{\prime}_{22}, \\ \notag
F^{\text{part}}_{2\bar{3}}
&= [ F_{1\bar{3}}  q_1 +  F_{2\bar{3}}  ] \cos{\theta_1}
= \frac{\tilde{N}^{\prime}_{12}}{\cos{\theta_1}} = 0.
\end{align}
Note that $q_1 = -\tan{\theta_1}$ under $q_2 = 0$.
Next we consider $F_{i\bar{j}} = P F^{\text{part}}_{i\bar{j}} P^{-1}$ using Eq. \eqref{part1}. We can obtain the following fluxes
\begin{align}
\begin{aligned}
F_{i\bar{j}} 
&= 
P F^{\text{part}}_{i\bar{j}} P^{-1}
\\ \notag
&=
\begin{bmatrix}
\cos{\theta_1} & -\sin{\theta_1} & 0 \\
\sin{\theta_1} & \cos{\theta_1} & 0 \\
0 & 0 & 1 \\
\end{bmatrix}
\begin{bmatrix}
\frac{\hat{N}^{\prime}}{\cos^2{\theta_1}} & 0 & 0 \\
0 & \frac{\tilde{N}^{\prime}_{11}}{\cos^2{\theta_1}} & \frac{\tilde{N}^{\prime}_{12}}{\cos{\theta_1}}\\
0 & \frac{\tilde{N}^{\prime}_{12}}{\cos{\theta_1}} & \tilde{N}^{\prime}_{22} \\    
\end{bmatrix}
\begin{bmatrix}
\cos{\theta_1} & \sin{\theta_1} & 0 \\
-\sin{\theta_1} & \cos{\theta_1} & 0 \\
0 & 0 & 1 \\
\end{bmatrix}
\\ \notag
&=
\begin{bmatrix}
\hat{N}^{\prime} + \tilde{N}^{\prime}_{11} \tan^2{\theta_1} 
& 
\hat{N}^{\prime} \tan{\theta_1} - \tilde{N}^{\prime}_{11} \tan{\theta_1} 
& 
0 
\\
\hat{N}^{\prime} \tan{\theta_1} - \tilde{N}^{\prime}_{11} \tan{\theta_1}
& 
\hat{N}^{\prime} \tan^2{\theta_1} + \tilde{N}^{\prime}_{11} 
& 
0
\\
0
&
0
&
\tilde{N}^{\prime}_{22} 
\\    
\end{bmatrix}
\\ \notag
&=
\begin{bmatrix}
\hat{N}^{\prime} + \tilde{N}^{\prime}_{11} q^2_1
& 
-\hat{N}^{\prime} q_1 + \tilde{N}^{\prime}_{11} q_1 
& 
0
\\
-\hat{N}^{\prime} q_1 + \tilde{N}^{\prime}_{11} q_1
& 
\hat{N}^{\prime} q^2_1 + \tilde{N}^{\prime}_{11} 
& 
0
\\
0
&
0
&
\tilde{N}^{\prime}_{22} 
\\    
\end{bmatrix}
.
\end{aligned}
\end{align}
In particular, when we focus on $(1,1)$, $(1,2)$, $(2,1)$, and $(2,2)$ components of the above $F_{i\bar{j}}$, we can realize a reduction from $T^6$ to $T^4$. 
That is,
we can find the following magnetic fluxes on $T^4$ with $q_1 = -\tan{\theta_1}$, $q_2 = 0$, $\hat{N}^{\prime}$, and $\tilde{N}^{\prime}_{11}$:
\begin{align}
\begin{aligned}
\label{fluxesT4}
F^{T^4}_{i\bar{j}}
=\hat{N}^{\prime}
\begin{bmatrix}
1 & -q_1 \\
-q_1 & q^2_1 \\
\end{bmatrix}
+
\tilde{N}^{\prime}_{11}
\begin{bmatrix}
 q^2_1  &  q_1  \\
 q_1 & 1 \\ 
\end{bmatrix}
.
\end{aligned}
\end{align}
According to the fluxes in Eq. \eqref{fluxesT4}, spinors $\psi_{+-}$ and $\psi_{-+}$ on magnetized $T^4$ are excited simultaneously when we consider the nonzero off-diagonal components of the fluxes.
We will make use of the fluxes to analyze the number of wave functions with negative chirality on magnetized $T^4/\mathbb{Z}_N$ twisted orbifolds.

\subsection{Magnetized $T^4/\mathbb{Z}_N$ orbifold}

First we review the $T^4/\mathbb{Z}_N$ twisted orbifold and then analyze the number of zero-modes with negative chirality $(+, -)$, $(-, +)$. In the $T^4$ torus, we can introduce complex structure moduli $\Omega$ and magnetic fluxes $N$ as the following $2\times 2$ symmetric complex or integer matrices, respectively
\begin{align}
\Omega
=
\begin{bmatrix}
\omega_{11} & \omega_{12} \\
\omega_{12} & \omega_{22} \\
\end{bmatrix}
,\quad
N = 
\begin{bmatrix}
n_{11} & n_{12} \\
n_{12} & n_{22} \\
\end{bmatrix}
.
\end{align}
We then introduce the coordinates $\vec{z}=\vec{x} + \Omega \vec{y}$ ($\vec{x}, \vec{y} \in \mathbb{R}^2$). By introducing specific complex structures $\Omega$, we define the $T^4/\mathbb{Z}_N$ twisted orbifold.
According to Ref.~\cite{Kikuchi:2022psj}, the solutions to the Dirac equation are of the form 
\begin{align}
	\Psi(\vec{z},\vec{\bar{z}})=\begin{pmatrix}
		\psi_{++}\\\psi_{+-}\\\psi_{-+}\\\psi_{--}
	\end{pmatrix}
, 
\end{align}
where the four components are expressed as linear combinations in the following way
\begin{align}
\psi=\sum_{\vec{J}\in\Lambda_N}a_{\vec{J}}\cdot\psi^{\vec{J}}_N,\ \ \ a_{\vec{J}}\in\mathbb{C}.
\end{align}
$T^4/\mathbb{Z}_N$ twisted orbifolds are identified by $\mathbb{Z}_N$-twist. In other words, we can introduce the following identification for $\vec{z}$ on $T^4$ to construct $T^4/\mathbb{Z}_N$ orbifold:
\begin{align}
	\vec{z}\sim \Omega_{twist}\vec{z},
\end{align}
where $\Omega^N_{twist}$ represents the $\mathbb{Z}_N$-twist. 
It is taken to be a $2\times 2$ unitary matrix satisfying $\Omega^N_{twist}={\bf 1}_2$. 
We define the boundary condition for the spinor under a $\mathbb{Z}_N$-twist as
\begin{align}
\label{twistBC}
\Psi(\Omega_{twist}\vec{z},
\bar{\Omega}_{twist}\vec{\bar{z}}
 )=\mathcal{S}V(\vec{z},\vec{\bar{z}})\Psi(\vec{z},\vec{\bar{z}}), 
\end{align}
where $\mathcal{S}$ is the spinor representation of the twist and $V(\vec{z},\vec{\bar{z}})$ is the transformation function. Since $\mathcal{S}$ is chosen to leave $\psi_{++}$ invariant, the spinor representation is described as:
\begin{align}
	\mathcal{S}=\text{diag}\left(1,e^{2\pi ik_1/N},e^{2\pi ik_2/N},e^{-2\pi i(k_1+k_2)/N}\right),
    \label{spinor}
\end{align}
for the following factorizable twist with integers $k_1$, $k_2$
\begin{align}
\label{ZNtwist}
	\begin{pmatrix}
		z_1\\z_2
	\end{pmatrix}\rightarrow
	\begin{pmatrix}
		e^{2\pi ik_1/N}z_1\\e^{2\pi ik_2/N}z_2
	\end{pmatrix}.
\end{align}
The transformation function is written as
\begin{align}
	V(\vec{z},\vec{\bar{z}}) = e^{2\pi i\beta}.
\end{align}
Since we consider the identification for a $\mathbb{Z}_N$-twist, we find the following condition
\begin{align}
	\beta N\equiv 0 \mod{1}.
\end{align}
Taking $\beta=\frac{1}{N}$, we obtain $\mathbb{Z}_N$ sectors with the $\mathbb{Z}_N$ charges $e^{2\pi i \rho/N}$, $(\rho=0,1,...,N-1)$. 
We can use the modular transformations to define the transformation matrix $\rho$ and find the spectrum.
Then, we will make use of the $Sp(4,\mathbb{Z})$ generators $S$ and $T_i$, $(i=1,2,3)$ when we construct lattices of $T^4/\mathbb{Z}_N$, $(N=3,4,6)$.

\subsection{Parity transformations}
Under parity transformation, a two component spinor in Minkowski space transforms as:
\begin{align}
	\begin{pmatrix}\psi_1\\\psi_2\end{pmatrix}\rightarrow\begin{pmatrix}0&1\\1&0\end{pmatrix}\begin{pmatrix}\psi_1\\\psi_2\end{pmatrix}=\begin{pmatrix}\psi_2\\\psi_1\end{pmatrix},
\end{align}
which flips the chiralities. We would like to do the same for the 4D spinor on the torus. In order to flip chiralities, we need to parity transform one of the 2D spinors it is constructed from. However, since the flux is not diagonal, we must first diagonalize it. We express the flux in a new coordinate system defined as $\vec{w}=P\vec{z}$ as
\begin{align}
\begin{aligned}
	F&=\pi\left[N^T(\operatorname{Im}\Omega)^{-1}\right]_{i\bar{j}}(idz^i\wedge d\bar{z}^j)\\
	&=\pi\begin{pmatrix}x&z\\z&y\end{pmatrix}_{i\bar{j}}(idz^i\wedge d\bar{z}^j).
\end{aligned}
\end{align}
When the negative chirality wavefunction is non-zero, the determinant is negative, so we have $xy-z^2<0$. Let the matrix $P$ diagonalize the flux. The eigenvalues are
\begin{align}
\lambda_{1}, \lambda_2&=\frac12\left[x+y\pm\sqrt{(x+y)^2-4(xy-z^2)}\right].
\end{align}
We know that the flux is not positive definite in this case, so there is at least one negative eigenvalue. Since the determinant is negative, the square root, which is positive, is larger than the trace. Thus, we conclude $\lambda_1>0$ and $\lambda_2<0$. This gives us a flux
\begin{align}
	F=\pi\begin{pmatrix}|\lambda_1|&\\&-|\lambda_2|\end{pmatrix}_{i\bar{j}}(idw^i\wedge d\bar{w}^j).
\end{align}
If we think of the $T^4$ spinor as a product of two spinors on $T^2$, we can parity transform one of them to change the overall chirality. This is equivalent to changing to $w^2\rightarrow\bar{w}^2$. The transformation matrix for the flux is thus
\begin{align}
	U=\begin{pmatrix}1&0\\0&-1\end{pmatrix},
\end{align}
and the flux in the new coordinate system $w'$is
\begin{align}
	F=\pi\begin{pmatrix}|\lambda_1|&\\&|\lambda_2|\end{pmatrix}_{i\bar{j}}(idw'^i\wedge d\bar{w'}^j).
\end{align}
We then rotate back into the $z$ coordinates
\begin{align}
	F=\pi\left[P\begin{pmatrix}|\lambda_1|&\\&|\lambda_2|\end{pmatrix}P^T\right]_{i\bar{j}}(idz'^i\wedge d\bar{z'}^j).
\end{align}
The whole procedure is defined as by $R=PUP^T$, which gives the flux in the new coordinates as
\begin{align}
	F=\pi[N^T(R\operatorname{Im}\Omega)^{-1}]_{ij}(idz'^i\wedge d\bar{z}'^j),
\end{align}
where
\begin{align}
\begin{aligned}\label{R_transformations}
\vec{z}'&=\operatorname{Re}\vec{z}+iR\operatorname{Im}\vec{z},\\
\Omega'&=\operatorname{Re}\Omega+iR\operatorname{Im}\Omega.
\end{aligned}
\end{align}
Since both eigenvalues are positive, we can find $N\operatorname{Im}\Omega'>0$. 
This means we can describe the spinor in $z^{\prime}$ system as follows 
\begin{align}
	\Psi'=\begin{pmatrix}
		\psi_{++}'\\
		0\\0\\0
	\end{pmatrix}.
\end{align}
The wave functions on $T^4$ are then the same as the positive chirality case:
\begin{align}
	\psi^{\vec{J}}=\mathcal{N}e^{\pi i[N\vec{z}']^T(\operatorname{Im}\Omega')^{-1}\operatorname{Im}\vec{z}'}
	\theta\begin{bmatrix}\vec{J}^T N^{-1}\\0\end{bmatrix}(N\vec{z}',N\Omega'),
\end{align}
if we change $\vec{z},\Omega$ according to Eq.\eqref{R_transformations}.
We could also consider $\lambda_2>0$ and $\lambda_1<0$. This would mean $U\rightarrow-U$, and therefore $\Omega'\rightarrow\bar{\Omega}'$, $\vec{z}'\rightarrow\bar{\vec{z}}$, which just exchanges the zero modes in each $\mathbb{Z}_N$ sector.

\subsection{Magnetized $T^4/\mathbb{Z}_2$}
We begin by analyzing the $\mathbb{Z}_2$-twist defined by the following identification
\begin{align}
	\vec{z}\sim-\vec{z}.
\end{align}
The transformation function is given by
\begin{align}
V(\vec{z},\vec{\bar{z}})=e^{2\pi i\rho/2},\qquad \rho=(0,1).
\end{align}
Thus, we have two sectors with eigenvalues $\pm1$.
Under the $\mathbb{Z}_2$-twist
\begin{align}
\begin{aligned}
\begin{pmatrix}
	z_1\\z_2
\end{pmatrix}\rightarrow
\begin{pmatrix}
	-z_1\\-z_2
\end{pmatrix}
\end{aligned}
,
\end{align}
the spinor representation $\mathcal{S}$ is given by
\begin{align}
\mathcal{S}
=\begin{pmatrix}
1\\&-1\\&&-1\\&&&1
\end{pmatrix}
.
\end{align}
The wave functions are transformed as
\begin{align}
	\psi(-z)=(-1)^{\rho+1}\psi(z).
\end{align}
In  $T^4/\mathbb{Z}_2$, the complex structure moduli $\Omega$ and the fluxes $N$ are given by 
\begin{align}
\Omega
= 
\begin{pmatrix}
\omega_{11}&\omega_{12}\\
\omega_{12}&\omega_{22} \\
\end{pmatrix}
,
\quad
N=\begin{pmatrix}
n_1&m\\m&n_2\end{pmatrix},\ n_1,n_2,m\in\mathbb{Z}
\end{align}
satisfying with the $F$-term SUSY condition $(N\Omega)^T=N\Omega$.
To compute the zero mode spectrum, let us find how the wave functions on $T^4$ transform under $\mathbb{Z}_2$.
We define the even and odd modes,
\begin{align}
\begin{aligned}
	\psi_{even}(-z)=\psi_{even}(z),\\
	\psi_{odd}(-z)=-\psi_{odd}(z),
\end{aligned}
\end{align}
for the positive chirality . 
These are constructed as a linear combination of the wave functions on $T^4$
\begin{align}
\begin{aligned}
\psi_{even}\propto \psi(z)+\psi(-z),\\
\psi_{odd}\propto \psi(z)-\psi(-z).
\end{aligned}
\end{align}
We take $\rho=0$ for even modes and $\rho=1$ for odd modes.
However, when we focus on the negative chirality, we have to add a factor $s = -1$ derived from the $\mathcal{S}$
\begin{align}
\begin{aligned}
\psi_{even}(-z)=s^{-1}\psi_{even}(z) = -\psi_{even} (z),\\
\psi_{odd}(-z)=-s^{-1}\psi_{odd}(z) = \psi_{odd} (z).
\end{aligned}
\end{align}
We can therefore obtain the spectrum for negative chirality from the positive one by switching the odd and even sectors. 
The even and odd-modes with negative chirality are transformed as
\begin{align}
\begin{aligned}
	\psi_{even}(-z)=-\psi_{even}(z),\\
	\psi_{odd}(-z)=\psi_{odd}(z),
\end{aligned}
\end{align}
which means the spectrum is interchanged comparing with the positive chirality.
\subsection{Magnetized $T^4/\mathbb{Z}_4$}
We begin with magnetized $T^4/\mathbb{Z}_4$.(See Ref. \cite{Kikuchi:2022psj}.) We can construct this orbifold by the algebraic relation $S^4={\bf{1}}_4$. The $S$-invariant $\Omega_S$ is given by
\begin{align}
\Omega_S = 
\begin{pmatrix}
i&0\\0&i
\end{pmatrix}
.
\end{align}
We get the following identification:
\begin{align}
\vec{z}\sim \Omega_S \vec{z}.
\end{align}
The transformation function is
\begin{align}
V(\vec{z},\vec{\bar{z}})&=e^{2\pi ik/4},\ \ \ \ k=(0,1,2,3),
\end{align}
according to the $\mathbb{Z}_N$-twist boundary condition.
Under the $S$,
the wave functions are transformed as
\begin{align}
\psi^{\vec{J}}_N(\Omega_S\vec{z},\Omega_S)
=\frac{1}{\sqrt{\det{N}}}\sum_{\vec{K}\in\Lambda_N}
    e^{2\pi i\vec{J}^TN^{-1}\vec{K}}\psi^{\vec{K}}_N(\vec{z},\Omega_S),
    \label{z4_spinor_trans}
\end{align}
where the $N$ is given by
\begin{align}
	N=\begin{pmatrix}
		n_1&m\\m&n_2
	\end{pmatrix},\quad n_1,n_2,m\in\mathbb{Z}.
\end{align}
The trace of the transformation matrix from eq.\eqref{z4_spinor_trans} is
\begin{align}
	\text{tr} \rho(S)&=\frac{1}{\sqrt{\det N}}\sum_{\vec{K}\in\Lambda_N}
	e^{2\pi i\vec{K}^TN^{-1}\vec{K}}.
\end{align}
We can make use of the $2g$-dimensional Landsberg-Schaar relation {\footnote{The detail of the $2g$-dimensional Landsberg-Schaar is written in Ref. \cite{Kikuchi:2023awe}.}} given by:
\begin{align}
\label{Landsberg-Schaar}
	\frac{1}{\sqrt{|\det N|}}
	\sum_{K\in\Lambda_N}e^{\pi i K^TN^{-1}BK}
	=
	\frac{e^{\pi i \left(\frac{g}{4}+\frac{n_-^N}{2}-\frac{n_-^B}{2}\right)}}{\sqrt{|\det B|}}
	\sum_{K\in\Lambda_B} e^{-\pi i K^TNB^{-1}K},
\end{align}
where $n^N_{-}$, $n^B_{-}$ denote the number of negative eigenvalues of $N$, $B$, respectively.
When we analyze the degenerated number of the wave functions in each
$\mathbb{Z}_N$ sector, $D_{k}$, the spectrum can be determined by the following formulae:
\begin{align}
\begin{aligned}
	\det N &= D_{+1}+D_{-1}+D_{+i}+D_{-i},\\
	\text{tr} \rho &= D_{+1}-D_{-1}+iD_{+i}-iD_{-i},\\
	N_{Z_2+}-N_{Z_2-}&=D_{+1}+D_{-1}-D_{+i}-D_{-i}\\
			&=\begin{cases}
				1; \text{ if }\det N\equiv1 \text{ (mod } 2),\\
				2; \text{ if }\det N\equiv0 \text{ (mod } 2), \text{ gcd }(n_i,m)\equiv 1 \text{ (mod } 2) \text{ for }i=1 \text{ or } 2,\\
				4; \text{ gcd}(n_i,m)\equiv 0 \text{ (mod } 2 \text{) for }i=1,2,
			\end{cases}
\end{aligned}
\end{align}
and for negative chirality
\begin{align}
\begin{aligned}
	|\det N| &= D_{+1}+D_{-1}+D_{+i}+D_{-i},\\
	tr\rho &= D_{+1}-D_{-1}+iD_{+i}-iD_{-i},\\
	-(N_{Z_2+}-N_{Z_2-})&=D_{+1}-D_{+i}+D_{-1}-D_{-i},\\
	&=\begin{cases}
	-1; \text{ if }\det N\equiv1 \text{ (mod } 2),\\
	-2; \text{ if }\det N\equiv0 \text{ (mod } 2), \text{ gcd }(n_i,m)\equiv 1 \text{ (mod } 2) \text{ for }i=1 \text{ or } 2,\\
	-4; \text{ gcd}(n_i,m)\equiv 0 \text{ (mod } 2 \text{) for }i=1,2.
	\end{cases}
\end{aligned}
\end{align}
Let $tr\rho=a+ib$, where $a,b\in\mathbb{Z}$. Then, we can express the four eigenvalue degeneracies as
\begin{align}
\begin{aligned}
	D_{+1} = \frac{|\det N|+\xi+2a}{4},\\
	D_{+i} = \frac{|\det N|+\xi-2a}{4},\\
	D_{+i} = \frac{|\det N|-\xi+2a}{4},\\
	D_{-i} = \frac{|\det N|-\xi-2a}{4},
\end{aligned}
\end{align}
where $\xi$ is the difference between the degeneracy number in $Z_2$. In other words, 
$\xi = 1,2,4$ and $\xi = -1,-2,-4$ 
for positive and negative chirality, respectively. 
We can find three-generation models in the following ranges.
For positive chirality: $8\leq \det N\leq 20$, however, since $\det N = 20$ is not possible, the upper limit is $16$.
For negative chirality: $-16 \leq \det N \leq -4$.
\subsubsection{Positive chirality}
We begin with the $\psi_{++}$ spinor, where we require the following positive-definite condition:
\begin{align}
	\det N>0,\quad \text{tr} N>0.
\end{align}
After we substitute $B=2I$, $\det B=4$, $g=2$, $n_-^B=0$, and $n_-^N = 0$, 
into Eq. \eqref{Landsberg-Schaar},
we find
\begin{align}
\begin{aligned}
	tr\rho_{+}(S)=\frac{i}{2}\sum_{K\in\Lambda_B}e^{-\frac{\pi i}{2}K^TNK}
			=\frac{i}{2}\left[1+(-i)^{n_1}+(-i)^{n_2}+(-1)^m(-i)^{n_1+n_2}\right].
\end{aligned}
\end{align}
\subsubsection{Negative chirality}
The $S$ transformation matrix for $\psi_{+-}$ is 
\begin{align}
\rho(S)=s^{-1}\sqrt{\det(N^{-1}\Omega'/i)}\sum_{\vec{K}\in\Lambda_N}\exp\left\{2\pi i\vec{J}^TN^{-1}\vec{K}\right\},
\end{align}
where the spinor representation factor is denoted by $s$, and $\det{R} = -1$.
In $z^{\prime}$ system, the corresponding spinor representation can be written as (See appendix H.4.1)
\begin{align}
	V\begin{pmatrix}
		i&0&0&0\\
		0&\sin^2{\phi}_0-\cos^2{\phi}_0&-\sin{\phi}_0\cos{\phi}_0&0\\
		0&-\sin{\phi}_0\cos{\phi}_0&\cos^2{\phi}_0-\sin^2{\phi}_0&0\\
		0&0&0&i
	\end{pmatrix}
.
\end{align}
We obtain
\begin{align}
\text{tr} \rho_-(S)
=\frac{1}{s\sqrt{-\det N}}\sum_{\vec{K}\in\Lambda_N}e^{2\pi i\vec{K}^TN^{-1}\vec{K}}.
\end{align}
To calculate the trace, we use the Landsberg-Schaar relation in Eq.\eqref{Landsberg-Schaar} with $B=2I$, $\det B=4$, $g=2$ and $n_+^B=0$. 
Since we have considered the negative chirality $(+-)$, $(-+)$, we have $n_-^N=1$ and get
\begin{align}
\begin{aligned}
tr\rho_-(S)	
=\frac{-1}{s\sqrt{\det B}}\sum_{\vec{K}\in\Lambda_B}e^{-\frac{\pi i}{2}K^TNK}
=(is^{-1})tr\rho_+ (S) = \text{tr}\rho_+ (S).
\end{aligned}
\end{align}

\subsubsection{Trace classification}
The traces are classified by the mod 4 and 2 structure of the flux components $n_i$, $m$ respectively, as shown in Table~\ref{trace_table_4}.
\begin{table}
    \begin{subtable}{0.5\textwidth}
    \centering
    \begin{tabular}{ | c || c | c | c | c |}
		\hline
 		\backslashbox{$n_1$}{$n_2$} & 0 & 1 & 2 & 3\\
  		\hline\hline  
 		0 & $2i$ & $1+i$ & $0$ & $-1+i$\\\hline
 		1 & $1+i$ & $1$ & $0$ & $i$\\\hline
 		2 & $0$ & $0$ & $0$ & $0$\\\hline
 		3 & $-1+i$ & $i$ & $0$ & $-1$\\\hline
	\end{tabular} 
    \caption{Traces for $m\equiv0$ (mod 2).}
    \end{subtable}%
    \begin{subtable}{0.5\textwidth}
    \centering
   	\begin{tabular}{ | c || c | c | c | c |}
		\hline
 		\backslashbox{$n_1$}{$n_2$} & 0 & 1 & 2 & 3\\
  		\hline\hline  
 		0 & $i$ & $i$ & $i$ & $i$\\\hline
 		1 & $i$ & $1+i$ & $1$ & $0$\\\hline
 		2 & $i$ & $1$ & $-i$ & $-1$\\\hline
 		3 & $i$ & $0$ & $-1$ & $-1+i$\\\hline
	\end{tabular}
    \caption{Traces for $m\equiv1$ (mod 2).} 
    \end{subtable}
    \medskip
    \caption{All possible traces categorized by the mod 4 structure of the flux components.}
    \label{trace_table_4}
\end{table}

\newgeometry{top=1.5cm}
\begin{table}
    \begin{subtable}{0.5\textwidth}
    \centering
    \begin{tabular}{ c || c c c c }
		\backslashbox{$n_1$}{$n_2$} & 0 & 1 & 2 & 3 \\
		\hline\hline
		0 & $2i$ & $1+i$ & 0 & $-1+i$ \\
		1 &  & $\times$ & $\times$ & $\times$ \\
		2 &  &  & 0 & $\times$ \\
		3 &  &  &  & $\times$
	\end{tabular}
    \caption{
    Traces for $\det N\equiv 0$ (mod 4) with\\ $m\equiv0$ (mod 2).}
    \end{subtable}
    \begin{subtable}{0.5\textwidth}
    \centering
   	\begin{tabular}{ c || c c c c }
		\backslashbox{$n_1$}{$n_2$} & 0 & 1 & 2 & 3 \\
		\hline\hline
		0 & $\times$ & $\times$ & $\times$ & $\times$ \\
		1 &  & $1+i$ & $\times$ & $\times$ \\
		2 &  &  & $\times$ & $\times$ \\
		3 &  &  &  & $-1+i$
	\end{tabular}
    \caption{Traces for $\det N\equiv 0$ (mod 4) with\\ $m\equiv1$ (mod 2).} 
    \end{subtable}
    \medskip
    \begin{subtable}{0.5\textwidth}
    \centering
    \begin{tabular}{ c || c c c c }
		\backslashbox{$n_1$}{$n_2$} & 0 & 1 & 2 & 3 \\
		\hline\hline
		0 & $\times$ & $\times$ & $\times$ & $\times$ \\
		1 &  & 1 & $\times$ & $\times$ \\
		2 &  &  & $\times$ & $\times$ \\
		3 &  &  &  & $-1$
	\end{tabular}
    \caption{Traces for $\det N\equiv 1$ (mod 4) with\\ $m\equiv0$ (mod 2).}
    \end{subtable}
    \begin{subtable}{0.5\textwidth}
    \centering
   	\begin{tabular}{ c || c c c c }
		\backslashbox{$n_1$}{$n_2$} & 0 & 1 & 2 & 3 \\
		\hline\hline
		0 & $\times$ & $\times$ & $\times$ & $\times$ \\
		1 &  & $\times$ & 1 & $\times$ \\
		2 &  &  & $\times$ & $-1$ \\
		3 &  &  &  & $\times$
	\end{tabular}
    \caption{Traces for $\det N\equiv 1$ (mod 4) with\\ $m\equiv1$ (mod 2).} 
    \end{subtable}

    \medskip

    \begin{subtable}{0.5\textwidth}
    \centering
    \begin{tabular}{ c || c c c c }
		\backslashbox{$n_1$}{$n_2$} & 0 & 1 & 2 & 3 \\
		\hline\hline
		0 & $\times$ & $\times$ & $\times$ & $\times$ \\
		1 &  & $\times$ & 0 & $\times$ \\
		2 &  &  & $\times$ & 0 \\
		3 &  &  &  & $\times$
	\end{tabular}
    \caption{Traces for $\det N\equiv 2$ (mod 4) with\\ $m\equiv0$ (mod 2).}
    \end{subtable}
    \begin{subtable}{0.5\textwidth}
    \centering
   	\begin{tabular}{ c || c c c c }
		\backslashbox{$n_1$}{$n_2$} & 0 & 1 & 2 & 3 \\
		\hline\hline
		0 & $\times$ & $\times$ & $\times$ & $\times$ \\
		1 &  & $\times$ & $\times$ & 0 \\
		2 &  &  & $\times$ & $\times$ \\
		3 &  &  &  & $\times$
	\end{tabular}
    \caption{Traces for $\det N\equiv 2$ (mod 4) with\\ $m\equiv1$ (mod 2).} 
    \end{subtable}
    \medskip

    \begin{subtable}{0.5\textwidth}
    \centering
    \begin{tabular}{ c || c c c c }
		\backslashbox{$n_1$}{$n_2$} & 0 & 1 & 2 & 3 \\
		\hline\hline
		0 & $\times$ & $\times$ & $\times$ & $\times$ \\
		1 &  & $\times$ & $\times$ & $i$ \\
		2 &  &  & $\times$ & $\times$ \\
		3 &  &  &  & $\times$
	\end{tabular}
    \caption{Traces for $\det N\equiv 3$ (mod 4) with\\ $m\equiv0$ (mod 2).}
    \end{subtable}
    \begin{subtable}{0.5\textwidth}
    \centering
   	\begin{tabular}{ c || c c c c }
		\backslashbox{$n_1$}{$n_2$} & 0 & 1 & 2 & 3 \\
		\hline\hline
		0 & $i$ & $i$ & $i$ & $i$ \\
		1 &  & $\times$ & $\times$ & $\times$ \\
		2 &  &  & $-i$ & $\times$ \\
		3 &  &  &  & $\times$
	\end{tabular}
    \caption{Traces for $\det N\equiv 3$ (mod 4) with\\ $m\equiv1$ (mod 2).} 
    \end{subtable}
    \medskip
    \caption{Possible traces for different $\det N$ mod 4 structures.}
    \label{possible_traces_4}
\end{table}
\restoregeometry
Table~\ref{possible_traces_4} is found by checking the mod 4 condition on the flux components in the following way: 
\begin{align}
\begin{aligned}
	n_1n_2-m^2 &\equiv 0 \text{ (mod 4)},\\
	n_1n_2 &\equiv m^2 \text{ (mod 4)},\\
	n_1n_2&\equiv\begin{cases}
		0 \text{ (mod 4); if $m\equiv 0$ (mod 2),}\\
		1 \text{ (mod 4); if $m\equiv 1$ (mod 2),}
	\end{cases}
\end{aligned}
\end{align}
for $\det N\equiv 0$ (mod 4),
\begin{align}
\begin{aligned}
	n_1n_2&\equiv\begin{cases}
		1 \text{ (mod 4); if $m\equiv 0$ (mod 2)},\\
		2 \text{ (mod 4); if $m\equiv 1$ (mod 2)},
	\end{cases}
\end{aligned}
\end{align}
for $\det N\equiv 1$ (mod 4), 
\begin{align}
\begin{aligned}
	n_1n_2&\equiv\begin{cases}
		2 \text{ (mod 4); if $m\equiv 0$ (mod 2),}\\
		3 \text{ (mod 4); if $m\equiv 1$ (mod 2),}
	\end{cases}
\end{aligned}
\end{align}
for  $\det N\equiv 2$ (mod 4), 
\begin{align}
\begin{aligned}
	n_1n_2&\equiv\begin{cases}
		3 \text{ (mod 4); if $m\equiv 0$ (mod 2),}\\
		0 \text{ (mod 4); if $m\equiv 1$ (mod 2),}
	\end{cases}
\end{aligned}
\end{align}
for $\det N\equiv 3$ (mod 4).

\newpage
There is further restriction on the existence of determinants. The conditions are shown in Tables~\ref{traces_mod_4_p} and~\ref{traces_mod_4_n}. 
The whole spectrum is shown in Tables~\ref{z4_spectrum_p} and~\ref{z4_spectrum_n}. We only show determinants up to 16 since that is the highest modulo condition - the spectrum pattern repeats after that except for $\text{tr}\rho=-1+i$, which needs to be checked case by case. We note that Table~\ref{z4_spectrum_p} agrees with \cite{Kikuchi:2022psj} as we expect.
\begin{table}[!htb]
\centering
    \begin{minipage}{.7\linewidth}
      \centering
        \begin{tabular}{|c|c|c|}
\hline
tr$\rho$ & $\det N>0$ conditions & $\xi$\\
\hline\hline
$i$ & 3 (mod 4) &1\\\hline
$-i$ & 3 (mod 8) &1\\\hline
0 & 2 (mod 4) &2\\\hline
1 & 1 (mod 4), except $\det N=1$ &1\\\hline
$-1$ & 1 (mod 4) &1\\\hline
$2i$ & 0, \quad 12 (mod 16) &4\\\hline
0 & 0 (mod 4) &4\\\hline
$1+i$ & 0 (mod 4) &2\\\hline
$-1+i$ & except $\det N=4$, $\det N=4(4x)^2$, $x\in\mathbb{Z}$ &2\\\hline
\end{tabular}
\caption{Traces for $\det N>0$.}
\label{traces_mod_4_p}
    \end{minipage}\\
    \begin{minipage}{.5\linewidth}
      \centering
        \begin{tabular}{|c|c|c|}
\hline
tr$\rho$ & $\det N<0$ conditions & $\xi$\\
\hline\hline
$i$ & 3 (mod 4) &$-1$\\\hline
$-i$ & 3 (mod 8) &$-1$\\\hline
0 & 2 (mod 4) &$-2$\\\hline
1 & 1 (mod 4) &$-1$\\\hline
$-1$ & 1 (mod 4) &$-1$\\\hline
$2i$ & 0,\quad 12 (mod 16) &$-4$\\\hline
0 & 0 (mod 4) &$-4$\\\hline
$1+i$ & 0 (mod 4) & $-2$\\\hline
$-1+i$ & 0 (mod 4) &$-2$\\\hline
\end{tabular}
\caption{Traces for $\det N<0$.}
\label{traces_mod_4_n}
    \end{minipage} 
\end{table}
\begin{table}[!htb]
    \begin{minipage}{.6\linewidth}
      \centering
\begin{tabular}{ | c ||  c | c | c | c | c | }
\hline
$\det N$  & trace & $D_{+1}$ & $D_{+i}$ & $D_{-1}$ & $D_{-i}$ \\\hline\hline 
1  & $1$    & 1 & 0 & 0 & 0 \\\hline
2  & $0$    & 1 & 0 & 1 & 0 \\\hline
3  & $i$    & 1 & 1 & 1 & 0 \\\hline
3  & $-i$   & 1 & 0 & 1 & 1 \\\hline
4  & $0$    & 2 & 0 & 2 & 0 \\\hline
4  & $1+i$  & 2 & 1 & 1 & 0 \\\hline
5  & $1$    & 2 & 1 & 1 & 1 \\\hline
5  & $-1$   & 1 & 1 & 2 & 1 \\\hline
6  & $0$    & 2 & 1 & 2 & 1 \\\hline
7  & $i$    & 2 & 2 & 2 & 1 \\\hline
8  & $0$    & \fbox{3} & 1 & \fbox{3} & 1 \\\hline
8  & $-1+i$ & 2 & 2 & \fbox{3} & 1 \\\hline
8  & $1+i$  & \fbox{3} & 2 & 2 & 1 \\\hline
9  & $1$    & \fbox{3} & 2 & 2 & 2 \\\hline
9  & $-1$   & 2 & 2 & \fbox{3} & 2 \\\hline
10 & $0$    & \fbox{3} & 2 & \fbox{3} & 2 \\\hline
11 & $i$    & \fbox{3} & \fbox{3} & \fbox{3} & 2 \\\hline
11 & $-i$   & \fbox{3} & 0 & \fbox{3} & \fbox{3} \\\hline
12 & $2i$   & 4 & \fbox{3} & 4 & 1 \\\hline
12 & $0$    & 4 & 2 & 4 & 2 \\\hline
12 & $-1+i$ & \fbox{3} & \fbox{3} & 4 & 2 \\\hline
12 & $1+i$  & 4 & \fbox{3} & \fbox{3} & 2 \\\hline
13 & $1$    & 4 & \fbox{3} & \fbox{3} & \fbox{3} \\\hline
13 & $-1$   & \fbox{3} & \fbox{3} & 4 & \fbox{3} \\\hline
14 & $0$    & 4 & \fbox{3} & 4 & \fbox{3} \\\hline
15 & $i$    & 4 & 4 & 4 & \fbox{3} \\\hline
16 & $2i$   & 5 & 4 & 5 & 2 \\\hline
16 & $0$    & 5 & \fbox{3} & 5 & \fbox{3} \\\hline
16 & $1+i$  & 5 & 4 & 4 & \fbox{3} \\\hline
\end{tabular}
\caption{$T^4/\mathbb{Z}_4$ spectrum for $\det N>0$.}
\label{z4_spectrum_p}
\end{minipage}%
    \begin{minipage}{.5\linewidth}
      \centering
\begin{tabular}{ | c ||  c | c | c | c | c | }
\hline
$\det N$  & trace & $D_{+1}$ & $D_{+i}$ & $D_{-1}$ & $D_{-i}$ \\\hline\hline  
$-1$  & $i$    & 0 & 1 & 0 & 0 \\\hline
$-2$  & $0$    & 0 & 1 & 0 & 1 \\\hline
$-3$  & $1$    & 1 & 1 & 0 & 1 \\\hline
$-3$ & $-1$   & 0 & 1 & 1 & 1 \\\hline
$-4$  & $2i$   & 0 & \fbox{3} & 0 & 1 \\\hline
$-4$  & $0$    & 0 & 2 & 0 & 2 \\\hline
$-4$  & $1+i$  & 1 & 2 & 0 & 1 \\\hline
$-4$  & $-1+i$ & 0 & 2 & 1 & 1 \\\hline
$-5$  & $i$    & 1 & 2 & 1 & 1 \\\hline
$-5$  & $-i$   & 1 & 1 & 1 & 2 \\\hline
$-6$  & $0$    & 1 & 2 & 1 & 2 \\\hline
$-7$  & $1$    & 2 & 2 & 1 & 2 \\\hline
$-7$  & $-1$   & 1 & 2 & 2 & 2 \\\hline
$-8$  & $0$    & 1 & \fbox{3} & 1 & \fbox{3} \\\hline
$-8$  & $1+i$  & 2 & \fbox{3} & 1 & 2 \\\hline
$-8$  & $-1+i$ & 1 & \fbox{3} & 2 & 2 \\\hline
$-9$  & $i$    & 2 & \fbox{3} & 2 & 2 \\\hline
$-10$ & $0$    & 2 & \fbox{3} & 2 & \fbox{3} \\\hline
$-11$ & $1$    & 3 & \fbox{3} & 2 & \fbox{3} \\\hline
$-11$ & $-1$   & 2 & \fbox{3} & \fbox{3} & \fbox{3} \\\hline
$-12$ & $0$    & 2 & 4 & 2 & 4 \\\hline
$-12$ & $1+i$  & \fbox{3} & 4 & 2 & \fbox{3} \\\hline
$-12$ & $-1+i$ & 2 & 4 & \fbox{3} & \fbox{3} \\\hline
$-13$ & $i$    & \fbox{3} & 4 & \fbox{3} & \fbox{3} \\\hline
$-13$ & $-i$   & \fbox{3} & \fbox{3} & \fbox{3} & 4 \\\hline
$-14$ & $0$    & \fbox{3} & 4 & \fbox{3} & 4 \\\hline
$-15$ & $1$    & 4 & 4 & \fbox{3} & 4 \\\hline
$-15$ & $-1$   & \fbox{3} & 4 & 4 & 4 \\\hline
$-16$ & $2i$   & \fbox{3} & 6 & \fbox{3} & 4 \\\hline
$-16$ & $0$    & \fbox{3} & 5 & \fbox{3} & 5 \\\hline
$-16$ & $1+i$  & 4 & 5 & \fbox{3} & 4 \\\hline
$-16$ & $-1+i$ & \fbox{3} & 5 & 4 & 4 \\\hline
\end{tabular}
\caption{$T^4/\mathbb{Z}_4$ spectrum for $\det N<0$.}
\label{z4_spectrum_n}
\end{minipage} 
\end{table}
\clearpage

\subsection{Magnetized $T^4/\mathbb{Z}_3$}
We consider the algebraic relation $(ST_1T_2)^3=I$. This makes the following complex structure
\begin{align}
	\Omega_{(ST_1T_2)} = \begin{pmatrix}
		\omega&0\\0&\omega
	\end{pmatrix},\qquad \omega=e^{2\pi i/3}.
\end{align}
The identification is
\begin{align}
	\vec{z} \sim \Omega_{(ST_1T_2)} \vec{z},
\end{align}
which leads to the transformation function
\begin{align}
	V(\vec{z},\vec{\bar{z}})=e^{2\pi i\rho/3},\qquad\rho=(0,1,2).
\end{align}
The $\text{tr} \rho_+ (ST_1 T_2)$ is given by
\begin{align}
\text{tr} \rho_+(ST_1T_2)&=\sqrt{\det(iN^{-1}\Omega_{(ST_1T_2)}^{-1})}\sum_{\vec{K}\in\Lambda_N} e^{3\pi i\vec{K}^TN^{-1}\vec{K}}.
\label{trace_z3}
\end{align}
According to the $T_i$ symmetry, we should define the fluxes $N$ so that the diagonal elements of $N$ can be even
\begin{align}
	N=
	\begin{pmatrix}
		2n_1&m\\m&2n_2
	\end{pmatrix},\quad n_1,n_2,m\in\mathbb{Z}.
\end{align}
\subsubsection{Positive chirality}
When we consider the $\psi_{++}$ spinor, i.e.
\begin{align}
	\det{N} > 0,\quad \text{tr} N>0,
\end{align}
the trace becomes
\begin{align}
	\text{tr}\rho_+(ST_1T_2)&=\frac{i\bar{\omega}}{\sqrt{\det N}}\sum_{\vec{K}\in\Lambda_N} e^{3\pi i\vec{K}^TN^{-1}\vec{K}},
\end{align}
where we take the positive square root.
By use of Eq.\eqref{Landsberg-Schaar}, when we substitute $B=3I$, $\det B=9$, $g=2$, $n_-^B=0$, and $n_-^N=0$, 
the trace simplifies to
\begin{align}
	\text{tr}\rho_+(ST_1T_2)
    &=\frac{-\bar{\omega}}{3}\left[1+2(\bar{\omega}^{n_1}+\bar{\omega}^{n_2}+\bar{\omega}^{n_1+n_2}(\bar{\omega}^m+\bar{\omega}^{2m}))\right].
\end{align}
We check possible $\det{N}$ numerically. The first few solutions for $4n_1n_2-m^2\in\mathbb{Z}_{+}$ are:
\begin{align}
	\det N = 3,4,7,8,11,12,15,16,19,20,23,24,27,28.
\end{align}
\subsubsection{Negative chirality}
By use of Eq.\eqref{trace_z3}, we find
\begin{align}	\text{tr}\rho_-(ST_1T_2)&=s^{-1}\sqrt{\det(iN^{-1}\Omega_{ST_1T_2}^{-1})}\sum_{\vec{K}\in\Lambda_N}e^{3\pi i\vec{K}^TN^{-1}\vec{K}}.
\end{align}
Substituting back, we obtain the entire transformation function as
\begin{align}
	\text{tr}\rho_-(ST_1T_2)	&=\frac{s^{-1}}{\sqrt{-\det N}}\sum_{\vec{K}\in\Lambda_N} e^{3\pi i\vec{K}^TN^{-1}\vec{K}}.
    \label{trace_withspinor_before}
\end{align}
The formulae for the spectrum calculation become
\begin{align}
	&D_{1}+D_{\omega}+D_{\bar{\omega}}=|\det N|,\label{determinant_z3}\\
	&D_{1}+\omega D_{\omega}+\bar{\omega} D_{\bar{\omega}}=tr\rho(ST_1T_2).
    \label{trace_z31}
\end{align}
We can simplify Eq. \eqref{trace_withspinor_before} using Eq. \eqref{Landsberg-Schaar} with $g=2$. 
Since we take $B=3{\bf 1}_2$, $n_-^B=0$. 
Also, we find $n_-^N=1$ and 
\begin{align}
\begin{aligned}
	\text{tr}\rho_-(ST_1T_2)	&=-\frac{s^{-1}}{3}[1+2(\bar{\omega}^{n_1}+\bar{\omega}^{n_2}+\bar{\omega}^{n_1+n_2}(\bar{\omega}^m+\bar{\omega}^{2m}))]\\
					&=(s\bar{\omega})^{-1}\text{tr}\rho_+.
\end{aligned}
\end{align}
The spinor representation in $z'$ coordinates becomes 
\begin{align}
	V\begin{pmatrix}
		\omega&0&0&0\\		0&\sin^2{\phi}_0+\bar{\omega}\cos^2{\phi}_0&(\bar{\omega}-1)\sin{\phi}_0\cos{\phi}_0&0\\
		0&(\bar{\omega}-1)\sin{\phi}_0\cos{\phi}_0&\cos^2{\phi}_0+\bar{\omega}\sin^2{\phi}_0&0\\
		0&0&0&\omega
	\end{pmatrix}
    .
\end{align}
Thus, we find $\text{tr}\rho_-=\text{tr}\rho_+$.
\subsubsection{Trace classification}
There are only 18 possible cases. We list them all below:
\newpage
\begin{enumerate}
	\item $m\equiv0$ (mod 3) $\rightarrow-\frac{\bar{\omega}}{3}[1+2(\bar{\omega}^{n_1}+\bar{\omega}^{n_2}+2\bar{\omega}^{n_1+n_2})],$
	\begin{enumerate}
		\item $n_1\equiv0$ (mod 3) $\rightarrow-\bar{\omega}(1+2\bar{\omega}^{n_2}),$
		\begin{enumerate}
			\item $n_2\equiv0$ (mod 3) $\rightarrow3(1+\omega),$
			\item $n_2\equiv1$ (mod 3) $\rightarrow 2+\bar{\omega},$
			\item $n_2\equiv2$ (mod 3) $\rightarrow\bar{\omega}+2\omega,$
		\end{enumerate}
		\item $n_1\equiv1$ (mod 3) $\rightarrow-\frac{\bar{\omega}}{3}[1+2(\bar{\omega}+\bar{\omega}^{n_2}(1+2\bar{\omega}))],$
		\begin{enumerate}
			\item $n_2\equiv0$ (mod 3) $\rightarrow 1-\omega,$
			\item $n_2\equiv1$ (mod 3) $\rightarrow\bar{\omega},$
			\item $n_2\equiv2$ (mod 3) $\rightarrow1+\omega,$
		\end{enumerate}
		\item $n_1\equiv2$ (mod 3) $\rightarrow-\frac{\bar{\omega}}{3}[1+2(\omega+\bar{\omega}^{n_2}(1+2\omega))],$
		\begin{enumerate}
			\item $n_2\equiv0$ (mod 3) $\rightarrow\bar{\omega}+2\omega,$
			\item $n_2\equiv1$ (mod 3) $\rightarrow 1+\omega,$
			\item $n_2\equiv2$ (mod 3) $\rightarrow\bar{\omega},$
		\end{enumerate}
	\end{enumerate}
	\item $m\not\equiv0$ (mod 3) $\rightarrow-\frac{\bar{\omega}}{3}[1+2(\bar{\omega}^{n_1}+\bar{\omega}^{n_2}-\bar{\omega}^{n_1+n_2})],$
	\begin{enumerate}
		\item $n_1\equiv0$ (mod 3) $\rightarrow 1+\omega,$
		\item $n_1\equiv1$ (mod 3) $\rightarrow-\frac{\bar{\omega}}{3}[1+2(\bar{\omega}+\bar{\omega}^{n_2}(1-\bar{\omega}))],$
		\begin{enumerate}
			\item $n_2\equiv0$ (mod 3) $\rightarrow 1+\omega,$
			\item $n_2\equiv1$ (mod 3) $\rightarrow 1-\omega,$
			\item $n_2\equiv2$ (mod 3) $\rightarrow \bar{\omega},$
		\end{enumerate}
		\item $n_1\equiv2$ (mod 3) $\rightarrow-\frac{\bar{\omega}}{3}[1+2(\omega+\bar{\omega}^{n_2}(1-\omega))],$
		\begin{enumerate}
			\item $n_2\equiv0$ (mod 3) $\rightarrow 1+\omega,$
			\item $n_2\equiv1$ (mod 3) $\rightarrow \bar{\omega},$
			\item $n_2\equiv2$ (mod 3) $\rightarrow \omega-1.$
		\end{enumerate}
	\end{enumerate}
\end{enumerate}
We find the traces shown in Table~\ref{trace_table_3}. Using these, we can express the degeneracy of each sector as in Table~\ref{traces_3}. The degeneracy also gives us a condition on the determinant, since it has to be an integer.
\begin{table}
    \begin{subtable}{0.5\textwidth}
    \centering
    \begin{tabular}{ | c || c | c | c |}
		\hline
 		\backslashbox{$n_1$}{$n_2$} & 0 & 1 & 2\\
  		\hline\hline  
 		0 & $3+3\omega$ & $2+\bar{\omega}$ & $2\omega+\bar{\omega}$\\\hline
 		1 & $2+\bar{\omega}$ & $\bar{\omega}$ & $1+\omega$\\\hline
 		2 & $2\omega+\bar{\omega}$ & $1+\omega$ & $\bar{\omega}$\\\hline
	\end{tabular}
    \caption{Traces for $m\not\equiv0$ (mod 3).}
    \end{subtable}%
    \begin{subtable}{0.5\textwidth}
    \centering
   	\begin{tabular}{ | c || c | c | c |}
		\hline
 		\backslashbox{$n_1$}{$n_2$} & 0 & 1 & 2\\
  		\hline\hline  
 		0 & $1+\omega$ & $1+\omega$ & $1+\omega$\\\hline
 		1 & $1+\omega$ & $2+\bar{\omega}$ & $\bar{\omega}$\\\hline
 		2 & $1+\omega$ & $\bar{\omega}$ & $2\omega+\bar{\omega}$\\
 		\hline
	\end{tabular}
    \caption{Traces for $m\neq0$ (mod 3).} 
    \end{subtable}
    \medskip
    \caption{All possible traces categorized by the mod 3 structure of the flux components.}
    \label{trace_table_3}
\end{table}
\begin{table}
\begin{center}
\begin{tabular}{| c | c | c | c | c |}\hline
${\rm tr}\rho$ & $D_1$ & $D_\omega$ & $D_{\bar{\omega}}$ & $\det N$\\\hline\hline  
$3+3\omega$ & $\frac{|\det N|}{3}+1$ & $\frac{|\det N|}{3}+1$ & $\frac{|\det N|}{3}-2$ & 0 (mod 3)\\\hline
$\bar{\omega}$ & $\frac{|\det N|-1}{3}$ & $\frac{|\det N|-1}{3}$ & $\frac{|\det N|+2}{3}$ & 1 (mod 3)\\\hline
$1+\omega$ & $\frac{|\det N|+1}{3}$ & $\frac{|\det N|+1}{3}$ & $\frac{|\det N|-2}{3}$ & 2 (mod 3)\\\hline
$2+\bar{\omega}$ & $\frac{|\det N|}{3}+1$ & $\frac{|\det N|}{3}-1$ & $\frac{|\det N|}{3}$ & 0 (mod 3)\\\hline
$2\omega+\bar{\omega}$ & $\frac{|\det N|}{3}-1$ & $\frac{|\det N|}{3}+1$ & $\frac{|\det N|}{3}$ & 0 (mod 3)\\\hline
\end{tabular}
\end{center}
\caption{Degeneracy for different traces.}
\label{traces_3}
\end{table}
From Table~\ref{traces_3}, we conclude three-generation models appear in the range of $6\leq\det N\leq15$.\\

Similarly to the $\mathbb{Z}_4$ case, we therefore find which traces are possible for different determinants mod $3$.\\

For $\det N \equiv 1$ (mod $3$), we express the determinant as:
\begin{align}
\begin{aligned}
	4n_1n_2-m^2&\equiv1\text{ (mod 3)}\\
	4n_1n_2&\equiv1+m^2\text{ (mod 3)}\\
	4n_1n_2&\equiv\begin{cases}
		1 \text{ (mod 3); if $m\equiv 0$ (mod 3),}\\
		2 \text{ (mod 3); if $m\equiv 1$ (mod 3),}
	\end{cases}
\end{aligned}
\end{align}
since if $m\equiv 2$ (mod 3), we have $m^2\equiv 1$ (mod 3). \\
For $\det N \equiv 2$ (mod 3), we express the determinant as:
\begin{align}
\begin{aligned}
	4n_1n_2&\equiv\begin{cases}
		2 \text{ (mod 3); if $m\equiv 0$ (mod 3),}\\
		0 \text{ (mod 3); if $m\equiv 1$ (mod 3).}
	\end{cases}
\end{aligned}
\end{align}\\
For $\det N \equiv 0$ (mod 3), we express the determinant as:
\begin{align}
\begin{aligned}
	4n_1n_2&\equiv\begin{cases}
		0 \text{ (mod 3); if $m\equiv 0$ (mod 3),}\\
		1 \text{ (mod 3); if $m\equiv 1$ (mod 3).}
	\end{cases}
\end{aligned}
\end{align}\\
The result is summarized in Table~\ref{possibletracesN}. 
\begin{table}
    \begin{subtable}{0.5\textwidth}
    \centering
    \begin{tabular}{ c || c c c }
		\backslashbox{$n_1$}{$n_2$} & 0 & 1 & 2 \\
		\hline\hline
		0 & $\times$ & $\times$ & $\times$ \\
		1 &  & $\bar{\omega}$ & $\times$ \\
		2 &  &  & $\bar{\omega}$ \\
	\end{tabular}
    \caption{ Traces for $\det N\equiv1$ (mod 3) with\\ $m\equiv0$ (mod 3). }
    \end{subtable}%
    \begin{subtable}{0.5\textwidth}
    \centering
   	\begin{tabular}{ c || c c c }
		\backslashbox{$n_1$}{$n_2$} & 0 & 1 & 2 \\
		\hline\hline
		0 & $\times$ & $\times$ & $\times$ \\
		1 &  & $\times$ & $\bar{\omega}$ \\
		2 &  &  & $\times$ \\
	\end{tabular}
    \caption{Traces for $\det N\equiv1$ (mod 3) with\\ $m\not\equiv0$ (mod 3).} 
    \end{subtable}
    \medskip
    \begin{subtable}{0.5\textwidth}
    \centering
    \begin{tabular}{ c || c c c }
		\backslashbox{$n_1$}{$n_2$} & 0 & 1 & 2 \\
		\hline\hline
		0 & $\times$ & $\times$ & $\times$ \\
		1 &  & $\times$ & $1+\omega$ \\
		2 &  &  & $\times$ \\
	\end{tabular}
    \caption{Traces for $\det N\equiv2$ (mod 3) with\\ $m\equiv0$ (mod 3).}
    \end{subtable}%
    \begin{subtable}{0.5\textwidth}
    \centering
   	\begin{tabular}{ c || c c c }
		\backslashbox{$n_1$}{$n_2$} & 0 & 1 & 2 \\
		\hline\hline
		0 & $1+\omega$ & $1+\omega$ & $1+\omega$ \\
		1 &  & $\times$ & $\times$ \\
		2 &  &  & $\times$ \\
	\end{tabular}
    \caption{Traces for $\det N\equiv2$ (mod 3) with\\ $m\not\equiv0$ (mod 3).} 
    \end{subtable}
    \medskip
    \begin{subtable}{0.5\textwidth}
    \centering
    \begin{tabular}{ c || c c c }
		\backslashbox{$n_1$}{$n_2$} & 0 & 1 & 2 \\
		\hline\hline
		0 & $3+3\omega$ & $2+\bar{\omega}$ & $2\omega+\bar{\omega}$ \\
		1 &  & $\times$ & $\times$ \\
		2 &  &  & $\times$ \\
	\end{tabular}
    \caption{Traces for $\det N\equiv0$ (mod 3) with\\ $m\equiv0$ (mod 3).}
    \end{subtable}%
    \begin{subtable}{0.5\textwidth}
    \centering
   	\begin{tabular}{ c || c c c }
		\backslashbox{$n_1$}{$n_2$} & 0 & 1 & 2 \\
		\hline\hline
		0 & $\times$ & $\times$ & $\times$ \\
		1 &  & $2+\bar{\omega}$ & x \\
		2 &  &  & $2\omega+\bar{\omega}$ \\
	\end{tabular}
    \caption{Traces for $\det N\equiv0$ (mod 3) with\\ $m\not\equiv0$ (mod 3).} 
    \end{subtable}
    \medskip
    \caption{All possible traces categorized by the mod 3 structure of the flux components.}
\label{possibletracesN}
\end{table}
\newpage
The fluxes are those with $\det N \equiv 0,3$ (mod 12) for tr$\rho=2+\bar{\omega}$ and tr$\rho=2\omega+\bar{\omega}$. For tr$\rho=3+3\omega$, the possible determinants are $\det N \equiv 0,9$ (mod 36). 
We can list the whole spectrum as shown in Tables~\ref{z3_spectrum_p} and~\ref{z3_spectrum_n}. Table~\ref{z3_spectrum_p} agrees with \cite{Kikuchi:2022psj}.
\begin{table}[!htb]
    \begin{minipage}{.6\linewidth}
      \centering
\begin{tabular}{ | c || c | c | c | c | c | c |}
\hline
$\det N$ & $\text{tr}\rho$ & $D_1$ & $D_\omega$ & $D_{\bar{\omega}}$\\\hline\hline  
3 & $2+\bar{\omega}$ & 2 & 0 & 1\\\hline
4 & $\bar{\omega}$ & 1 & 1 & 2\\\hline
7 & $\bar{\omega}$ & 2 & 2 & \fbox{3}\\\hline
8 & $1+\omega$ & \fbox{3} & \fbox{3} & 2\\\hline
11 & $1+\omega$ & 4 & 4 & \fbox{3}\\\hline
12 & $2+\bar{\omega}$ & 5 & \fbox{3} & 4\\\hline
12 & $2\omega+\bar{\omega}$ & \fbox{3} & 5 & 4\\\hline
15 & $2+\bar{\omega}$ & 6 & 4 & 5\\\hline
15 & $2\omega+\bar{\omega}$ & 4 & 6 & 5\\\hline
16 & $\bar{\omega}$ & 5 & 5 & 6\\\hline
19 & $\bar{\omega}$ & 6 & 6 & 7\\\hline
20 & $1+\omega$ & 7 & 7 & 6\\\hline
23 & $1+\omega$ & 8 & 8 & 7\\\hline
24 & $2+\bar{\omega}$ & 9 & 7 & 8\\\hline
24 & $2\omega+\bar{\omega}$ & 7 & 9 & 8\\\hline
27 & $3+3\omega$ & 10 & 10 & 7\\\hline
27 & $2+\bar{\omega}$ & 10 & 8 & 9\\\hline
28 & $\bar{\omega}$ & 9 & 9 & 10\\\hline
\end{tabular}
\caption{$T^4/\mathbb{Z}_3$ spectrum for $\det N>0$.}
\label{z3_spectrum_p}
\end{minipage}%
    \begin{minipage}{.5\linewidth}
      \centering
\begin{tabular}{ | c || c | c | c | c | c | c |}
\hline
$
{\det  N}
$ & $\text{tr}\rho$ & $D_1$ & $D_\omega$ & $D_{\bar{\omega}}$\\\hline\hline  
$-9$ &$3+3\omega$& 4 & 4 & 1\\\hline
$-9$ &$2+\bar{\omega}$& 4 & 2 & \fbox{3}\\\hline
$-9$ &$2\omega+\bar{\omega}$& 2 & 4 & \fbox{3}\\\hline
$-12$ &$2+\bar{\omega}$& 5 & \fbox{3} & 4\\\hline
$-12$ &$2\omega+\bar{\omega}$& \fbox{3} & 5 & 4\\\hline
$-21$  &$2+\bar{\omega}$& 8&6&7\\\hline
$-21$ &$2\omega+\bar{\omega}$& 6&8&7\\\hline
$-24$ &$2+\bar{\omega}$& 9&7&8\\\hline
$-24$ &$2\omega+\bar{\omega}$& 7&9&8\\\hline
$-33$  &$2+\bar{\omega}$& 12&10&11\\\hline
$-33$ &$2\omega+\bar{\omega}$& 10&12&11\\\hline
\end{tabular}
\caption{$T^4/\mathbb{Z}_3$ spectrum for $\det N<0$.}
\label{z3_spectrum_n}
\end{minipage} 
\end{table}
\clearpage
\subsection{Magnetized $T^4/\mathbb{Z}_6$}
To construct $T^4/\mathbb{Z}_6$, we focus on
the algebraic relation $((T_1T_2)^{-1}S)^6 = {\bf{1}}_4$.
According to Ref. \cite{Kikuchi:2022psj},
the complex structure are fixed as
\begin{align}
	 \Omega_{(T_1T_2)^{-1}S} = \begin{pmatrix}
		\omega&0\\0&\omega
	\end{pmatrix} .
\end{align}
The following identification corresponds to the $\mathbb{Z}_6$ twist:
\begin{align}
	\vec{z}\sim\kappa\vec{z},\ \ \ \ \kappa=e^{\frac{\pi i}{3}}.
\end{align}
The transformation function and the modular transformation are given by
\begin{align}
\begin{aligned}
&V(\vec{z},\vec{\bar{z}})=e^{\pi i\rho/3},\ \ \ \ \ \rho=(0,1,2,3,4,5),\\
    &\text{tr}\rho_{+}=\sqrt{\det (iN^{-1} \Omega_{(T_1T_2)^{-1}S})}\sum_{\vec{K}\in\Lambda_N}e^{\pi i\vec{K}^TN^{-1}\vec{K}},
\end{aligned}
\end{align}
where we have used $\vec{J}=\vec{K}\in\Lambda_N$.
\subsubsection{Positive chirality}
The trace becomes
\begin{align}
	{\rm tr}\rho_{+}=\frac{e^{\pi i/6}}{\sqrt{\det N}}\sum_{\vec{K}\in\Lambda_N}e^{\pi i\vec{K}^TN^{-1}\vec{K}}
    = 
    \omega e^{-2\pi i(n_1K_1^2+n_2K_2^2+mK_1K_2)}.
\end{align}
The flux is the same as in the $Z_3$ case. 
We use eq.\eqref{Landsberg-Schaar} with $B=I$, $n_-^B=0$, $n_-^N=0$ and $g=2$.
Therefore we find ${\rm tr}\rho=\omega=\kappa^2$.
Also, we have the following formulae
\begin{align}
\begin{split}
&D_1+D_\kappa+D_{\kappa^2}+D_{\kappa^3}+D_{\kappa^4}+D_{\kappa^5}=\det N,\\
&D_1+\kappa D_\kappa+\kappa^2D_{\kappa^2}+\kappa^3D_{\kappa^3}+\kappa^4D_{\kappa^4}+\kappa^5D_{\kappa^5}={\rm tr}\rho_+=\omega.\label{Z6 trace}
\end{split}
\end{align}
Since $Z_6=Z_2\otimes Z_3$, we have the following $Z_2$ and $Z_3$ properties for some of the eigenvalues
\begin{align}
\begin{split}
	D_1+D_{\kappa^2}+D_{\kappa^4}&=N_{Z_2,+},\\
	D_1+D_{\kappa^3}&=N_{Z_3,1},\\
	D_{\kappa}+D_{\kappa^4}&=N_{Z_3,\omega},\\
       D_{\kappa^2}+D_{\kappa^5}&=N_{Z_3,\bar{\omega}}.
\end{split}
\end{align}
We find the following relation
\begin{align}
	D_1-D_{\kappa^3}=D_{\kappa^4}-D_{\kappa}=D_{\kappa^2}-D_{\kappa^5}-1,
\end{align}
where we have made use of the relations
\begin{align}
	\kappa^3=-1, \qquad
	\kappa^4=-\kappa=\bar{\omega}, \qquad
	\kappa^5=-\kappa^2=-\omega
\end{align}
to rewrite the trace as
\begin{align}
	(D_1-D_{\kappa^3})+\omega(D_{\kappa^2}-D_{\kappa^5})+\bar{\omega}(D_{\kappa^4}-D_{\kappa})=\omega.
\end{align}
Since the spectrum must correspond to that of $Z_3$, and there is no additional conditions on the fluxes, we will have the same $\det N$ values available.
The calculation is straightforward by rewriting
\begin{align*}
	&(2D_1-N_{Z_3,1})+\omega(2D_{\kappa^2}-N_{Z_3,\bar{\omega}})+\bar{\omega}(2D_{\kappa^4}-N_{Z_3,\omega})=\omega,
\end{align*}
and using $1+\omega+\bar{\omega}=1$ as well as the $Z_2$ spectrum constraint.
\subsubsection{Negative chirality}
The trace is now 
\begin{align}
\text{tr}\rho_-((T_1T_2)^{-1}S)&=s^{-1}\sqrt{\det (iN^{-1}\Omega^{' -1 }_{(T_1T_2)^{-1}S})}\sum_{\vec{K}\in\Lambda_N}e^{\pi i\vec{K}^TN^{-1}\vec{K}}.
\end{align}
Then, the trace becomes
\begin{align}
	\text{tr}\rho_-((T_1T_2)^{-1}S)&=\frac{-s^{-1}}{\sqrt{-\det N}}\sum_{\vec{K}\in\Lambda_N}e^{\pi i\vec{K}^TN^{-1}\vec{K}}.
\end{align}
Using Eq. \eqref{Landsberg-Schaar} with $B=I,n_-^B=0,n_-^N=1$ and $g=2$, we get
\begin{align}
	\text{tr}\rho_=((T_1T_2)^{-1}S)&=s^{-1}e^{-2\pi i (n_1K^2_1+n_2K^2_2+mK_1K_2)}=s^{-1}.
\end{align}
The spinor representation in ${z}'$ becomes
\begin{align}
	V\begin{pmatrix}
		\kappa&0&0&0\\
		0&\sin^2{\phi}_0+\kappa^2cos^2{\phi}_0&(\kappa^2-1)\sin{\phi}_0\cos{\phi}_0&0\\
		0&(\kappa^2-1)\sin{\phi}_0\cos{\phi}_0&\cos^2{\phi}_0+\kappa^2\sin^2{\phi}_0&0\\
		0&0&0&\kappa
	\end{pmatrix}
    .
\end{align}
Thus, $\text{tr}\rho_-=\kappa^5=\omega$.
We can calculate the spectrum using the following formulae:
\begin{align}
\begin{split}
&D_1+D_\kappa+D_{\kappa^2}+D_{\kappa^3}+D_{\kappa^4}+D_{\kappa^5}=|\det N|,\\
&D_1+\kappa D_\kappa+\kappa^2D_{\kappa^2}+\kappa^3D_{\kappa^3}+\kappa^4D_{\kappa^4}+\kappa^5D_{\kappa^5}=\tr\rho_-=\omega,
\end{split}
\end{align}
as well as
\begin{align}
\begin{aligned}
	D_1+D_{\kappa^2}+D_{\kappa^4}&=N_{Z_2,+},\\
	D_1+D_{\kappa^3}&=N_{Z_3,1},\\
	D_{\kappa}+D_{\kappa^4}&=N_{Z_3,\omega},\\
D_{\kappa^2}+D_{\kappa^5}&=N_{Z_3,\bar{\omega}},
\end{aligned}
\end{align}
or
\begin{align}
	&(2D_1-N_{Z_3,1})+\omega(2D_{\kappa^2}-N_{Z_3,\bar{\omega}})+\bar{\omega}(2D_{\kappa^4}-N_{Z_3,\omega})=\omega,
\end{align}
where $N_{Z_2,+}$ is the even spectrum for negative chirality (odd sector for positive chirality).
\subsubsection{Spectrum}
The spectra are given in Tables~\ref{z6_spectrum_p} and~\ref{z6_spectrum_n} (in agreement with \cite{Kikuchi:2022psj}).
\begin{table}[!htb]
    \begin{minipage}{.6\linewidth}
      \centering
\begin{tabular}{ | c || c | c | c | c | c | c | c | c |}
\hline
$\det N$  & $D_1$ & $D_\kappa$ & $D_{\kappa^2}$ & $D_{\kappa^3}$ & $D_{\kappa^4}$ & $D_{\kappa^5}$\\\hline\hline  
3 & 1 & 0 & 1 & 1 & 0 & 0\\\hline
4 & 1 & 0 & 2 & 0 & 1 & 0\\\hline
7 & 1 & 1 & 2 & 1 & 1 & 1\\\hline
8 & 2 & 1 & 2 & 1 & 2 & 0\\\hline
11 & 2 & 2 & 2 & 2 & 2 & 1\\\hline
12 & \fbox{3} & 1 & \fbox{3} & 2 & 2 & 1\\\hline
12 & 2 & 2 & \fbox{3} & 1 & \fbox{3} & 1\\\hline
15 & \fbox{3} & 2 & \fbox{3} & \fbox{3} & 2 & 2\\\hline
15 & 2 & \fbox{3} & \fbox{3} & 2 & \fbox{3} & 2\\\hline
16 & \fbox{3} & 2 & 4 & 2 & \fbox{3} & 2\\\hline
19 & \fbox{3} & \fbox{3} & 4 & \fbox{3} & \fbox{3} & \fbox{3}\\\hline
20 & 4 & \fbox{3} & 4 & \fbox{3} & 4 & 4\\\hline
23 & 4 & 4 & 4 & 4 & 4 & \fbox{3}\\\hline
24 & 5 & \fbox{3} & 5 & 4 & 4 & \fbox{3}\\\hline
24 & 4 & 4 & 5 & \fbox{3} & 5 & \fbox{3}\\\hline
27 & 5 & 5 & 4 & 5 & 5 & \fbox{3}\\\hline
27 & 5 & 4 & 5 & 5 & 4 & 4\\\hline
28 & 5 & 4 & 6 & 4 & 5 & 6\\\hline
\end{tabular}
\caption{$T^4/\mathbb{Z}_6$ spectrum for $\det N>0$.}
\label{z6_spectrum_p}
\end{minipage}%
    \begin{minipage}{.5\linewidth}
\centering
\begin{tabular}{ | c || c | c | c | c | c | c | c | c |}
\hline
$\det N$  & $D_1$ & $D_\kappa$ & $D_{\kappa^2}$ & $D_{\kappa^3}$ & $D_{\kappa^4}$ & $D_{\kappa^5}$\\\hline\hline  
$-9$ & 2 & 2 & 1 & 2 & 2 & 0\\\hline
$-9$ & 2 & 1 & 2 & 2 & 1 & 1\\\hline
$-9$ & 2 & 1 & 2 & 2 & 1 & 2\\\hline
$-12$ & 2 & 2 & 2 & 3 & 1 & 2\\\hline
$-12$ & 1 & \fbox{3} & 2 & 2 & 2 & 2\\\hline
$-21$ & 4 & \fbox{3} & 4 & 4 & \fbox{3} & \fbox{3}\\\hline
$-21$ & \fbox{3} & 4 & 4 & \fbox{3} & 4 & \fbox{3}\\\hline
$-24$ & 4 & 4 & 4 & 5 & \fbox{3} & 4\\\hline
$-24$ & \fbox{3} & 5 & 4 & 4 & 4 & 4\\\hline
$-33$ & 6 & 6 & 6 & 7 & 5 & 6\\\hline
$-33$ & 5 & 7 & 6 & 6 & 6 & 6\\\hline
\end{tabular}
\caption{$T^4/\mathbb{Z}_6$ spectrum for $\det N<0$.}
\label{z6_spectrum_n}
\end{minipage} 
\end{table}

\newpage

\subsection{Magnetized $T^6/\mathbb{Z}_{12}$}
\subsubsection{Diagonalization of $T^6/\mathbb{Z}_{12}$}
We review magnetized $T^6/\mathbb{Z}_{12}$ in Ref. \cite{Kikuchi:2023awm}. 
$T^6/\mathbb{Z}_{12}$ is introduced by the following algebraic relation
\begin{align}
(ST_1 T_2 T^{-1}_3 T_5 T_6)^{12} = {\bf{1}}_6,
\end{align}
and then we can describe $\mathbb{Z}_{12}$-twist by $ST_1 T_2 T^{-1}_3 T_5 T_6$. Note that since $T^6/\mathbb{Z}_{12}$ has $E_6$ root lattice, all of the scale factors $2\pi r_i$ become $1$.
The integer fluxes $N$ and complex structure moduli $\Omega_{12}$ are given by
\begin{align}
N
=
\begin{bmatrix}
n_{11} & n_{12} & n_{13} \\
n_{12} & n_{11} & n_{13} \\
n_{13} & n_{13} & n_{11} + n_{12} -2n_{13} \\
\end{bmatrix}
,\quad
\Omega_{12} 
= 
\begin{bmatrix}
-\frac{1}{2} -\frac{\sqrt{3}}{6} i & \frac{\sqrt{3}}{3} i & -\frac{1}{2} +\frac{\sqrt{3}}{6} i \\
 \frac{\sqrt{3}}{3} i & -\frac{1}{2} -\frac{\sqrt{3}}{6} i & -\frac{1}{2} +\frac{\sqrt{3}}{6} i \\
-\frac{1}{2} +\frac{\sqrt{3}}{6} i & -\frac{1}{2} +\frac{\sqrt{3}}{6} i & \frac{1}{2} -\frac{\sqrt{3}}{6} i \\
\end{bmatrix}
,
\end{align}
where $z$ is identified to $\Omega_{12} z$.
We can verify that the lattice corresponds to $E_6$ root lattice. Here, the integer fluxes $N$ satisfy the $F$-term SUSY condition $(N\Omega_{12})^T = N\Omega_{12}$, and we impose $n_{11}\equiv n_{12} \equiv n_{13} ~ (\text{mod}~{2})$ because of the modular $T$ symmetry. 
When we consider the diagonalization of $ F_{i\bar{j}} = \pi [N^T (\text{Im}\Omega_{12})^{-1}]_{ij}$, 
we can find the following diagonalization matrix $P$ that corresponds to the  $SO(3)$ rotation
\begin{align}
\label{P12}
P
&=
\begin{bmatrix}
-\frac{1}{\sqrt{2}} & \frac{1-\sqrt{3}}{2\sqrt{3-\sqrt{3}}} & \frac{1+\sqrt{3}}{2\sqrt{3+\sqrt{3}}} \\
\frac{1}{\sqrt{2}} & \frac{1-\sqrt{3}}{2\sqrt{3-\sqrt{3}}} & \frac{1+\sqrt{3}}{2\sqrt{3+\sqrt{3}}} \\
0 & \frac{1}{\sqrt{3-\sqrt{3}}} & \frac{1}{\sqrt{3+\sqrt{3}}} \\
\end{bmatrix}
\\ \notag
&= 
\begin{bmatrix}
\cos{\theta_2} & 0 & \sin{\theta_2} \\
0 & 1 & 0 \\
-\sin{\theta_2} & 0 & \cos{\theta_2} \\
\end{bmatrix}
\begin{bmatrix}
\cos{\theta_1} & -\sin{\theta_1} & 0 \\
\sin{\theta_1} & \cos{\theta_1} & 0 \\
0 & 0 & 1 \\
\end{bmatrix}
\begin{bmatrix}
1 & 0 & 0 \\
0 & \cos{\theta_3} & -\sin{\theta_3}  \\
0 & \sin{\theta_3} & \cos{\theta_3}  \\
\end{bmatrix}
\\ \notag
&=
\begin{bmatrix}
\cos{\theta_1} \cos{\theta_2} & \sin{\theta_2} \sin{\theta_3} - \sin{\theta_1} \cos{\theta_2} \cos{\theta_3}& \sin{\theta_1} \cos{\theta_2} \sin{\theta_3} + \sin{\theta_2} \cos{\theta_3} \\
\sin{\theta_1} & \cos{\theta_1}\cos{\theta_3} & -\cos{\theta_1}\sin{\theta_3} \\
-\cos{\theta_1}\sin{\theta_2} & \sin{\theta_1}\sin{\theta_2}\cos{\theta_3} + \cos{\theta_2}\sin{\theta_3} & \cos{\theta_2} \cos{\theta_3} - \sin{\theta_1}\sin{\theta_2}\sin{\theta_3}\\
\end{bmatrix}
,
\end{align}
where $\det{P} = 1$ and note that the $P$ does not depend on the values of the background magnetic fluxes. Then $\Omega_{12}$ are diagonalized as
\begin{align}
P^{-1} \Omega_{12} P =\text{diag} \left(e^{-2\pi i/3}, e^{-\pi i/6}, e^{5\pi i/6} \right).
\end{align}
From Eq. \eqref{P12}, we can assign $\theta_1 = \frac{3}{4} \pi$, $\theta_2 = 0$ for simplicity, so we find $q_1 = 1$ and $q_2 = 0$. We can just obtain the wave functions $\psi_{+--} = \psi_{-+-}$ and $\psi_{--+} = 0$ in magnetized $T^6/\mathbb{Z}_{12}$.
That is, we can describe the following spinor generated by the fluxes on $T^6/\mathbb{Z}_{12}$
\begin{align}
\Psi_{+} 
=
\begin{bmatrix}
\psi_{+--} \\
\psi_{-+-} \\
\psi_{--+} \\
\end{bmatrix}
= 
\alpha_1
\begin{bmatrix}
\psi \\
\psi \\
0 \\
\end{bmatrix}
.
\end{align}
Meanwhile, we have defined the spinor in $z^{\prime}$ system, namely $\Psi^{\prime}$ with only $\psi^{\prime}_{+++}$. In what follows, we will make use of $\Psi^{\prime}$ to obtain the number of zero-modes in magnetized $T^6/\mathbb{Z}_{12}$. Also, $z^{\prime}$ is also transformed as $\Omega^{\prime} z^{\prime}$ under the transformation $z \rightarrow \Omega z$.
In $z^{\prime}$ system, 
$\psi^{\prime J}_{+++}$, ($J \in \Lambda_N$) is transformed under $ST_1 T_2 T^{-1}_3 T_5 T_6$ as 
\begin{align}
&\psi^{\prime J}_{+++} (\Omega^{\prime}_{\text{twist}} z^{\prime}, \Omega^{\prime}_{\text{twist}})  \\ \notag
&= 
\frac{\sqrt{\det{[-i(\Omega^{\prime}_{\text{twist}} + B)]}}}{\sqrt{|\det{N}|}}
\sum_{K \in \Lambda_N} 
e^{2\pi i J^T N^{-1} K} e^{\pi i K^T N^{-1} B K}  
\cdot \psi^{\prime K}_{+++} (z^{\prime}, \Omega^{\prime}_{\text{twist}} ),
\end{align}
where $B = B_1 + B_2 - B_3 + B_5 + B_6$ and $\Omega^{\prime}_{\text{twist}} = \text{Re}\Omega_{12} + i R \text{Im}\Omega_{12}$. 
In $z^{\prime}$ system, the $\mathbb{Z}_N$-twist boundary condition of spinor $\Psi^{\prime} (z^{\prime})$ is given by
\begin{align}
\Psi^{\prime}(\Omega^{\prime} z^{\prime}) = \mathcal{S}^{\prime} V^{\prime} \Psi^{\prime}(z^{\prime}),
\end{align}
where $\mathcal{S}^{\prime}$, $V^{\prime}$ denote the spinor representation of the $\mathbb{Z}_N$-twist and transformation function in $z^{\prime}$ system, respectively.
Note that since we can also identify $z^{\prime}$ to $\Omega^{\prime} z^{\prime}$ where $\Omega^{\prime N} = {\bf{1}}_3$, and only $\psi^{\prime}_{+++}$ is excited in $z^{\prime}$ system, $\psi^{\prime}_{+++}$ is invariant under the $\mathcal{S}^{\prime}$ if we define the following $V^{\prime}$
\begin{align}
V^{\prime} = e^{2\pi i \beta^{\prime}}, \quad \text{where} \quad \beta^{\prime} N \equiv 0 \mod{1} \quad \text{and $\beta^{\prime} \neq \beta$ in general.}
\end{align}
This is consistent with the Dirac equation under the $\mathbb{Z}_N$-twist in $z^{\prime}$ system.
Then, by use of the wave functions with positive chirality $(+++)$ and the $\mathbb{Z}_N$-twist boundary condition of the spinor in $z^{\prime}$ system, we can find the number of zero-modes in each $\mathbb{Z}_N$-sector when we take $U_1 = \text{diag}(+, -, -), U_2 = \text{= diag}(-, +, -), U_3 = \text{diag}(-, -, +)$.

\subsubsection{The number of zero-modes in magnetized $T^6/\mathbb{Z}_{12}$}
We analyze three-generation models satisfying both the $F$-term and the $D$-term SUSY conditions in magnetized $T^6/\mathbb{Z}_{12}$.
We can find some three-generation models when we make use of the $F$-term and $D$-term SUSY conditions in $z^{\prime}$, the $\mathbb{Z}_N$-twist boundary condition of $\psi^{\prime}_{+++}$, the modular transformation, and  $U$.
For example, when we take $U_1 = \text{diag}(+, -, -)$, we can find the following modular transformation 
\begin{align}
&\psi^{\prime J}_{+++} (\Omega^{\prime}_{\text{twist}} z^{\prime}, \Omega^{\prime}_{\text{twist}})  \\ \notag
&= 
\frac{e^{\frac{5}{12}\pi i}}{\sqrt{|\det{N}|}}
\sum_{K \in \Lambda_N} 
e^{2\pi i J^T N^{-1} K} e^{\pi i K^T N^{-1} B K}  
\cdot \psi^{\prime K}_{+++} (z^{\prime}, \Omega^{\prime}_{\text{twist}} ).
\end{align}
Then, we obtain the consistent fluxes $N$ and the $M$
\begin{align}
N = 
\begin{bmatrix}
-13 & 5 & -3 \\
 5 & -13 & -3 \\
-3 & -3 & -2 \\
\end{bmatrix}
,
\quad
M = 
\begin{bmatrix}
-5 & 13 & 3 \\
 13 & -5 & 3 \\
3 & 3 & 2 \\
\end{bmatrix}
,
\end{align}
where $\det{N} = 36$ and zero-modes number in $\mathbb{Z}_{N}$ sectors $n_i$, $(i=0,1,\cdots,11)$ is 
\begin{align}
[n_0, n_1, n_2, n_3, n_4, n_5, n_6, n_7, n_8, n_9, n_{10}, n_{11} ] = [9, 0, 4, 0, 9, 0, 2, 0, 9, 0, \fbox{3}, 0]
.
\end{align}
The others are
\begin{align}
N = 
\begin{bmatrix}
-3 & 3 & -1 \\
 3 & -3 & -1 \\
-1 & -1 &  2 \\
\end{bmatrix}
,
\quad
M = 
\begin{bmatrix}
-3 & 3 & 1 \\
 3 & -3 & 1 \\
1 & 1 &  -2 \\
\end{bmatrix}
,
\end{align}
where $\det{N} = 12$ and each zero-modes number is 
\begin{align}
[n_0, n_1, n_2, n_3, n_4, n_5, n_6, n_7, n_8, n_9, n_{10}, n_{11} ]=
[\fbox{3}, 0, 2, 0, \fbox{3}, 0, 0, 0, \fbox{3}, 0, 1, 0].
\end{align}
When we take $U_2 = \text{diag}(-, +, -)$, we can find  the modular transformation 
\begin{align}
&\psi^{\prime J}_{+++} (\Omega^{\prime}_{\text{twist}} z^{\prime}, \Omega^{\prime}_{\text{twist}})  \\ \notag
&= 
\frac{e^{-\frac{\pi}{12} i}}{\sqrt{|\det{N}|}}
\sum_{K \in \Lambda_N} 
e^{2\pi i J^T N^{-1} K} e^{\pi i K^T N^{-1} B K}  
\cdot \psi^{\prime K}_{+++} (z^{\prime}, \Omega^{\prime}_{\text{twist}} ).
\end{align}
Then, we obtain the consistent fluxes $N$ and the $M$
\begin{align}
N = 
\begin{bmatrix}
 1 & -5 & -1 \\
 -5 &  1 & -1 \\
 -1 & -1 & -2 \\
\end{bmatrix}
,
\quad
M = 
\begin{bmatrix}
 \sqrt{3}-3 & \sqrt{3}+3 & \sqrt{3} \\
 \sqrt{3}+3 &  \sqrt{3}-3 & \sqrt{3} \\
 \sqrt{3} & \sqrt{3} & 0 \\
\end{bmatrix}
,
\end{align}
where $\det{N} = 36$ and each zero-modes number is 
\begin{align}
[n_0, n_1, n_2, n_3, n_4, n_5, n_6, n_7, n_8, n_9, n_{10}, n_{11} ]= [5, 1, \fbox{3}, \fbox{3}, 4, 0, 5, \fbox{3}, \fbox{3}, 2, 4, \fbox{3}].
\end{align}
When we take $U_3 = \text{diag}(-, -, +)$, we find the modular transformation 
\begin{align}
&\psi^{\prime J}_{+++} (\Omega^{\prime}_{\text{twist}} z^{\prime}, \Omega^{\prime}_{\text{twist}})  \\ \notag
&= 
\frac{e^{-\frac{\pi}{12} i}}{\sqrt{|\det{N}|}}
\sum_{K \in \Lambda_N} 
e^{2\pi i J^T N^{-1} K} e^{\pi i K^T N^{-1} B K}  
\cdot \psi^{\prime K}_{+++} (z^{\prime}, \Omega^{\prime}_{\text{twist}} ).
\end{align}
Then, we obtain the consistent fluxes $N$ and the $M$
\begin{align}
N = 
\begin{bmatrix}
 5 & -1 &  1 \\
 -1 & 5 &  1 \\
 1 &  1 &  2 \\
\end{bmatrix}
,
\quad
M = 
\begin{bmatrix}
 \sqrt{3}-3 & \sqrt{3}+3 & \sqrt{3} \\
 \sqrt{3}+3 &  \sqrt{3}-3 & \sqrt{3} \\
 \sqrt{3} & \sqrt{3} & 0 \\
\end{bmatrix}
,
\end{align}
where $\det{N} = 36$ and each zero-modes number is 
\begin{align}
[n_0, n_1, n_2, n_3, n_4, n_5, n_6, n_7, n_8, n_9, n_{10}, n_{11} ]=
[5, \fbox{3}, \fbox{3}, 2, 4, \fbox{3}, 5, 1, \fbox{3}, \fbox{3}, 4, 0].
\end{align}
In conclusion, we have analyzed zero-modes counting in magnetized $T^{2g}/\mathbb{Z}_N$ models.
We have found some three-generation models in both $T^4/\mathbb{Z}_N$ and $T^6/\mathbb{Z}_{12}$ orbifolds using the wave functions with various chiralities. 

\section{Conclusion}
\label{conclusion}

We have studied and constructed magnetized $T^6$ model taking the chirality for spinor components into account. We have discovered the spinor wave functions with various chiralities under non-factorizable background magnetic fluxes $\pi N^T (\text{Im}\Omega)^{-1}$. 
When we define $M$ and impose the condition $N\text{Re}\Omega + iM\text{Im}\Omega \in \mathcal{H}^3$, we find the wave functions converge and satisfy both the Dirac equation and the boundary conditions in $T^6$. By use of the $R = N^{-1}M$ derived from the diagonalization for the magnetic fluxes, we define $z^{\prime}$ system where the background magnetic fluxes become positive-definite. In $z^{\prime}$, we find the Laplacian and its eigen equation, and the $D$-term SUSY condition. 
We also analyze the modular transformations of the wave functions under $Sp(6,\mathbb{Z})$ modular symmetry in $z^{\prime}$ system.
These play a crucial role in the analysis on the zero-mode number of the wave functions in magnetized $T^{2g}/\mathbb{Z}_N$. 
On the other hands, we have also discussed how to derive the Yukawa couplings taking both the chirality and the $SO(6)$ symmetry into account. 
We have found that each chirality for different sectors can be changed by linear combinations of fluxes if they are commutative. In other words, we can introduce the wave functions with different chiralities in each sector through these fluxes. Thus, we can calculate the Yukawa couplings in $T^6$, and we describe them specifically.
Lastly, we have researched the number of the zero-modes in magnetized $T^{2g}/\mathbb{Z}_N$ to confirm whether we can realize three-generation models or not. 
We have three-generation models when we consider the wave functions with various chiralities in magnetized $T^4/\mathbb{Z}_N$, $(N=2,3,4,6)$ and $T^6/\mathbb{Z}_{12}$.
It would be important to  verify the consistency with the Atiyah-Singer index theorem on orbfiolds \cite{Sakamoto:2020vdy,Kobayashi:2022tti,Aoki:2024rmf}, e.g. through the blow-up with localized flux and curvature \cite{Kobayashi:2019fma,Kobayashi:2019gyl,Kikuchi:2023clx}.
We would study it elsewhere.
It is also important to apply our results to realize quark and lepton masses and their mixing angles.

\appendix
\section*{Acknowledgement}
This work was supported by JPSP KAKENHI Grant Numbers, JP23K03375
(T.K.) and JP24KJ0249 (K.N.), and JST SPRING, Grant Number JPMJSP2119. (S.T.).
\section*{Appendix}

\section{Magnetized $T^{6}$ model with non-trivial scale factors}
In this paper, since we are interested in the theory of magnetized $T^{2g}/\mathbb{Z}_N$ orbifold with simply-laced Lie lattice, we have normalized the scale factors $2\pi r_i$, ($i=1,2,3$) as one.
In this Appendix, we consider magnetized $T^{6}$ model with the non-trivial scale factors. 
Also, we suppose $(\Omega^T \bar{\Omega})^T = \Omega^T \bar{\Omega}$. 
The following discussion is consistent with the discussion in the main body even if we introduce the asymmetric integer fluxes $N$ and complex structure moduli $\Omega$ under the $(1, 1)$-form background magnetic fluxes and the $F$-term SUSY condition $(N\Omega)^T = N\Omega$.
Specifically, when we consider arbitrary $2\pi r_i$, we can define complex coordinates on $\mathbb{C}^3$, $Z^{i}$, by using the complex coordinates on $T^6$, $z^i$, 
\begin{align}
\begin{aligned}
Z^{i} = 2\pi r_i z^i = 2\pi r_i (x^i + \Omega^{ik} y^k), \quad
\bar{Z}^{\bar{i}}  =  2\pi r_i \bar{z}^{\bar{i}} = 2\pi r_i (x^i + \bar{\Omega}^{ik} y^k).
\end{aligned}
\end{align}
The metric $g$ on $\mathbb{C}^3$ is defined as
\begin{align}
\begin{aligned}
g 
&=
\delta_{i, \bar{j}} dZ^i d\bar{Z}^{\bar{j}} \\ 
&= (2\pi r_i)^2  dz^i d\bar{z}^{\bar{i}} \\
&= 
(2\pi r_i)^2
\begin{bmatrix}
dx^T & dy^T    
\end{bmatrix}
\begin{bmatrix}
{\bf{1}}_3 & \text{Re}\Omega \\
(\text{Re}\Omega)^T & (\text{Re}\Omega)^T \text{Re}\Omega + (\text{Im}\Omega)^T \text{Im}\Omega
\end{bmatrix}
\begin{bmatrix}
 dx \\
 dy \\
\end{bmatrix}
.
\end{aligned}
\end{align}
In particular, we focus on not $z^i$ basis but $Z^i$ basis because we can describe various quantities taking $2\pi r_i$ into account such as the background magnetic fluxes, the Dirac operator, and the Laplacian. 

First, the background magnetic flux $F$ on $T^6$ is written as
\begin{align}
\begin{aligned}
F 
&= F_{i\bar{j}} (i dz^{i} \wedge d\bar{z}^{\bar{j}}) \\ 
&= \frac{F_{i\bar{j}}}{(2\pi r_i) (2\pi r_j)} [i (2\pi r_i) dz^i \wedge (2\pi r_j) d \bar{z}^{\bar{j}}] \\
&\equiv G_{i\bar{j}} (idZ^i \wedge d\bar{Z}^{\bar{j}}), \\
\end{aligned}
\end{align}
where $F_{i\bar{j}} = \pi [N^T (\text{Im}\Omega)^{-1}]_{ij}$, and $G_{i\bar{j}} =  \frac{F_{i\bar{j}}}{(2\pi r_i) (2\pi r_j)}$.
The above formula comes from the following Dirac quantization condition
\begin{align}
\int_{T^6} dz^i d\bar{z}^{\bar{j}} F_{i\bar{j}}
=
\int_{T^6} dZ^i d\bar{Z}^{\bar{j}} G_{i\bar{j}}
= 2\pi N^T.
\end{align}
Note that we use the same $N^T$ and $\Omega$ in both $F_{i\bar{j}}$ and $G_{i\bar{j}}$. 
We can represent $G_{i\bar{j}}$ as
\begin{align}
\begin{aligned}
&G_{i\bar{j}} 
\equiv
\frac{[\pi \hat{N}^T (\text{Im}\Omega)^{-1}]_{ij}}{(2\pi r_i)(2\pi r_j)} +\frac{[\pi  \tilde{N}^T (\text{Im}\Omega)^{-1}]_{ij}}{(2\pi r_i)(2\pi r_j)} \\
&\equiv
\hat{N}^{\prime}
\begin{bmatrix}
1 & -q_1 & -q_2 \\
-q_1 & q^2_{1} & q_1 q_2 \\
-q_2 & q_1 q_2 & q^2_2 \\
\end{bmatrix}
\\
&+
\begin{bmatrix}
\tilde{N}^{\prime}_{11} q^2_1 + 2\tilde{N}^{\prime}_{12}  q_1 q_2 + \tilde{N}^{\prime}_{22} q^2_2 & \tilde{N}^{\prime}_{11} q_1  + \tilde{N}^{\prime}_{12} q_2  &  \tilde{N}^{\prime}_{12} q_1 +  \tilde{N}^{\prime}_{22} q_2 \\
 \tilde{N}^{\prime}_{11} q_1+  \tilde{N}^{\prime}_{12} q_2 & \tilde{N}^{\prime}_{11} & \tilde{N}^{\prime}_{12} \\
\tilde{N}^{\prime}_{12} q_1  +  \tilde{N}^{\prime}_{22} q_2& \tilde{N}^{\prime}_{12} & \tilde{N}^{\prime}_{22} \\
\end{bmatrix}
,
\end{aligned}
\end{align}
where $N = \hat{N} + \tilde{N}$.
When we replace $F_{i\bar{j}}$ with $G_{i\bar{j}}$, we obtain $U(1)$ gauge fields $A_{Z^i}, A_{\bar{Z}^{\bar{i}}}$ from $G_{i\bar{j}}$ and the following covariant derivatives 
\begin{align}
\begin{aligned}
D_{Z^i} = \partial_{Z^i} - i A_{Z^i}, \quad
\bar{D}_{\bar{Z}^{\bar{i}}} = \partial_{\bar{Z}^{\bar{i}}} - i A_{\bar{Z}^{\bar{i}}}.
\end{aligned}
\end{align}
We also get
\begin{align}
[D_{Z^i}, \bar{D}_{\bar{Z}^{\bar{j}}}] = G_{i\bar{j}},\quad 
D_{Z^i} = \frac{1}{2\pi r_i} D_{i}, \quad 
\bar{D}_{\bar{Z}^{\bar{i}}} = \frac{1}{2\pi r_i} \bar{D}_{i}.
\end{align}
We find the Dirac equation by using the Dirac operator
\begin{align}
i\slashed{D}_{r_i} \equiv i (\Gamma^{Z^i} D_{Z^i} + \Gamma^{\bar{Z}^{\bar{i}}} \bar{D}_{\bar{Z}^{\bar{i}}}), \quad i \slashed{D}_{r_i} \Psi = 0.
\end{align}
On the other hand, we also define the Laplacian $\Delta_{r_i}$
\begin{align}
\Delta_{r_i} \equiv -2\sum_{i=1}^{3}  \{ D_{Z^i}, \bar{D}_{\bar{Z}^{{\bar{i}}}}  \}. 
\end{align}
When we perform the following replacements, we can construct magnetized $T^{6}$ model with $2\pi r_i$ by making use of the similar way discussed in the main body
\begin{align}
\begin{aligned}
&z^i \Rightarrow Z^i,\quad z^i +e^i \Rightarrow Z^i + e^i, \quad z^i + \Omega^{ij}e^j \Rightarrow Z^i + \Omega^{ij} e^j, \\
&F_{i\bar{j}} \Rightarrow G_{i\bar{j}}, \quad 
\slashed{D} \Rightarrow \slashed{D}_{r_i}, \quad
\Delta \Rightarrow \Delta_{r_i}.
\end{aligned}
\end{align}
Since we have used $Z^i$, $G_{i\bar{j}}$, $N$, when we consider the boundary condition of the wave functions $\Phi_r (Z, \bar{Z})$ on magnetized $T^6$ written in $Z$, we obtain
\begin{align}
\begin{aligned}
&\Phi_r (Z^i + e^i, \bar{Z}^i +e^i )
= e^{i [G_{i\bar{j}} \text{Im}Z^{j}]} \Phi_r (Z^i, \bar{Z}^i), \\
&\Phi_r (Z^i + \Omega^{ij} e^j, \bar{Z}^i + \bar{\Omega}^{ij} e_j )
= e^{i \text{Im} [\bar{\Omega}_{ji} G_{j\bar{k}} Z^k] } \Phi_r (Z^i, \bar{Z}^i).
\end{aligned}
\end{align}
Also, when we define $Z^{\prime i} = (2\pi r_i) z^{\prime i}$, we find the wave functions with various chiralities and the scale factors, $\Phi^J_r (Z, \bar{Z})$,
\begin{align}
\Phi^J_r (Z,\bar{Z}) = \mathcal{N}_r \cdot 
f_r (Z, \bar{Z}) \cdot {\Theta}_r (Z, \bar{Z}),
\end{align}
where $\mathcal{N}_r$ denotes the normalization constant and
\begin{align}
\begin{aligned}
f_r (Z,\bar{Z}) 
&\equiv 
\exp{ \left[ i \left[  Z^{i} \frac{[\pi \hat{N}^T (\text{Im}\Omega)^{-1}]_{ij}}{(2\pi r_i) (2\pi r_j)} \text{Im} Z^j +  \bar{Z}^{i} \frac{[\pi \tilde{N}^T (\text{Im}\Omega)^{-1}]_{ij}}{(2\pi r_i) (2\pi r_j)} \text{Im} Z^j  \right] \right]},  \\
{\Theta}_r (Z, \bar{Z})
&\equiv
\theta
\begin{bmatrix}
J^T N^{-1} \\
0 \\
\end{bmatrix}
(\hat{N} Z + \tilde{N} \bar{Z}, \hat{N}\Omega + \tilde{N}\bar{\Omega}) \\
&= \theta
\begin{bmatrix}
J^T N^{-1} \\
0 \\
\end{bmatrix}
(N Z^{\prime}, N\Omega^{\prime} )
,
\end{aligned}
\end{align}
with $J \in \Lambda_{N^T}$, $(N\Omega)^T = N\Omega$, and $N\Omega^{\prime} \in \mathcal{H}^3$.

\section{Boundary conditions of wave functions}
In magnetized $T^6$ model,
we can verify that the wave functions $\psi = \psi^{j}_{N,M}$ ($j = J^T N^{-1}$, $J \in \Lambda_{N^T}$) shown in Eq. \eqref{psi} satisfy the boundary conditions Eq. \eqref{boundary}. In the following, we impose both  $(N\Omega)^T = N\Omega$ and  $N\text{Re}\Omega + iM \text{Im}\Omega \in \mathcal{H}^3$.
\footnote{We can verify that even if we introduce the asymmetric $N$ and $\Omega$ under the $(1,1)$-form background magnetic fluxes and $(N\Omega)^T = N\Omega$.}
On the boundary conditions for $e_k$, $(k = 1,2,3)$ directions, we obtain
\begin{align}
\psi (z+e_k) 
&=\mathcal{N} \cdot e^{i\pi [N\text{Re}(z+e_k) + i M \text{Im}{z}]^T (\text{Im}\Omega)^{-1} \text{Im}z} \\ \notag
&\cdot \sum_{m\in \mathbb{Z}^3} e^{\pi i (m + j)^T (N\text{Re}\Omega + iM\text{Im}\Omega) (m+j)} e^{2\pi i (m+j)^T (N \text{Re} z +i M\text{Im}z)} \cdot e^{2\pi i (m+j)^T N e_k}\\ \notag
&= e^{i\pi e^T_k (N^T (\text{Im}\Omega)^{-1} \text{Im}z) }  \psi (z) \\ \notag
&= e^{i\chi_{e_k} (z)} \psi(z),
\end{align}
where we have used $e^{2\pi i (m+j)^T N e_k} = 1$ for any $e_k$, ($k=1,2,3$).
Next we consider the boundary conditions for $\Omega e_k$ directions.  For each component, $f(z,\bar{z})$, $\hat{\Theta} (z, \bar{z})$, we find
\begin{align}
f(z+\Omega e_k, \bar{z}+\bar{\Omega} e_k) 
&= e^{ i\pi [N\text{Re} (z +\Omega e_k) +i M \text{Im} (z + \Omega e_k)]^T (\text{Im}\Omega)^{-1} \text{Im} (z+\Omega e_k)} \\ \notag
&= e^{i \pi [[\hat{N} (z + \Omega e_k)]^T + [\tilde{N} (\bar{z} + \bar{\Omega} e_k)]^T  ] (\text{Im}\Omega)^{-1} \text{Im}(z + \Omega e_k)}  \\ \notag
&=e^{i\pi [\hat{N}z + \tilde{N} \bar{z}]^T e_k}\cdot e^{i \pi [ (\hat{N}\Omega + \tilde{N}\bar{\Omega}) e_k]^T (\text{Im}\Omega)^{-1} \text{Im}(z + \Omega e_k)} \cdot f(z, \bar{z}),
\end{align}
\begin{align}
\begin{aligned}
&\hat{\Theta} (z + \Omega e_k, \bar{z} +\bar{\Omega}e_k) \\
&= \sum_{m} e^{\pi i (m+j)^T (N\text{Re}\Omega + iM\text{Im}\Omega) (m+j)}e^{2\pi i (m+j)^T [\hat{N} (z + \Omega e_k) + \tilde{N} (\bar{z} + \bar{\Omega}e_k)] } \\ 
&= \sum_{m} e^{\pi i (m+j)^T (N\text{Re}\Omega + iM\text{Im}\Omega) (m+j)} e^{2\pi i (m+j)^T [\hat{N} z + \tilde{N} \bar{z}]} e^{2\pi i (m+j)^T (\hat{N} \Omega +\tilde{N} \bar{\Omega})e_k } \\ 
&= \sum_{m} e^{\pi i (m+j + e_k)^T (N\text{Re}\Omega + iM\text{Im}\Omega) (m+ j + e_k)} e^{2\pi i (m+j + e_k)^T [\hat{N} z + \tilde{N} \bar{z}]}  e^{2\pi i (m+j)^T (\hat{N} \Omega +\tilde{N} \bar{\Omega})e_k } \\
&\cdot e^{-\pi i e^T_k (N\text{Re}\Omega + iM\text{Im}\Omega)e_k} \cdot e^{-2\pi i (m+j)^T (\hat{N}\Omega + \tilde{N}\bar{\Omega}) e_k} \cdot e^{-2\pi i e^T_k (\hat{N}z + \tilde{N}\bar{z})} \\ 
&= e^{-\pi i e^T_k (N\text{Re}\Omega + iM\text{Im}\Omega)e_k}\cdot  e^{-2\pi i e^T_k (\hat{N}z + \tilde{N}\bar{z})}\cdot \hat{\Theta} (z, \bar{z}),
\end{aligned}
\end{align}
where we have used $(N\text{Re}\Omega + iM \text{Im}\Omega)^T = N\text{Re}\Omega + iM \text{Im}\Omega$.
We then find
\begin{align}
\begin{aligned}
&\psi (z + \Omega e_k) \\ \notag
&= e^{i\pi [\hat{N}z + \tilde{N} \bar{z}]^T e_k}\cdot e^{i \pi [ (\hat{N}\Omega + \tilde{N}\bar{\Omega}) e_k]^T (\text{Im}\Omega)^{-1} \text{Im}(z + \Omega e_k)} \cdot e^{-\pi i e^T_k (N\text{Re}\Omega + iM\text{Im}\Omega)e_k} \\ \notag
&\cdot  e^{-2\pi i e^T_k (\hat{N}z + \tilde{N}\bar{z})}\cdot \psi(z) \\ \notag
&= e^{-i\pi (N\text{Re} z + iM\text{Im} z)e_k} \cdot e^{\pi i (N\text{Re}\Omega + iM\text{Im}\Omega ) (\text{Im}\Omega)^{-1} \text{Im}z e_k } \cdot \psi(z) \\ \notag
&= e^{i\pi e^{T}_k [N(\text{Re}\Omega (\text{Im}\Omega)^{-1} \text{Im}z - \text{Re}z)]} \psi(z) \\ \notag
&= e^{i\chi_{\Omega e_k} (z)} \psi (z).
\end{aligned}
\end{align}
Therefore, the wave functions satisfy the boundary conditions in magnetized $T^6$.
\section{Dirac equation and its solutions in magnetized $T^6$}
When we consider the Dirac equation and the $(1, 1)$-form background magnetic fluxes $F = F_{z\bar{z}} (idz\wedge d\bar{z})$, we suppose $\psi_{i+} = \alpha_i \psi$, ($\alpha_i \neq 0 $ for $i=1,2,3$). According to Eq. \eqref{dirac4}, the Dirac equation is given by 
\begin{align}
\left\{ \,
\begin{aligned}
(\alpha_1 D_2 + \alpha_2 D_1)\psi &= 0, \\
(\alpha_1 D_3 - \alpha_3 D_1)\psi &= 0, \\
(\alpha_1 \bar{D}_1 - \alpha_2 \bar{D}_2 + \alpha_3 \bar{D}_3)\psi &= 0.
\end{aligned}
\right.
\end{align} 
We can prove that the wave functions $\psi^{j}_{N,M}$ in  Eq. \eqref{psi} satisfy the Dirac equation Eq. {\eqref{dirac4} by using  Eq. \eqref{flux}. 
Take an example for $(\alpha_1 \bar{D}_1 - \alpha_2 \bar{D}_2 + \alpha_3 \bar{D}_3)\psi = 0$.
We consider $\psi = \mathcal{N} \cdot f(z,\bar{z}) \cdot \hat{\Theta} (z, \bar{z})$, where both $f(z,\bar{z})$ and $\hat{\Theta} (z, \bar{z})$ depend on $z$ and $\bar{z}$. We then analyze $f(z,\bar{z})$ and $\hat{\Theta} (z, \bar{z})$, respectively.
For $k=1,2,3$, we consider covariant derivative $\bar{D}_k = \bar{\partial}_k - i \bar{A}_k$, where $\bar{\partial}_k = \partial_{\bar{z}_k}$ and $\bar{A}_k = A_{\bar{k}}$:
\begin{align}
\begin{aligned}
\bar{D}_k \psi 
&= (\bar{\partial}_k - i \bar{A}_k) \psi \\ 
&= \mathcal{N} \cdot \bar{\partial}_k (f\cdot \hat{\Theta}) - i \bar{A}_k \psi \\
&= \mathcal{N} [ (\bar{\partial}_k f) \cdot \hat{\Theta} + f\cdot (\bar{\partial}_k \hat{\Theta}) ] - i \bar{A}_k \psi,
\end{aligned}
\end{align}
\begin{align}
\begin{aligned}
\bar{\partial}_k f 
&= f \cdot i\pi \bar{\partial}_{k}[ (\hat{N}z)^T (\text{Im}\Omega)^{-1} \text{Im}z - (\tilde{N}\bar{z})^T (\text{Im}\Omega)^{-1} \text{Im}\bar{z}    ] \\
&= -f\cdot \pi i \left[\frac{1}{2i} [ (\hat{N}z)^T (\text{Im}\Omega)^{-1} + (\tilde{N}\bar{z})^T (\text{Im}\Omega)^{-1} ]_k + (\tilde{N}^T (\text{Im}\Omega)^{-1} \text{Im}\bar{z})_k    \right],
\end{aligned}
\end{align}
where we have used the definitions $\text{Im}z = (z - \bar{z})/2i$, $\text{Im}\bar{z} =- \text{Im} z$. Then, derivation of the theta-functions is described as follows:
\begin{align}
\begin{aligned}
\bar{\partial}_k \hat{\Theta} (z,\bar{z})
&= \sum_{m} e^{\pi i (m+j)^T (N\text{Re}\Omega + iM \text{Im}\Omega) (m+j)} \cdot \bar{\partial}_k (e^{2\pi i (m+j)^T (\hat{N} z +\tilde{N} \bar{z})}) \\ 
&= \sum_{m} e^{\pi i (m+j)^T (N\text{Re}\Omega + iM \text{Im}\Omega) (m+j)} \cdot e^{2\pi i (m+j)^T (\hat{N} z +\tilde{N} \bar{z})} \cdot \bar{\partial}_k (2\pi i (m+j)^T \tilde{N} \bar{z}) \\ 
&= 2\pi i \sum_{m} (m+j)^T_{\ell} \tilde{N}_{\ell k} e^{\pi i (m+j)^T (N\text{Re}\Omega + iM \text{Im}\Omega) (m+j)} \cdot e^{2\pi i (m+j)^T (\hat{N} z +\tilde{N} \bar{z})} \\
&\equiv 2\pi i \sum_{m} (m+j)^T_{\ell} \tilde{N}_{\ell k} \cdot g(z,\bar{z}),
\end{aligned}
\end{align}
\begin{align}
\begin{aligned}
&\alpha_1 \bar{\partial}_1 \hat{\Theta}  - \alpha_2 \bar{\partial}_2 \hat{\Theta} + \alpha_3 \bar{\partial}_3 \hat{\Theta} \\ 
&= 2\pi i \sum_{m} (m+j)^T_{\ell} (\alpha_1 \tilde{N}_{\ell 1} - \alpha_2 \tilde{N}_{\ell 2} + \alpha_3 \tilde{N}_{\ell 3} )\cdot g(z,\bar{z}) \\
&= 2\pi i \sum_{m} (m+j)^T_{k} (\text{Im}\Omega)_{kr} \\ 
&\cdot (\alpha_1 [\tilde{N}^T(\text{Im}\Omega)^{-1}]_{1r} - \alpha_2 [\tilde{N}^T(\text{Im}\Omega)^{-1}]_{2r} + \alpha_3 [\tilde{N}^T(\text{Im}\Omega)^{-1}]_{3r} )\cdot g(z,\bar{z}) \\
&= 0,
\end{aligned}
\end{align}
where we can directly verify $\alpha_1 [\tilde{N}^T(\text{Im}\Omega)^{-1}]_{1r} - \alpha_2 [\tilde{N}^T(\text{Im}\Omega)^{-1}]_{2r} + \alpha_3 [\tilde{N}^T(\text{Im}\Omega)^{-1}]_{3r} = 0$.
Thus we find 
\begin{align}
\begin{aligned}
&(\alpha_1 \bar{D}_1 - \alpha_2 \bar{D}_2 + \alpha_3 \bar{D}_3) \\
&= \mathcal{N} [ \alpha_1 \bar{\partial}_1 f - \alpha_2 \bar{\partial}_2 f + \alpha_3 \bar{\partial}_3 f ] \cdot \hat{\Theta} -i( \alpha_1 \bar{A}_1 - \alpha_2 \bar{A}_2 + \alpha_3 \bar{A}_3  )\psi \\
&= -\mathcal{N}\cdot f \cdot \pi i \{
 \alpha_1 \left[ \frac{1}{2i} [ z^T \hat{N}^T (\text{Im} \Omega)^{-1} + \bar{z}^T \tilde{N}^T (\text{Im}\Omega)^{-1}  ]_1 + [\tilde{N}^{T} (\text{Im}\Omega)^{-1} \text{Im} \bar{z}]_1    \right]  \\
&-\alpha_2 \left[ \frac{1}{2i} [ z^T \hat{N}^T (\text{Im} \Omega)^{-1} + \bar{z}^T \tilde{N}^T (\text{Im}\Omega)^{-1}  ]_2 + [\tilde{N}^{T} (\text{Im}\Omega)^{-1} \text{Im} \bar{z}]_2     \right] \\
&+\alpha_3 \left[ \frac{1}{2i} [ z^T \hat{N}^T (\text{Im} \Omega)^{-1} + \bar{z}^T \tilde{N}^T (\text{Im}\Omega)^{-1}  ]_3 + [\tilde{N}^{T} (\text{Im}\Omega)^{-1} \text{Im} \bar{z}]_3     \right]  \}
\cdot \hat{\Theta} \\
&-i(\frac{\pi}{2} i[\alpha_1 [z^T N^{T} (\text{Im}\Omega)^{-1}]_1 - \alpha_2 [z^T N^{T} (\text{Im}\Omega)^{-1}]_2 + \alpha_3 [z^T N^{T} (\text{Im}\Omega)^{-1}]_3] )\psi \\
&= -\frac{\pi}{2} \cdot \psi \cdot [\bar{z}^T_j (\alpha_1  [\tilde{N}^T (\text{Im}\Omega)^{-1}]_{j1} -\alpha_2  [\tilde{N}^T (\text{Im}\Omega)^{-1}]_{j2} + \alpha_3  [\tilde{N}^T (\text{Im}\Omega)^{-1}]_{j3}  ) \\
&- z^{T}_j (\alpha_1  [\tilde{N}^T (\text{Im}\Omega)^{-1}]_{j1} -\alpha_2  [\tilde{N}^T (\text{Im}\Omega)^{-1}]_{j2} + \alpha_3  [\tilde{N}^T (\text{Im}\Omega)^{-1}]_{j3} ) ] \\
&= 0,
\end{aligned}
\end{align}
where we can also verify
\begin{align}
\begin{aligned}
&\bar{z}^T_j (\alpha_1  [\tilde{N}^T (\text{Im}\Omega)^{-1}]_{j1} -\alpha_2  [\tilde{N}^T (\text{Im}\Omega)^{-1}]_{j2} + \alpha_3  [\tilde{N}^T (\text{Im}\Omega)^{-1}]_{j3}  ) \\ 
&- z^{T}_j (\alpha_1  [\tilde{N}^T (\text{Im}\Omega)^{-1}]_{j1} -\alpha_2  [\tilde{N}^T (\text{Im}\Omega)^{-1}]_{j2} + \alpha_3  [\tilde{N}^T (\text{Im}\Omega)^{-1}]_{j3} ) = 0
\end{aligned}
\end{align} 
when we use $q_1 = \alpha_2/\alpha_1$, $q_2 = -\alpha_3/\alpha_1$ and $F$.

In summary, we can prove $(\alpha_1 \bar{D}_1 - \alpha_2 \bar{D}_2 + \alpha_3 \bar{D}_3)\psi = 0$. Similarly, we can also verify $(\alpha_1 D_2 + \alpha_2 D_1)\psi = 0$ and
$(\alpha_1 D_3 - \alpha_3 D_1)\psi = 0$. Thus the wave functions $\psi$ satisfy not only the boundary conditions but the Dirac equation in magnetized $T^6$.

\section{$Sp(2g,\mathbb{Z})$ symplectic modular group}
Generally, the symplectic modular group $Sp(2g, \mathbb{Z})$ is defined by the following $2g \times 2g$ integer matrices:
\begin{align}
\gamma 
=
\begin{bmatrix}
A & B \\
C & D \\
\end{bmatrix}
,
\end{align}
satisfying
\begin{align}
\gamma J \gamma^T = J, 
\quad 
J
=
\begin{bmatrix}
O_g & {\bf{1}}_g \\
-{\bf{1}}_g & O_g \\
\end{bmatrix}
.
\end{align}
Note that $A, B, C, D$ are $g \times g$ integer symmetric matrices, and $O_g$, ${\bf{1}}_3$ correspond to $g\times g$ zero, unit matrices respectively.

We adopt the following generators called $S$ and $T_i$, ($i=1,2,\cdots, \frac{g}{2} (g+1)$)
\begin{align}
S = 
\begin{bmatrix}
O_g & {\bf{1}}_g \\
-{\bf{1}}_g & O_g \\
\end{bmatrix}
, \quad
T_i
= 
\begin{bmatrix}
{\bf{1}}_g & B_i \\
O_g & {\bf{1}}_g \\
\end{bmatrix}
.
\end{align}
In particular, in the case of $g=2$, we represent each $B_i$ as
\begin{align}
B_1
=
\begin{bmatrix}
1 & 0 \\
0& 0 \\
\end{bmatrix}
, 
B_2
=
\begin{bmatrix}
0 & 0 \\
0& 1 \\
\end{bmatrix}
, 
B_3
=
\begin{bmatrix}
0 & 1 \\
1& 0 \\
\end{bmatrix}
.
\end{align}
In the case of $g=3$, 
\begin{align}
B_1
=
\begin{bmatrix}
1 & 0 & 0\\
0& 0 & 0 \\
0 & 0 & 0 \\
\end{bmatrix}
, 
B_2
=
\begin{bmatrix}
0 & 0 & 0\\
0& 1 & 0 \\
0 & 0 & 0 \\
\end{bmatrix}
, 
B_3
=
\begin{bmatrix}
0 & 0 & 0\\
0& 0 & 0 \\
0 & 0 & 1 \\
\end{bmatrix}
, \\
B_4
=
\begin{bmatrix}
0 & 1 & 0\\
1 & 0 & 0 \\
0 & 0 & 0 \\
\end{bmatrix}
,
B_5
=
\begin{bmatrix}
0 & 0 & 0\\
0 & 0 & 1 \\
0 & 1 & 0 \\
\end{bmatrix}
,
B_6
=
\begin{bmatrix}
0 & 0 & 1 \\
0 & 0 & 0 \\
1 & 0 & 0 \\
\end{bmatrix}.
\end{align}
Under $Sp(2g,\mathbb{Z})$, complex coordinates $z$ and complex structure moduli $\Omega$ are transformed as
\begin{align}
&z \rightarrow (C\Omega + D)^{-1} z, \\
&\Omega \rightarrow (A\Omega + B)(C\Omega + D)^{-1}.
\end{align}

\section{Details of calculation for Yukawa couplings}
\subsection{Calculation for products of wave functions}
In this Appendix, we analyze the detail of Yukawa couplings. 
The product $\psi^{I}_{N_L, M_L} (z^{\prime}_L, \Omega^{\prime}_L)\cdot
\psi^{J}_{N_R, M_R} (z^{\prime}_R, \Omega^{\prime}_R)$ can be calculated as
\begin{align}
\begin{aligned}
&\psi^{I}_{N_L, M_L} (z^{\prime}_L, \Omega^{\prime}_L)\cdot
\psi^{J}_{N_R, M_R} (z^{\prime}_R, \Omega^{\prime}_R) \\ 
&\propto 
e^{i\pi [N_L z^{\prime}_L]^T (\text{Im}\Omega)^{-1} \text{Im} z} \cdot 
e^{i\pi [N_R z^{\prime}_R]^T (\text{Im}\Omega)^{-1} \text{Im} z}\cdot 
\Theta^{I}_{N_L}(z^{\prime}_L,\Omega^{\prime}_L) \cdot
\Theta^{J}_{N_R}(z^{\prime}_R,\Omega^{\prime}_R) \\
&= e^{i\pi [(N_L + N_R)\text{Re}z + i(M_L + M_R)\text{Im}z]^T (\text{Im}\Omega)^{-1} \text{Im} z} \cdot 
\Theta^{I}_{N_L}(z^{\prime}_L,\Omega^{\prime}_L) \cdot
\Theta^{J}_{N_R}(z^{\prime}_R,\Omega^{\prime}_R) \\
&= e^{i\pi [x^T (N_L + N_R) y + y^T [(N_L + N_R)\text{Re}\Omega + i (M_L + M_R) \text{Im}\Omega]y]} \cdot 
\Theta^{I}_{N_L}(z^{\prime}_L,\Omega^{\prime}_L) \cdot
\Theta^{J}_{N_R}(z^{\prime}_R,\Omega^{\prime}_R) \\
&= e^{i\pi [x^T (N_L + N_R) y + y^T [(N_L + N_R)\text{Re}\Omega + i (M_L + M_R) \text{Im}\Omega]y]} \\ \cdot 
&\sum_{\vec{\ell}_1, \vec{\ell}_2 \in \mathbb{Z}^g} e^{i\pi (\vec{\ell}_1 + \vec{i})^T (N_L \text{Re}\Omega + i M_L \text{Im}\Omega) (\vec{\ell}_1 + \vec{i})} \cdot e^{i\pi (\vec{\ell}_2 + \vec{j})^T (N_R \text{Re}\Omega + i M_R \text{Im}\Omega) (\vec{\ell}_2 + \vec{j})} \\ \cdot 
&e^{2i\pi (\vec{\ell}_1 + \vec{i})^T (N_L x + (N_L \text{Re}\Omega + i M_L \text{Im}\Omega)y)} \cdot e^{2i\pi (\vec{\ell}_2 + \vec{j})^T (N_R x + (N_R \text{Re}\Omega + i M_R \text{Im}\Omega)y)} \\
&\equiv e^{i\pi [x^T (N_L + N_R) y + y^T [(N_L + N_R)\text{Re}\Omega + i (M_L + M_R) \text{Im}\Omega]y]} \\ \cdot 
&\sum_{\vec{\ell}_1, \vec{\ell}_2 \in \mathbb{Z}^g} e^{i\pi (i) [\vec{L}^T \hat{Q}_{N, M} \vec{L} ]} \cdot e^{2\pi i [\vec{L}^T Q_{N} \vec{X}]} \cdot 
e^{2 \pi i (i) [\vec{L}^T \hat{Q}_{N, M} \vec{Y} ]}
.
\end{aligned}
\end{align}
In Eqs. \eqref{hoge1}, \eqref{hoge2} and \eqref{hoge3}, we have defined 
\begin{align}
\vec{L} = 
\begin{bmatrix}
 \vec{i} + \vec{\ell}_1 \\
 \vec{j} + \vec{\ell}_2 \\
\end{bmatrix}
,
\vec{X} =
\begin{bmatrix}
\vec{x} \\
\vec{x} \\
\end{bmatrix}
,
\vec{Y} =
\begin{bmatrix}
\vec{y} \\
\vec{y} \\
\end{bmatrix}
,
\end{align}
and the following $2g \times 2g$ matrices:
\begin{align}
\begin{aligned}
&Q_{N} =
\begin{bmatrix}
N_L & O \\
O & N_R \\
\end{bmatrix}
,\\ 
&\hat{Q}_{N, M}
=
\begin{bmatrix}
M_L \text{Im} \Omega - iN_L \text{Re}\Omega & O \\
O & M_R \text{Im} \Omega - iN_R \text{Re}\Omega \\
\end{bmatrix}
\equiv
\begin{bmatrix}
M^{\prime}_L & O \\
O & M^{\prime}_R \\
\end{bmatrix}
.
\end{aligned}
\end{align}
From these definitions, we can describe the following quantities
\begin{align}
\begin{aligned}
&\vec{L}^T \hat{Q}_{N,M} \vec{L} 
= \vec{L}^T ((T^{\prime})^{-1} T^{\prime}) \hat{Q}_{N,M} ((T^{\prime})^{-1}T^{\prime})^{T} \vec{L} \\
&= (\vec{L}^T (T^{\prime})^{-1}) ((T^{\prime}) \hat{Q}_{N,M} (T^{\prime})^{T}) ( ((T^{\prime})^{-1})^T \vec{L}), \\
&\vec{L}^T Q_N \vec{X} = \vec{L}^T  ((T^{\prime})^{-1} T^{\prime}) Q_{N} ((T^{\prime})^{-1}T^{\prime})^{T} \vec{X} \\
&= ( \vec{L}^T  (T^{\prime})^{-1} ) (T^{\prime} Q_{N} (T^{\prime})^{T} ) (((T^{\prime})^{-1})^T \vec{X}), \\
&\vec{L}^T  \hat{Q}_{N,M}  \vec{Y} = \vec{L}^T  ((T^{\prime})^{-1} T^{\prime})  \hat{Q}_{N,M}  ((T^{\prime})^{-1}T^{\prime})^{T} \vec{Y} \\
&= (\vec{L}^T  (T^{\prime})^{-1} )( T^{\prime} \hat{Q}_{N,M} (T^{\prime})^{T} ) (((T^{\prime})^{-1})^T \vec{Y}),
\end{aligned}
\end{align}
and define the following matrices $\hat{Q}^{\prime}_{N,M}$, $Q^{\prime}_N$, 
\begin{align}
\begin{aligned}
&\hat{Q}^{\prime}_{N,M} 
\equiv T^{\prime} \hat{Q}_{N,M} (T^{\prime})^T 
=
\begin{bmatrix}
M^{\prime}_L + M^{\prime}_R & (M^{\prime}_L N^{-1}_L - M^{\prime}_R N_R^{-1}) \alpha^T, \\
\alpha (N^{-1}_L M^{\prime}_L  - N^{-1}_R M^{\prime}_R ) & \alpha (N^{-1}_L M^{\prime}_L N^{-1}_L  + N^{-1}_R M^{\prime}_R N^{-1}_R ) \alpha^T
\end{bmatrix}
,\\
&Q^{\prime}_N 
\equiv T^{\prime} Q_N (T^{\prime})^T 
=
\begin{bmatrix}
N_L + N_R & O \\
O & \alpha (N^{-1}_L + N^{-1}_R) \alpha^T \\
\end{bmatrix}
,
\end{aligned}
\end{align}
and the vectors $\vec{L}^T (T^{\prime})^{-1}$, $((T^{\prime})^{-1})^T \vec{X}$, and $((T^{\prime})^{-1})^T \vec{Y}$:
\begin{align}
\vec{L}^T (T^{\prime})^{-1}
=
\begin{bmatrix}
(\vec{i} + \vec{\ell}_1)(N^{-1}_L + N^{-1}_R)^{-1} N^{-1}_R + (\vec{j} + \vec{\ell}_2)(N^{-1}_L + N^{-1}_R)^{-1} N^{-1}_L \\ 
[(\vec{i} + \vec{\ell}_1) - (\vec{j} + \vec{\ell}_2)] (N^{-1}_L + N^{-1}_R)^{-1} \alpha^{-1} 
\end{bmatrix}
^{T},
\end{align}
\begin{align}
((T^{\prime})^{-1})^T \vec{L}
=
\begin{bmatrix}
N^{-1}_R (N^{-1}_L + N^{-1}_R)^{-1} (\vec{i} + \vec{\ell}_1) + N^{-1}_L (N^{-1}_L + N^{-1}_R)^{-1}  (\vec{j} + \vec{\ell}_2) \\ 
 (\alpha^{-1})^T (N^{-1}_L + N^{-1}_R)^{-1}  [(\vec{i} + \vec{\ell}_1) - (\vec{j} + \vec{\ell}_2)] 
\end{bmatrix}
,
\end{align}
\begin{align}
((T^{\prime})^{-1})^T \vec{X} 
= 
\begin{bmatrix}
x \\
0 \\
\end{bmatrix}
,\text{and}\quad
((T^{\prime})^{-1})^T \vec{Y} 
= 
\begin{bmatrix}
y \\
0 \\
\end{bmatrix}
.
\end{align}
According to Ref. \cite{Antoniadis:2009bg}, we notice that both the matrix $Q^{\prime}_N$ and the vectors $\vec{L}^T (T^{\prime})^{-1}$, $((T^{\prime})^{-1})^T \vec{L}$ are the same as ones with $\Omega = i 1_g$. 
In what follows, we consider the calculations for Yukawa couplings with the value of $\alpha = |N_L||N_R|1_g$. 
Then the product $\psi^{I}_{N_L, M_L} (z^{\prime}_L, \Omega^{\prime}_L)\cdot
\psi^{J}_{N_R, M_R} (z^{\prime}_R, \Omega^{\prime}_R)$ is rewritten as follows:
\begin{align}
\begin{aligned}
\label{process}
&\psi^{I}_{N_L, M_L} (z^{\prime}_L, \Omega^{\prime}_L)\cdot
\psi^{J}_{N_R, M_R} (z^{\prime}_R, \Omega^{\prime}_R) \\
&\propto e^{i\pi [x^T (N_L + N_R) y + y^T [(N_L + N_R)\text{Re}\Omega + i (M_L + M_R) \text{Im}\Omega]y]} \\ \cdot 
&\sum_{\vec{\ell}_1, \vec{\ell}_2 \in \mathbb{Z}^g} e^{i\pi (i) [\vec{L}^T \hat{Q}_{N, M} \vec{L} ]} \cdot e^{2\pi i [\vec{L}^T Q_{N} \vec{X}]} \cdot 
e^{2 \pi i (i) [\vec{L}^T \hat{Q}_{N, M} \vec{Y} ]} \\
&= 
e^{i\pi [x^T (N_L + N_R) y + y^T [(N_L + N_R)\text{Re}\Omega + i (M_L + M_R) \text{Im}\Omega]y]} \cdot \\
&\sum_{\vec{\ell}_1, \vec{\ell}_2 \in \mathbb{Z}^g} e^{i\pi (i) [(\vec{L}^T T^{-1}) (T \hat{Q}_{N,M} T^{T}) ( (T^{-1})^T \vec{L})]} \cdot e^{2\pi i [( \vec{L}^T  T^{-1} ) (T Q_{N} T^{T} ) ((T^{-1})^T \vec{X})]} \\
&\cdot e^{2 \pi i (i) [(\vec{L}^T  T^{-1} )( T \hat{Q}_{N,M} T^{T} ) ((T^{-1})^T \vec{Y}) ]} \\
&= e^{i\pi [x^T (N_L + N_R) y + y^T [(N_L + N_R)\text{Re}\Omega + i (M_L + M_R) \text{Im}\Omega]y]} \cdot \\
&\sum_{\vec{\ell}_1, \vec{\ell}_2 \in \mathbb{Z}^g} 
e^{i\pi (i) [((\vec{i} + \vec{\ell}_1)N_L + (\vec{j} + \vec{\ell}_2)N_R )(N_L + N_R)^{-1} (M^{\prime}_L + M^{\prime}_R)  (N_L + N_R)^{-1} (N_L (\vec{i} + \vec{\ell}_1) + N_R (\vec{j} + \vec{\ell}_2) )]} \cdot \\
&e^{2\pi i  [ ((\vec{i} + \vec{\ell}_1) N_L + (\vec{j} + \vec{\ell}_2) N_R ) (N_L + N_R)^{-1}] (N_L + N_R) x }\cdot e^{2\pi i (i) [ (\vec{i} + \vec{\ell}_1) N_L + (\vec{j} + \vec{\ell}_2) N_R ] (N_L + N_R)^{-1} (M^{\prime}_L + M^{\prime}_R )y} \cdot \\
&e^{2\pi i (i) [(\vec{i} + \vec{\ell}_1) - (\vec{j} + \vec{\ell}_2)](N^{-1}_L + N^{-1}_R)^{-1} \alpha^{-1}    \alpha (N^{-1}_L M^{\prime}_L - N^{-1}_R M^{\prime}_R)y } \cdot \\
&e^{\pi i (i) [((\vec{i} + \vec{\ell}_1) N_L + (\vec{j} + \vec{\ell}_2) N_R ) (N_L + N_R)^{-1} (M^{\prime}_L N^{-1}_L - M^{\prime}_R N^{-1}_R )] \alpha^T (\alpha^{-1})^T N_R (N_L + N_R)^{-1} N_L [(\vec{i} - \vec{j}) + (\vec{\ell}_1 - \vec{\ell}_2)] } \cdot \\
&e^{\pi i (i) [(\vec{i} - \vec{j}) + (\vec{\ell}_1 - \vec{\ell}_2)]N_L (N_L + N_R)^{-1} N_R \alpha^{-1} \alpha (N^{-1}_L M^{\prime}_L  - N^{-1}_R M^{\prime}_R  ) (N_L + N_R)^{-1} ( N_L (\vec{i} + \vec{\ell}_1)  +  N_R (\vec{j} + \vec{\ell}_2) ) } \cdot \\
&e^{\pi i (i) [(\vec{i} - \vec{j}) + (\vec{\ell}_1 - \vec{\ell}_2)](N^{-1}_L + N^{-1}_R)^{-1} \alpha^{-1} \alpha (N^{-1}_L M^{\prime}_L N^{-1}_L  + N^{-1}_R M^{\prime}_R  N^{-1}_R ) \alpha^T (\alpha^{-1})^T (N^{-1}_L + N^{-1}_R)^{-1}  [(\vec{i} - \vec{j}) + (\vec{\ell}_1 - \vec{\ell}_2)]} \\ 
&= e^{i\pi [x^T (N_L + N_R) y + y^T [(N_L + N_R)\text{Re}\Omega + i (M_L + M_R) \text{Im}\Omega]y]} \cdot \\
&\sum_{\vec{\ell}_3, \vec{\ell}_4 \in \mathbb{Z}^g} \sum_{\vec{m}}  
e^{i\pi (i) [ [(\vec{i}N_L + \vec{j} N_R + \vec{m}N_L )(N_L + N_R)^{-1} + \vec{\ell}_3] (M^{\prime}_L + M^{\prime}_R)  [(N_L + N_R)^{-1}  (N_L\vec{i} +N_R \vec{j}  + N_L\vec{m} ) + \vec{\ell}_3]]} \cdot \\
&e^{2\pi i  [ ( \vec{i}N_L + \vec{j} N_R + \vec{m}N_L  ) (N_L + N_R)^{-1} + \vec{\ell}_3 ] (N_L + N_R) x } \cdot e^{2\pi i (i) [ ( \vec{i}N_L + \vec{j} N_R + \vec{m}N_L  ) (N_L + N_R)^{-1} + \vec{\ell}_3  ]  (M^{\prime}_L + M^{\prime}_R )y} \cdot \\
&e^{2\pi i (i) [(\vec{i} - \vec{j} + \vec{m})  \frac{N_L (N_L + N_R)^{-1} N_R}{{|N_L|} {|N_R|}} + \vec{\ell}_4 ] [(|{N}_L| |{N}_R|) (N^{-1}_L M^{\prime}_L - N^{-1}_R M^{\prime}_R ) ] y} \cdot \\
&e^{i\pi (i) [ [(\vec{i}N_L + \vec{j} N_R + \vec{m}N_L )(N_L + N_R)^{-1} + \vec{\ell}_3] [(|{N}_L| |{N}_R|) ( M^{\prime}_L N^{-1}_L  - M^{\prime}_R N^{-1}_R  )] (\frac{N_R (N_L + N_R)^{-1} N_L }{|{N_L}| |{N_R}|} (\vec{i} - \vec{j} + \vec{m})   + \vec{\ell}_4 )  } \cdot \\
&e^{i\pi (i)  [(\vec{i} - \vec{j} + \vec{m})  (\frac{N_L (N_L + N_R)^{-1} N_R }{|{N_L}| |{N_R}|}  ) + \vec{\ell}_4  ] [(|{N}_L| |{N}_R|) (  N^{-1}_L M^{\prime}_L - N^{-1}_R M^{\prime}_R )]  [(N_L + N_R)^{-1} (N_L \vec{i} + N_R \vec{j}  + N_L \vec{m})+ \vec{\ell}_3] } \cdot \\
&e^{i\pi (i)  [(\vec{i} - \vec{j} + \vec{m})  (\frac{(N^{-1}_L + N^{-1}_R)^{-1} }{|{N_L}| |{N_R}|} )  + \vec{\ell}_4  ] [(|{N}_L| |{N}_R|)^2 (  N^{-1}_L M^{\prime}_L N^{-1}_L - N^{-1}_R M^{\prime}_R N^{-1}_R )] [\frac{(N^{-1}_L + N^{-1}_R)^{-1} }{|{N_L}| |{N_R}|} (\vec{i} - \vec{j} + \vec{m})   + \vec{\ell}_4 ]  },
\end{aligned}
\end{align}
where we have used the relation $(N^{-1}_L + N^{-1}_R)^{-1} = N_L (N_L + N_R)^{-1} N_R = N_R (N_L + N_R)^{-1} N_L$ under $[N_L, N_R] = 0$.
For $g=3$, on the other hand, the wave functions in Higgs sector become
\begin{align}
\begin{aligned}
\label{Higgssector}
&[\psi^{\vec{K}}_{N_H = N_L + N_R, M_H} (z^{\prime}_H, \Omega^{\prime}_H)]^{*} \\
&=
\mathcal{N}_k \cdot e^{-i \pi [x^T (N_L + N_R) y + y^T [(N_L + N_R)\text{Re}\Omega - i M_H \text{Im}\Omega]y ]} \cdot \Theta^{-K}_{-(N_L + N_R)} (\bar{z}^{\prime}_H, \bar{\Omega}^{\prime}_H)\\
&=
\mathcal{N}_k \cdot e^{-i \pi [x^T (N_L + N_R) y + y^T [(N_L + N_R)\text{Re}\Omega - i M_H \text{Im}\Omega]y ]} \cdot \\ 
&\sum_{\ell^{\prime}_3 \in \mathbb{Z}^3} e^{-i\pi (\ell^{\prime}_3+k)^T [(N_L + N_R)\text{Re}\Omega -iM_H \text{Im}\Omega] (\ell^{\prime}_3+k)}\cdot e^{-2\pi i (\ell^{\prime}_3+k)^T [(N_L + N_R)(x+(\text{Re}\Omega)y) - i M_H (\text{Im}\Omega)y]},
\end{aligned}
\end{align}
where $k= K N^{-1}_H = (-K) (-N^{-1}_H)$, $K \in \Lambda_{N_H}$ with Eq. \eqref{Higgstheta}, 
we can see the integration over $x$ in $C$
\begin{align}
\begin{aligned}
&\int_{0}^{1} d^3 x e^{i\pi x^T [N_L + N_R - (N_L + N_R)] y} \cdot \\
&\sum_{\vec{\ell_3}, \vec{\ell_4}, \vec{\ell^{\prime}}_3 \in \mathbb{Z}^{3}}  \sum_{\vec{m}} e^{2\pi i  [ ( \vec{i}N_L + \vec{j} N_R + \vec{m}N_L  ) (N_L + N_R)^{-1} + \vec{\ell}_3 ]^T (N_L + N_R) x } e^{-2\pi i (\vec{\ell^{\prime}}_3 + \vec{k} )^T (N_L + N_R) x }
\end{aligned}
\end{align}
and the following rule
\begin{align}
\begin{aligned}
\label{selection}
&\vec{\ell}_3 = \vec{\ell^{\prime}}_3, \\ 
&( \vec{i}N_L + \vec{j} N_R + \vec{m}N_L  )(N_L + N_R)^{-1}  = \vec{k}.
\end{aligned}
\end{align}
This is what is called the selection rule.
Note that we have used $N_H = N_L + N_R$ derived from the gauge symmetry.
Making use of Eqs. \eqref{process}, \eqref{Higgssector}, and \eqref{selection}, the $C$ in Eq. \eqref{C1} is calculated as follows
\begin{align}
\begin{aligned}
\label{CCC}
C 
&=
|-2i\text{Im}\Omega| \cdot \mathcal{N}_i \mathcal{N}_j \mathcal{N}_k \cdot \\
&\int_{0}^{1} d^3 y [ e^{-\pi [y^T [M_L + M_R + M_H ] (\text{Im}\Omega) y  ]}\cdot \\ &\sum_{\vec{\ell}_3, \vec{\ell}_4 \in \mathbb{Z}^3} \sum_{\vec{m}}  
e^{i\pi (i) [ [\vec{k}+ \vec{\ell}_3]^T [M^{\prime}_L + M^{\prime}_R + (M^{\prime}_H)^{*}]  [\vec{k} + \vec{\ell}_3]]} \cdot 
e^{2\pi i (i) [ \vec{k} + \vec{\ell}_3  ]^T  [M^{\prime}_L + M^{\prime}_R +(M^{\prime}_H)^{*}]y} \cdot \\
&e^{2\pi i (i) [\vec{h} + \vec{\ell}_4 ]^T [(|{N}_L| |{N}_R|) (N^{-1}_L M^{\prime}_L - N^{-1}_R M^{\prime}_R ) ] y} \cdot \\
&e^{i\pi (i) [ [\vec{k} + \vec{\ell}_3]^T [(|{N}_L| |{N}_R|) ( M^{\prime}_L N^{-1}_L  - M^{\prime}_R N^{-1}_R  )] [\vec{h} + \vec{\ell}_4 ]  } \cdot \\
&e^{i\pi (i)  [\vec{h} + \vec{\ell}_4  ]^T [(|{N}_L| |{N}_R|) (  N^{-1}_L M^{\prime}_L - N^{-1}_R M^{\prime}_R )]  [\vec{k}+ \vec{\ell}_3] } \cdot \\
&e^{i\pi (i)  [\vec{h} + \vec{\ell}_4  ]^T [(|{N}_L| |{N}_R|)^2 (  N^{-1}_L M^{\prime}_L N^{-1}_L - N^{-1}_R M^{\prime}_R N^{-1}_R )] [\vec{h} + \vec{\ell}_4 ]  } ],
\end{aligned}
\end{align}
where $(M^{\prime}_H)^{*} \equiv M_H \text{Im}\Omega + i (N_L + N_R)\text{Re}\Omega$ and $\vec{h} \equiv \frac{N_R (N_L + N_R)^{-1} N_L }{|{N_L}| |{N_R}|} (\vec{i} - \vec{j} + \vec{m})$.
\subsection{Calculation for $C$}
We have found the integral part of the Yukawa couplings.
In particular, when we take $\alpha = \text{L.C.M.} (|N_L|, |N_R|) 1_g \equiv L_R 1_g$ and $g=3$ under the condition that $|N_L|$ and $|N_R|$ are not coprime, we obtain the integral part of the Yukawa couplings as follows
\begin{align}
\label{lcmYukawa123}
\begin{aligned}
C
&=|-2i \text{Im}\Omega|\cdot \mathcal{N}_i \mathcal{N}_j \mathcal{N}_k \\ \notag
&\cdot \int_{0}^{1}  d^3 {y} \left[ e^{-\pi [y^T [M_L + M_R + M_H ] (\text{Im}\Omega) y  ]} \cdot \sum_{\vec{\ell_3}, \vec{\ell_4} \in \mathbb{Z}^3}  \sum_{\vec{p}, \vec{\tilde{p}}}  e^{\pi i (i) [\vec{\bf{K}} + \vec{\bf{L_1}}]^T \cdot {\bf{\hat{Q^{\prime}  }  } }^{L_R}_{N, M^{\prime} } \cdot    [\vec{\bf{K}} + \vec{\bf{L_1}}] } \cdot e^{2\pi i (i)  [\vec{\bf{K}} + \vec{\bf{L_1}}]^T \vec{\bf{{Y}^{\prime}}}^{L_R}_{N, M^{\prime}}   }    \right],
\end{aligned}
\end{align}
where
\begin{align}
\begin{aligned}
&\vec{\bf{L_1}} = 
\begin{bmatrix}
\vec{\ell_3} \\
\vec{\ell_4} \\
\end{bmatrix}
,
\quad
\vec{\bf{K}}
=
\begin{bmatrix}
\vec{k} \\
\frac{N_R (N_L + N_R)^{-1} N_L}{L_R } (\vec{i} - \vec{j} + \vec{\tilde{m}}) 
\end{bmatrix}
,
\end{aligned}
\end{align}
and
\begin{align}
\begin{aligned}
\vec{\bf{{Y}^{\prime }}}^{L_R}_{N, M^{\prime}} 
\equiv
\begin{bmatrix}
[M^{\prime}_L + M^{\prime}_R +(M^{\prime}_H)^{*}] {y} \\
[L_R (M^{\prime}_L N^{-1}_{L} - M^{\prime}_R N^{-1}_{R} )] y
\end{bmatrix}
=
\begin{bmatrix}
[(M_L + M_R + M_H) \text{Im} \Omega] {y} \\
L_R  [(R_L - R_R)  \text{Im} \Omega ] y
\end{bmatrix}
\end{aligned}
,
\end{align}
\begin{align}
\begin{aligned}
{\bf{\hat{Q^{\prime}  }  } }^{L_R}_{N, M^{\prime}} 
\equiv 
\begin{bmatrix}
M^{\prime}_L + M^{\prime}_R +(M^{\prime}_H)^{*} & L_R  (M^{\prime}_L N^{-1}_L - M^{\prime}_R N_R^{-1})  \\
L_R  (N^{-1}_L M^{\prime}_L  - N^{-1}_R M^{\prime}_R ) & (L_R)^2 (N^{-1}_L M^{\prime}_L N^{-1}_L  + N^{-1}_R M^{\prime}_R N^{-1}_R )
\end{bmatrix}
\end{aligned}
,
\end{align}
where $\text{Re} {\bf{\hat{Q^{\prime}  }  } }^{L_R}_{N, M^{\prime}} > 0$,
with the values of $M^{\prime}_L = M_L \text{Im} \Omega - iN_L \text{Re}\Omega$, $M^{\prime}_R = M_R \text{Im} \Omega - iN_R \text{Re}\Omega$, and $(M^{\prime}_H)^{*} = M_H \text{Im}\Omega + i (N_L + N_R)\text{Re}\Omega$, similar to the calculations under $\alpha = |N_L||N_R| 1_g$. Also, we have replaced $\vec{m}$ with the conditions $\vec{m} N_L (N_L + N_R)^{-1} \in \mathbb{Z}$ and $\vec{\tilde{m}} + \vec{p} (N_L + N_R) N^{-1}_L + \vec{\tilde{p}} (N_L + N_R) N^{-1}_R$, where $\vec{\tilde{m}}$ is the solution for $(N_L + N_R) \vec{k} = N_L \vec{i} + N_R \vec{j} + N_L \vec{m}$, and  $\vec{p} \in \Lambda_{L_R N^{-1}_R }, \vec{\tilde{p}} \in \Lambda_{L_R N^{-1}_L }$.

In general, when we impose Eq. \eqref{condition1}, we obtain the $C$
\begin{align}
C 
= 
&\frac{|-2i\text{Im} \Omega| \cdot \mathcal{N}_i\mathcal{N}_j \mathcal{N}_k}{\sqrt{|(M_L + M_R + M_H) \text{Im}\Omega|}}  
\cdot \sum_{\vec{p}, \vec{\tilde{p}}} 
\theta
\begin{bmatrix}
 \frac{N_R (N_L + N_R)^{-1} N_L}{L_R} (\vec{i}-\vec{j}+\vec{\tilde{m}}) \\
 \vec{0} \\
\end{bmatrix}
(\vec{0}, \tilde{\Omega}_{Y} ) \\ \notag
&\cdot \delta_{ (N_L + N_R) \vec{k}, N_L \vec{i} + N_R \vec{j} + N_L \vec{\tilde{m}}}
,
\end{align}
where 
\begin{align}
\tilde{\Omega}_Y 
\equiv 
&(L_R)^2 [(N^{-1}_L + N^{-1}_R)\text{Re}\Omega \\ \notag
&+ i[(R_L N^{-1}_L + R_R N^{-1}_R ) - (R_L - R_R) (M_L + M_R + M_H)^{-1} (R_L - R_R)]\text{Im}\Omega].
\end{align}
We also make use of the following Gaussian integral
\begin{align}
\label{Gaussian123}
\int_{-\infty}^{\infty} e^{-\frac{1}{2} {\bf{x}}^T A {\bf{x}} + b^T {\bf{x}}}  d{\bf{x}}
= 
\sqrt{\frac{(2\pi)^n}{ |A| }} \cdot e^{\frac{1}{2} b^T A^{-1}b},
\end{align}
where $A$ is a symmetric positive-definite matrix. Lastly, note that the Riemann-Theta functions have both uniform convergence and absolutely convergent.
Then, we would like to prove the above $C$.
From Eq. \eqref{lcmYukawa123}, we get
\begin{align}
\label{lcmYukawa1}
\begin{aligned}
&C
=|-2i \text{Im}\Omega|\cdot \mathcal{N}_i \mathcal{N}_j \mathcal{N}_k  \\
&\cdot\int_{0}^{1}  d {\bf{y}} \left[ e^{-\pi [{\bf{y}}^T [M_L + M_R + M_H ] (\text{Im}\Omega) {\bf{y}}  ]} \ \sum_{\vec{\ell_3}, \vec{\ell_4} \in \mathbb{Z}^3}  \sum_{\vec{p}, \vec{\tilde{p}}}  e^{-\pi [\vec{\bf{K}} + \vec{\bf{L_1}}]^T  {\bf{\hat{Q^{\prime}  }  } }^{L_R}_{N, M^{\prime} }    [\vec{\bf{K}} + \vec{\bf{L_1}}] } \cdot e^{-2\pi  [\vec{\bf{K}} + \vec{\bf{L_1}}]^T \vec{\bf{{Y}^{\prime}}}^{L_R}_{N, M^{\prime}}   }    \right],
\end{aligned}
\end{align}
where
\begin{align}
&(\vec{\bf{K}} + \vec{\bf{L_1}})^T 
= 
\begin{bmatrix}
(\vec{k} + \vec{\ell_3})^T, &
\left( \frac{N_R (N_L + N_R)^{-1} N_L}{L_R } (\vec{i} - \vec{j} + \vec{\tilde{m}}) +\vec{\ell_4} \right)^T
\end{bmatrix}
,\\ \notag
&\vec{\bf{K}} + \vec{\bf{L_1}} 
= 
\begin{bmatrix}
\vec{k} + \vec{\ell_3} \\
\frac{N_R (N_L + N_R)^{-1} N_L}{L_R } (\vec{i} - \vec{j} + \vec{\tilde{m}}) +\vec{\ell_4}
\end{bmatrix}
,
\end{align}
\begin{align}
&{\bf{\hat{Q^{\prime}  }  } }^{L_R}_{N, M^{\prime}} 
=  \\ \notag
&\begin{bmatrix}
(M_L + M_R + M_H)\text{Im}\Omega & L_R  (R_L - R_R)\text{Im}\Omega  \\
L_R  (R_L - R_R)\text{Im}\Omega & (L_R)^2 [(R_L N^{-1}_L + R_R N^{-1}_R)(\text{Im}\Omega) - i (\text{Re}\Omega)(N^{-1}_L + N^{-1}_R)]
\end{bmatrix}
,
\end{align}
\begin{align}
\vec{\bf{{Y}^{\prime }}}^{L_R}_{N, M^{\prime}} 
&=
\begin{bmatrix}
[(M_L + M_R + M_H) \text{Im} \Omega] {\bf{y}} \\
L_R  [(R_L - R_R)  \text{Im} \Omega ] {\bf{y}}
\end{bmatrix}
.
\end{align}
In what follows, note that we obtain accurate values of the integral part of the $C$ when we change the order of the integral and the series, and the finite sum and the infinite sum guaranteed by both uniform convergence and absolutely convergent of the Riemann Theta functions:
\begin{align}
\begin{aligned}
&\int_{0}^{1}  d {\bf{y}} \left[ e^{-\pi [{\bf{y}}^T [M_L + M_R + M_H ] (\text{Im}\Omega) {\bf{y}}  ]} \ \sum_{\vec{\ell_3}, \vec{\ell_4} \in \mathbb{Z}^3}  \sum_{\vec{p}, \vec{\tilde{p}}}  
e^{-\pi [\vec{\bf{K}} + \vec{\bf{L_1}}]^T  {\bf{\hat{Q^{\prime}  }  } }^{L_R}_{N, M^{\prime} }    [\vec{\bf{K}} + \vec{\bf{L_1}}] } 
\cdot 
e^{-2\pi  [\vec{\bf{K}} + \vec{\bf{L_1}}]^T \vec{\bf{{Y}^{\prime}}}^{L_R}_{N, M^{\prime}}   }    \right] \\ \notag
&= \sum_{\vec{p}, \vec{\tilde{p}}} \sum_{\vec{\ell_4} \in \mathbb{Z}^3} e^{-\pi \left[  \frac{N_R N^{-1}_H N_L}{L_R} (\vec{i} -\vec{j} + \vec{\tilde{m}}) +\vec{\ell}_4   \right]^T  (L_R)^2 [(R_LN^{-1}_L + R_R N^{-1}_R)\text{Im}\Omega -i(N^{-1}_L + N^{-1}_R)\text{Re}\Omega]\left[  \frac{N_R N^{-1}_H N_L}{L_R} (\vec{i} -\vec{j} + \vec{\tilde{m}}) +\vec{\ell}_4   \right]} \\ \notag
&\cdot \sum_{\vec{\ell}_3 \in \mathbb{Z}^3} \int_{0}^{1} d {\bf{y}} 
\Bigg[ e^{-\pi [{\bf{y}} + \vec{k} + \vec{\ell}_3 ]^T 
(M_L + M_R + M_H ) (\text{Im}\Omega) [{\bf{y}} + \vec{k} + \vec{\ell}_3 ]  } 
 \\
&\qquad \quad \quad \qquad \times e^{-2\pi \left[L_R[(R_L - R_R)\text{Im}\Omega]^T \left[  \frac{N_R N^{-1}_H N_L}{L_R} (\vec{i} -\vec{j} + \vec{\tilde{m}}) +\vec{\ell}_4   \right]\right]^T  [{\bf{y}} + \vec{k} + \vec{\ell}_3 ]  }    \Bigg] \\ \notag
&=
\sum_{\vec{p}, \vec{\tilde{p}}} \sum_{\vec{\ell_4} \in \mathbb{Z}^3} e^{-\pi \left[  \frac{N_R N^{-1}_H N_L}{L_R} (\vec{i} -\vec{j} + \vec{\tilde{m}}) +\vec{\ell}_4   \right]^T  (L_R)^2 [(R_LN^{-1}_L + R_R N^{-1}_R)\text{Im}\Omega -i(N^{-1}_L + N^{-1}_R)\text{Re}\Omega]\left[  \frac{N_R N^{-1}_H N_L}{L_R} (\vec{i} -\vec{j} + \vec{\tilde{m}}) +\vec{\ell}_4   \right]} \\ \notag
&\cdot \int_{-\infty}^{\infty} d {\bf{y}^{\prime}} 
\left[ e^{-\pi {\bf{y}}^{\prime T} 
(M_L + M_R + M_H ) (\text{Im}\Omega) {\bf{y}}^{\prime}  } 
\cdot 
e^{-2\pi \left[L_R[(R_L - R_R)\text{Im}\Omega]^T \left[  \frac{N_R N^{-1}_H N_L}{L_R} (\vec{i} -\vec{j} + \vec{\tilde{m}}) +\vec{\ell}_4   \right]\right]^T  {\bf{y}}^{\prime}  }    \right] \\ \notag
&=
\frac{1}{\sqrt{|(M_L + M_R + M_H)\text{Im}\Omega|}}
\sum_{\vec{p}, \vec{\tilde{p}}}
\theta
\begin{bmatrix}
\frac{N_R N^{-1}_H N_L}{L_R} (\vec{i} -\vec{j} + \vec{\tilde{m}}) \\
\vec{0}
\end{bmatrix}
(\vec{0}, \tilde{\Omega}_Y)
,
\end{aligned}
\end{align}
where ${\bf{y}}^{\prime} \equiv  {\bf{y}} + \vec{k} + \vec{\ell}_3$ and
\begin{align}
\tilde{\Omega}_Y 
\equiv 
&(L_R)^2 [(N^{-1}_L + N^{-1}_R)\text{Re}\Omega \\ \notag
&+ i[(R_L N^{-1}_L + R_R N^{-1}_R ) - (R_L - R_R) (M_L + M_R + M_H)^{-1} (R_L - R_R)]\text{Im}\Omega].
\end{align}
We have made use of the Gaussian integral in Eq. \eqref{Gaussian123}.
By use of the above values and the selection rules, we finally find the $C$
\begin{align}
C 
=
&\frac{|-2i\text{Im}\Omega|\cdot \mathcal{N}_i \mathcal{N}_j \mathcal{N}_k }{\sqrt{|(M_L+M_R+M_H)\text{Im}\Omega|}}
\cdot
\sum_{\vec{p}, \vec{\tilde{p}}} 
\theta
\begin{bmatrix}
  \frac{N_R N_H^{-1} N_L}{L_R } (\vec{i} - \vec{j} + \vec{\tilde{m}}) \\
  \vec{0}
\end{bmatrix}
(\vec{0}, \tilde{\Omega}_Y ) \\ \notag
&\cdot
\delta_{N_H \vec{k}, N_L \vec{i} + N_R \vec{j} + N_L \vec{\tilde{m}}}
.
\end{align}

\section{Existence of  fluxes with negative determinants in $T^4/\mathbb{Z}_{4}$}

We search for the existence of fluxes with specific determinants.
The results (from Table~\ref{traces_mod_4_n}) are: 
\begin{itemize}
  \item when $tr\rho=i$, $\det N = 3$ mod 4,
  \item when $tr\rho=-i$, $\det N = 3$ mod 8,
  \item when $tr\rho = 0$, $\det N =2$ mod 4,
  \item when $tr\rho = 1$, $\det N = 1$ mod 4,
  \item when $tr\rho = -1$, $\det N = 1$ mod 4,
  \item when $tr\rho = 2i$, $\det N =0,12$ mod 16,
  \item when $tr\rho = 0$, $\det N = 0$ mod 4,
  \item when $tr\rho = 1+i$, $\det N = 0$ mod 4,
  \item when $tr\rho = -1+i$, $\det N = 0$ mod 4.
\end{itemize}

  We will use the general form of the flux below to analyze the modulus of each component:
  \begin{align}
    N=
    \begin{pmatrix}
      4a+k_1&&2c+k_3\\2c+k_3&&4b+k_2
    \end{pmatrix},
  \end{align}
  where $a,b,c,k_i\in\mathbb{Z}$. The determinant is expressed as
  \begin{align}
    \begin{split}
      \det N &= (4a+k_1)(4b+k_2)-(2c+k_3)^2 \\
      &= (16ab+4ak_2+4bk_1)-4c^2-4ck_3-k_3^2+ k_1k_2 <0.
    \end{split}\label{determinant_Z4}
  \end{align}
  For $tr\rho=i,0,\pm1,\pm1+i$, all the fluxes with an appropriate determinant can be constructed.
  Note that we are only considering fluxes with negative determinants.
  Thus we just need to pick one of the possible $(k_1,k_2,k_3)$ (given in Table~\ref{possible_traces_4}) and fix two of $a,b,c$ to prove the mod $4$ condition.
  \begin{proof}
    Let $tr\rho = i$, $(k_1,k_2,k_3)=(0,1,1)$, $b=c=0$. Following eq.(~\ref{determinant_Z4}), the determinant is expressed as:
    \begin{align}
      \det N = 4a-1.
    \end{align}
    Since $a\in\mathbb{Z}$, we take $a<1$ to obtain any $\det N = 3$ mod $4$ for any negative determinant with the corresponding mod $4$ structure.
  \end{proof}
  \begin{proof}
    Let $tr\rho = 1$, $(k_1,k_2,k_3)=(1,1,0)$, $b=c=0$. The determinant is
    \begin{align}
      \det N = 4a+1.
    \end{align}
    We take $a<0$ and obtain $\det N = 1$ mod $4$.
  \end{proof}
  \begin{proof}
    Let $tr\rho = -1$, $(k_1,k_2,k_3)=(3,3,0)$, $b=-1,c=0$. The determinant is
    \begin{align}
      \det N = -4(a+1)+1.
    \end{align}
    With $a>-1$ we obtain $\det N = 1$ mod $4$.
  \end{proof}
  \begin{proof}
    Let $tr\rho = 1+i$, $(k_1,k_2,k_3)=(0,1,0)$, $b=c=0$. The determinant is
    \begin{align}
      \det N = 4a.
    \end{align}
    With $a<0$ we obtain $\det N = 0$ mod $4$.
  \end{proof}
  \begin{proof}
    Let $tr\rho = -1+i$, $(k_1,k_2,k_3)=(0,3,0)$. The determinant is
    \begin{align}
      \det N = -4a.
    \end{align}
    With $a>0$ we obtain $\det N = 0$ mod $4$.
  \end{proof}
  \begin{proof}
    Let $tr\rho = 0$, $(k_1,k_2,k_3)=(1,2,0)$, $a=c=0$. The determinant is
    \begin{align}
      \det N = 4b+2.
    \end{align}
    With $b<0$ we obtain $\det N = 2$ mod $4$.
  \end{proof}
  \begin{proof}
    Let $tr\rho = 0$, $(k_1,k_2,k_3)=(0,2,0)$. First, take $b=0, c=1$. The determinant is
    \begin{align}
      \det N = 8a-1.
    \end{align}
    Next, take $b=c=0$. The determinant is
    \begin{align}
      \det N = 8a.
    \end{align}
    Combining the two and taking $a<1$ we get $\det N = 0$ mod $4$.
  \end{proof}
  The other cases do not allow all mod $4$ values. However, each only has one possible $(k_1,k_2,k_3)$. Below, we prove which ones are allowed for each case.
  \begin{proof}
    Let $tr\rho = -i$, $(k_1,k_2,k_3)=(2,2,1)$, $b=0$. The determinant is
    \begin{align}
      \det N = 8a-4(c(c+1))+3.
    \end{align}
    The bracket in the second term is always even so we take $c=0$. Then with $a<0$ we get $\det N = 3$ mod $8$.
  \end{proof}
  \begin{proof}
    Let $tr\rho = 2i$, $(k_1,k_2,k_3)=(0,0,0)$, $b=1$. The determinant is
    \begin{align}
      \det N = 16a-4c^2.
    \end{align}
    For $c=0$ we get $\det N = 16a$. For $c=1$ we get $\det N = 16a-4$. With $a<1$ we get $\det N = 0,12$ mod $16$ since $4c^2=0,1$ mod $4$ for any $c$.
  \end{proof}

\section{Existence of fluxes with negative determinants in $T^4/\mathbb{Z}_{3}$}

  From Section 6.6.3 we have the following
  \begin{itemize}
    \item when $tr\rho = 3 + 3\omega$, $\det N = 0,9$ mod $36$,
    \item when $tr\rho = 2 + \bar{\omega}$, $\det N = 0,3$ mod $12$,
    \item when $tr\rho = 2\omega + \bar{\omega}$, $\det N = 0,3$ mod $12$.
  \end{itemize}
  The flux is written in terms of the mod 3 structure as
  \begin{align}
    \begin{split}
      N=
      \begin{pmatrix}
        2(3a+k_1)&3c+k_3\\3c+k_3&2(3b+k_2)
      \end{pmatrix},
    \end{split}
  \end{align}
  so the determinant is expressed as
  \begin{align}
    \begin{split}
      \det N &= 4(9ab+3ak_2+3bk_1+k_1k_2)-9c^2-6ck_3-k_3^2\\
      &= 3(4[3ab+ak_2+bk_1]-3c^2-2ck_3)+4k_1k_2-k_3^2 <0.
    \end{split}
  \end{align}

  As in the previous section, we prove the above by considering $(k_1,k_2,k_3)$.
  \begin{proof}
    Let $tr\rho = 3+3\omega$, $(k_1,k_2,k_3)=(0,0,0)$, $b=1$. The determinant is
    \begin{align}
      \det N = 36a - 9c^2.
    \end{align}
    Since $c^2=0,1$ mod $4$ we first take $c=0$. Then $\det N = 36a$. Next, take $c=1$. Then $\det N = 36a-9$. Together, for $a<1$, we have $\det N = 0,9$ mod $36$
  \end{proof}
  For $tr\rho = 2+\bar{\omega}$ we have three different mod 3 structures possible. We prove for each of them.
  \begin{proof}
    Let $tr\rho = 2+\bar{\omega}$, $(k_1,k_2,k_3)=(0,1,0)$, $b=0$. The determinant is
    \begin{align}
      \det N = 3(4a-3c^2).
    \end{align}
    This is $\det N = 0$ mod $3$ if we can show $4a-3c^2=0,1,2,3$ mod $4$.
    The two terms are either $0,0$ mod $4$ or $0,1$ mod $4$.
    Thus we have $\det N = 0,3$ mod $12$
  \end{proof}
  \begin{proof}
    Let $tr\rho = 2+\bar{\omega}$, $(k_1,k_2,k_3)=(1,1,2)$, $b=0$. The determinant is
    \begin{align}
      \det N = 3(4a-3c^2-4c).
    \end{align}
    The two terms are $0,0,0$ mod $4$ or $0,1,0$ mod $4$.
    We again only have $\det N = 0,3$ mod $12$.
  \end{proof}
  \begin{proof}
    Let $tr\rho = 2+\bar{\omega}$, $(k_1,k_2,k_3)=(1,1,1)$, $b=0$. The determinant is
    \begin{align}
      \det N = 3(4a-3c^2-2c+1).
    \end{align}
    The terms are $0,0,0,1$ mod $4$ or $0,1,2,1$ mod $4$.
    We again only have $\det N = 0,3$ mod $12$.
  \end{proof}
  \begin{proof}
    Let $tr\rho = 2\omega + \bar{\omega}$, $(k_1,k_2,k_3)=(0,2,0)$
    The determinant is
    \begin{align}
      \det N = 3(4[3ab+2a]).
    \end{align}
    This is $\det N = 0$ mod $12$. (Take $b=-1,a>0$)
  \end{proof}
  \begin{proof}
    Let $tr\rho = 2\omega + \bar{\omega}$, $(k_1,k_2,k_3)=(2,2,1)$, $b=-1$. The determinant is
    \begin{align}
      \det N = 3(-4a-3c^2-2c-3).
    \end{align}
    The terms are either $0,0,0,1$ mod $4$ or $0,1,2,1$ mod $4$.
    This means we have $\det N = 0,3$ mod $12$.
  \end{proof}
  \begin{proof}
    Let $tr\rho = 2\omega + \bar{\omega}$, $(k_1,k_2,k_3)=(2,2,2)$, $b=-1$. The determinant is
    \begin{align}
      \det N = 3(-4a-3c^2-4c-4).
    \end{align}
    The terms are either $0,0,0,0$ mod $4$ or $0,1,0,0$ mod $4$.
    This means we have $\det N = 0,3$ mod $12$.
  \end{proof}

\section{Spinor transformation in $\mathbb{C}^N$}
\subsection{Introduction}

Let $\Psi(z,\bar{z})$ be a spinor on $T^6$ with eight components. Under a $\mathbb{Z}_{N}$ twist $z'=\Omega z$ we have
\begin{align}
	\Psi(\Omega z,\bar{\Omega}\bar{z})=V\mathcal{S}\Psi(z,\bar{z}),
\end{align}
where $\mathcal{S}$ is the spinor representation of the transformation and $V$ is some overall phase which we ignore for now. The spinor transformation is defined on $\mathbb{R}$ as
\begin{align}
	\mathcal{S}=\exp\left\{\frac{i}{2}\omega_{ij}\Sigma^{ij}\right\},
 \label{Spinor1}
\end{align}
where $\Sigma^{ij}=\frac{i}{4}\left[\Gamma^i,\Gamma^j\right]$ and $\omega_{ij}$ are the rotation angles. 

\subsection{Definitions}

Let $X^\mu$ denote a real coordinate system $\mathbb{R}^{2N}$ with a Euclidean metric. In this space, we define rotation as
\begin{align}
	X'^\mu = M^\mu{}_\nu X^\nu.
\end{align} 
We have a Clifford algebra 
\begin{align}
	\{\Gamma^\mu,\Gamma^\nu\}=2\delta^{\mu\nu}.
\end{align}
Next we define new coordinates on $\mathbb{C}^{N}$ in the following way
\begin{align}
\begin{split}
	z^1 &= X^1 + iX^2,\\\label{coordinates}
    z^2 &= X^3 + iX^4,
\end{split}
\end{align}
and similarly for the other coordinates.
We define rotations
\begin{align}
	z'^a = \theta^a{}_b z^b.
\end{align}
Our manifold has the metric $h^{a\bar{b}} = 2\delta^{a\bar{b}}$. Then we define the Clifford algebra as
\begin{align}
	\{\gamma^a,\gamma^{\bar{b}}\} = 2 h^{a\bar{b}} = 4 \delta^{a\bar{b}}.
\end{align}

\subsection{Spinor representation}
In real space, spinors transform with 
\begin{align}
	\mathcal{S}=e^{\frac{i}{2}\omega_{\mu\nu}\Sigma^{\mu\nu}},
\end{align}
where
\begin{align}
	\Sigma^{\mu\nu}=\frac{i}{4}[\Gamma^\mu,\Gamma^\nu]
\end{align}
and $\omega_{\mu\nu}$ is rotation matrix. In other words, rotation $M$ can be written by
\begin{align}
\begin{split}
	M 	&= e^{i\sum_i\theta_i T_i}\\
		&\approx I + \sum_i\theta_i T_i\\
		&= I + \omega,
\end{split}
\end{align}
where $\theta_i$ are the rotation angles and $T_i$ are the corresponding generators.\\
We would like to find this expression for complex coordinate system. First we define
\begin{align}
	X^\mu = e^\mu{}_az^a + e^\mu{}_b\bar{z}^{\bar{b}}.
\end{align}

\subsubsection{Vielbeins}

We can read off from eq.\eqref{coordinates} that the vielbeins are
\begin{align}
\begin{split}
	e^\mu{}_a &= \frac12\begin{pmatrix}
		1&0&0\\
		-i&0&0\\
		0&1&0\\
		0&-i&0\\
		0&0&1\\
		0&0&-i
	\end{pmatrix},
	\qquad e^\mu{}_{\bar{a}} = \frac12\begin{pmatrix}
		1&0&0\\
		i&0&0\\
		0&1&0\\
		0&i&0\\
		0&0&1\\
		0&0&i
	\end{pmatrix},\\
	e^a{}_\mu &= \begin{pmatrix}
		1&i&0&0&0&0\\
		0&0&1&i&0&0\\
		0&0&0&0&1&i
	\end{pmatrix},
	\qquad e^{\bar{a}}{}_\mu = \begin{pmatrix}
		1&-i&0&0&0&0\\
		0&0&1&-i&0&0\\
		0&0&0&0&1&-i
	\end{pmatrix}.
\end{split}
\end{align}
Since these matrices are not square, we need to be careful when inverting. The inner products are
\begin{align}
\begin{split}
	e^a{}_\mu e^\mu{}_a &= I_{N\times N},\\
	e^a{}_\mu e^\mu{}_{\bar{a}} &= 0,\\
	e^{\bar{a}}{}_\mu e^\mu{}_{\bar{a}} &= I_{N\times N},\\
	e^{\bar{a}}{}_\mu e^\mu{}_a &= 0,
\end{split}
\end{align}
and
\begin{align}
\begin{split}
	e^\mu{}_a e^\mu{}_a &= I_{2N \times 2N}-\sigma^2\otimes I_{N\times N},\\
	e^\mu{}_{\bar{a}} e^{\bar{a}}{}_\mu &=I_{2N \times 2N}+\sigma^2\otimes I_{N\times N},\\
	e^\mu{}_a e^{\bar{a}}{}_\mu &= (\sigma^3+i\sigma^1)\otimes I_{N \times N},\\
	e^\mu{}_{\bar{a}} e^a{}_\mu &= (\sigma^3-i\sigma^1)\otimes I_{N \times N}.
\end{split}
\end{align}

\subsubsection{Gamma matrices}
Assume $\Gamma^\mu = e^\mu_a \gamma^a + e^\mu{}_{\bar{a}} \gamma^{\bar{a}}$. Then the Clifford algebra can be written by
\begin{align}
\begin{split}
	\{\Gamma^\mu,\Gamma^\nu\} 	&= 	\{e^\mu_a \gamma^a, e^\nu_b \gamma^b\} +
									\{e^\mu{}_{\bar{a}} \gamma^{\bar{a}}, e^\nu_b \gamma^b\} + 
									\{e^\mu_a \gamma^a, e^\nu{}_{\bar{b}} \gamma^{\bar{b}}\} +
									\{e^\mu{}_{\bar{a}} \gamma^{\bar{a}}, e^\nu{}_{\bar{b}} \gamma^{\bar{b}}\}\\
								&=  (e^\mu_ae^\nu{}_{\bar{b}}+e^\mu{}_{\bar{b}}e^\nu_a)\{\gamma^a,\gamma^{\bar{b}}\}\\
								&=  2\delta^{\mu\nu}.
\end{split}
\end{align}
This is indeed the correct Clifford algebra.

\subsubsection{Spinor in complex space}
We can now find the expression for the spinor in complex space
\begin{align}
\begin{split}
	\mathcal{S}	&=	\exp\left\{
				-\frac18\omega_{\mu\nu}	
				[\Gamma^\mu,\Gamma^\nu]
			\right\}\\
		&=	\exp\left\{
				-\frac18\omega_{\mu\nu}	\left(
				[e^\mu{}_a\gamma^a + e^\mu{}_{\bar{a}}\gamma^{\bar{a}},e^\nu{}_b\gamma^b + e^\nu{}_{\bar{b}}\gamma^{\bar{b}}]
			\right)\right\}\\
		&=	\exp\left\{
				-\frac18\omega_{\mu\nu}\left(
				e^\mu{}_a e^\nu{}_b					[\gamma^a,\gamma^b]+
				e^\mu{}_a e^\nu{}_{\bar{b}}			[\gamma^a,\gamma^{\bar{b}}]+
				e^\mu{}_{\bar{a}} e^\nu{}_b			[\gamma^{\bar{a}},\gamma^b]+
				e^\mu{}_{\bar{a}} e^\nu{}_{\bar{b}}	[\gamma^{\bar{a}},\gamma^{\bar{b}}]
			\right)\right\}.
\end{split}
\end{align}
However, we can show
\begin{align}
\begin{split}
	&\omega_{\mu\nu} e^\mu{}_{\bar{a}} e^\mu{}_b \Sigma^{\bar{a}b}  \\
	&=(-\omega_{\nu\mu}) e^\nu{}_{\bar{a}} e^\mu{}_b (-\Sigma^{b\bar{a}}) \quad \text{  (rename the indices)}\\
	&=\omega_{\mu\nu} e^\mu{}_{\bar{b}} e^\nu{}_a \Sigma^{a\bar{b}}.
\end{split}
\end{align}
Then the spinor becomes
\begin{align}
\begin{split}
	\mathcal{S}	&=	\exp\left\{
				\frac{i}{2}\omega_{\mu\nu}\left(
					e^\mu{}_a e^\nu{}_b \Sigma^{ab}+
					e^\mu{}_{\bar{a}} e^\nu{}_{\bar{b}} \Sigma^{\bar{a}\bar{b}}+
					2e^\mu{}_a e^\nu{}_{\bar{b}} \Sigma^{a\bar{b}}
				\right)
			\right\}\\
		&=	\exp\left\{\frac{i}{2}
				\omega_{ab}\Sigma^{ab}+
				\omega_{\bar{a}\bar{b}}\Sigma^{\bar{a}\bar{b}}+
				2\omega_{a\bar{b}}\Sigma^{a\bar{b}}
			\right\}.
\end{split}
\end{align}

\subsubsection{Rotations}
We would like to find the explicit $\omega$ values. Let us assume the following holomorphic transformations in complex coordinates
\begin{align}
\begin{split}
	z'^a	&=	\theta^a{}_b z^b,\\
	\bar{z}'^a	&=	\theta^{\bar{a}}{}_{\bar{b}} \bar{z}^{\bar{b}}.
\end{split}
\end{align}
In infinitesimal form this is
\begin{align}
\begin{split}
	\theta^a{}_b	&\approx	I + \omega^a{}_b,\\
	\theta^{\bar{a}}{}_{\bar{b}}	&\approx	I + \omega^{\bar{a}}{}_{\bar{b}},
\end{split}
\end{align}
so that for a discrete rotation
\begin{align}
\begin{split}
	\theta^a{}_b	&=	e^{i\omega^a{}_b},\\
	\theta^{\bar{a}}{}_{\bar{b}}	&=	e^{i\omega^{\bar{a}}{}_{\bar{b}}}.
\end{split}
\end{align}
Then, we can lower the indices
\begin{align}
\begin{split}
	\omega_{ab}&=h_{a\bar{c}}\omega^{\bar{c}}{}_b,\\
	\omega_{\bar{a}\bar{b}}&=h_{c\bar{b}}\omega^c{}_{\bar{b}},\\
	\omega_{a\bar{b}}&= h_{a\bar{c}}\omega^{\bar{c}}{}_{\bar{b}},\\
	\omega_{\bar{a}b}&= h_{c\bar{a}}\omega^c{}_b.
\end{split}
\end{align}
Only the last two are non-zero, because the transformations should be holomorphic. Additionally, since both $h_{a\bar{b}}$ and $\omega^a{}_b$ are symmetric under the interchange of the indices, $\omega_{a\bar{b}}$ is also symmetric.\\
Finally we find 
\begin{align}
	\mathcal{S}&=	\exp\left\{i
				\omega_{a\bar{b}}\Sigma^{a\bar{b}}
			\right\}.\label{complex}
\end{align}

\subsubsection{Spinor on $T^4/\mathbb{Z}_N$ with diagonal twists}

We consider twists of the form 
\begin{align}
	\Phi = \text{diag}\left(e^{i{\phi}_1},e^{i{\phi}_2}\right)= \{ e^{i\omega_{a\bar{a}}} \}.
\end{align}
Then we only need to consider $\Sigma^{a\bar{a}}$. The Gamma functions are defined in ref.~\cite{Kikuchi:2022psj} as
\begin{align}
\begin{split}
	\Gamma^1&=\sigma^Z\otimes\sigma^3,\\
	\Gamma^{\bar{1}}&=\sigma^{\bar{Z}}\otimes\sigma^3,\\
	\Gamma^2&=I_{2\times2}\otimes\sigma^Z,\\
	\Gamma^{\bar{2}}&=I_{2\times2}\otimes\sigma^{\bar{Z}}.
\end{split}
\end{align}
We find the commutators:
\begin{align}
\begin{split}
	[\Gamma^1,\Gamma^{\bar{1}}]&=[\sigma^Z\otimes\sigma^3,\sigma^{\bar{Z}}\otimes\sigma^3]\\
								&=\sigma^Z\sigma^{\bar{Z}}\otimes I_{2\times2}-\sigma^{\bar{Z}}\sigma^Z\otimes I_{2\times2}\\
								&=[\sigma^Z,\sigma^{\bar{Z}}]\otimes I_{2\times2}\\
								&=4\sigma^3\otimes I_{2\times2}.\\
	[\Gamma^2,\Gamma^{\bar{2}}]&=I_{2\times2}\otimes[\sigma^Z,\sigma^{\bar{Z}}]\\
								&=4I_{2\times2}\otimes\sigma^3.
\end{split}
\end{align}
The spinor is expressed as
\begin{align}
\begin{split}
	\mathcal{S}&=\exp\left\{i
					\omega_{a\bar{a}}\Sigma^{a\bar{a}}
			\right\}\\
				&=\exp\left\{
					-i{\phi}_1(\sigma^3\otimes I_{2\times2})-i{\phi}_2(I_{2\times2}\otimes\sigma^3)
			\right\}\\
				&=\exp\left\{
					-i{\phi}_1
					\begin{pmatrix}
						1&&&\\
						&1&&\\
						&&-1&\\
						&&&-1
					\end{pmatrix}
					-i{\phi}_2
					\begin{pmatrix}
						1&&&\\
						&-1&&\\
						&&1&\\
						&&&-1
					\end{pmatrix}
			\right\}\\
				&=\exp \left\{i
					\begin{pmatrix}
						-{\phi}_1-{\phi}_2&&&\\
						&-{\phi}_1+{\phi}_2&&\\
						&&{\phi}_1-{\phi}_2&\\
						&&&{\phi}_1+{\phi}_2
					\end{pmatrix}
			\right\}\\
				&=	\begin{pmatrix}
						e^{i{\phi}}&&&\\
						&e^{i({\phi}+2{\phi}_2)}&&\\
						&&e^{i({\phi}+2{\phi}_1)}&\\
						&&&e^{-i{\phi}}
					\end{pmatrix}\\
				&=	e^{i{\phi}}\begin{pmatrix}
						1&&&\\
						&e^{2i{\phi}_2}&&\\
						&&e^{2i{\phi}_1}&\\
						&&&e^{-2i({\phi}_1+{\phi}_2)}
				\end{pmatrix}
                .
\end{split}
\end{align}
By including the overall phase in the transformation function we obtain the spinor transformation as
\begin{align}
	\mathcal{S}=\text{diag}\left(1,e^{2\pi ik_1/N},e^{2\pi ik_2/N},e^{-2\pi i(k_1+k_2)/N}\right).
\end{align}

\subsection{Spinor representation under parity transformation}
We need to define the spinor representation of the parity transformation. Define the $R$ transformation acting on a spinor in the following way: $\Psi'=\Lambda_{\frac12}\Psi$.

The boundary condition in the parity transformed coordinates is expressed as
\begin{align}
	\Psi'	&\rightarrow S'V'\Psi'=\Lambda_{\frac12}SV\Lambda_{\frac12}^{-1}\Psi.
\end{align}
We can find $\Lambda_{\frac12}$ by considering the parity invariance of the Dirac equation. First, we note that since the $R$ transformation is defined as $R=P^TUP$, we can express the corresponding spinor representation as
\begin{align}
	\Lambda_{\frac12}=P^{-1}_R U P_R,
\end{align}
where $P_R$ is the spinor representation of rotation. To find the parity transformation matrix, we first transform the Dirac equation to the $\vec{w}=P\vec{z}$ coordinate system. Note that $\Gamma^{w}=\Gamma^{z}$. Since it must be invariant, we find:
\begin{align}	(\Gamma^{z^i}D_{z^i}+\Gamma^{\bar{z}^i}D_{\bar{z}^i})\Psi=0\rightarrow	(\Gamma^{z^i}D_{w^i}+\Gamma^{\bar{z}^i}D_{\bar{w}^i})P_R\Psi=0.
\end{align}
We denote $P_R\Psi=\Psi'$. Since we are in the $\vec{w}$ system now, we can directly find the parity transformation matrix. We note that under parity $w^1\rightarrow w^1$ and $w^2\rightarrow \bar{w}^2$. Then the Dirac equation transforms like
\begin{align}
\begin{aligned}
	&(\Gamma^{z^1}D_{w^1}+\Gamma^{\bar{z}^1}D_{\bar{w}^1}+\Gamma^{z^2}D_{w^2}+\Gamma^{\bar{z}^2}D_{\bar{w}^2})\Psi'=0	\\\rightarrow&U_S(\Gamma^{z^1}D_{w^1}+\Gamma^{\bar{z}^1}D_{\bar{w}^1}+\Gamma^{z^2}D_{\bar{w}^2}+\Gamma^{\bar{z}^2}D_{w^2})\Psi'=0.
\end{aligned}
\end{align}
To make this invariant, we need to find a matrix $U_S$ that satisfies the following conditions:
\begin{align}
\begin{aligned}
	\{U_S,\Gamma^{z^1}\}=0,\\
	\{U_S,\Gamma^{\bar{z}^1}\}=0,\\
	U_S\Gamma^{z^2}=-\Gamma^{\bar{z}^2}U_S,\\
	U_S\Gamma^{\bar{z}^2}=-\Gamma^{z^2}U_S.
\end{aligned}
\end{align}
We find
\begin{align}
	U_S=\begin{pmatrix}
		0&0&1&0\\
		0&0&0&1\\
		-1&0&0&0\\
		0&-1&0&0
	\end{pmatrix}.
\end{align}
This gives us
\begin{align}
	-(\Gamma^{z^1}D_{w^1}+\Gamma^{\bar{z}^1}D_{\bar{w}^1}+\Gamma^{\bar{z}^2}D_{\bar{w}^2}+\Gamma^{z^2}D_{w^2})U_S\Psi'=0.
\end{align}
Rotating back to the $\vec{z}$ system, we find the full transformation of the spinor to be
\begin{align}
	\Psi\xrightarrow{R}P^{-1}_RU_SP_R\Psi.
\end{align}
We can then express the boundary conditions as:
\begin{align}
	S'V'=-P^{-1}_RU_SP_RSVP_R^{-1}U_SP_R,
\end{align}
where we have used $U_S^{-1}=-U_S$.
We could equivalently redefine the gamma matrices $\Gamma^{z^2}\leftrightarrow\Gamma^{\bar{z}^2}$, which would change the chirality matrix but keep the order of spinor components. Specifically, the chirality matrix can be written as
\begin{align}
	\Gamma^5=\text{diag}(-1,+1,+1,-1).
\end{align}
The final spectrum is the same with both approaches.

\subsection{$T^4/Z_N$ with diagonal twists}
We would like to derive the spinor representation of the $R$ transformation. The spinor representation of SO(4) in real space is
\begin{align}
	S=e^{\left\{\frac12i\omega_{ij}\Sigma^{ij}\right\}}.
\end{align}
Let us denote the axes $\operatorname{Re}z^1$, $\operatorname{Im}z^1$, $\operatorname{Re}z^2$, $\operatorname{Im}z^2$ by $i=1,2,3,4$. Then only the $13$ and $24$ components are non-zero. We want the commutators
\begin{align}
	\Sigma^{ij}=\frac14i[\Gamma^i,\Gamma^j],
\end{align}
where the gamma matrices in real space are
\begin{align}
\begin{aligned}
	\Gamma^1&=\frac{1}{2}(\Gamma^{z^1}+\Gamma^{\bar{z}^1})=\sigma^1\otimes\sigma^3,\\
	\Gamma^2&=\frac{1}{2i}(\Gamma^{z^1}-\Gamma^{\bar{z}^1})=\sigma^2\otimes \sigma^3,\\
	\Gamma^3&=\frac{1}{2}(\Gamma^{z^2}+\Gamma^{\bar{z}^2})=I\otimes\sigma^1,\\
	\Gamma^4&=\frac{1}{2i}(\Gamma^{z^2}-\Gamma^{\bar{z}^2})=I\otimes\sigma^2.
\end{aligned}
\end{align}
Then we find the non-zero components of $\Sigma^{ij}$
\begin{align}
\begin{aligned}
	\Sigma^{13}	&=-\frac12\sigma^1\otimes\sigma^2,\\
	\Sigma^{24}	&=\frac12\sigma^2\otimes\sigma^1.
\end{aligned}
\end{align}
Since $\omega_{13}=-\omega_{31}={\phi}_0$ and $\omega_{24}=-\omega_{42}={\phi}_0$ we find
\begin{align}
	P_R=e^{\left\{i{\phi}_0(\sigma^2\otimes\sigma^1-\sigma^1\otimes\sigma^2)\right\}}=\begin{pmatrix}
		1&0&0&0\\
		0&\cos{\phi}_0&-\sin{\phi}_0&0\\
		0&\sin{\phi}_0&\cos{\phi}_0&0\\
		0&0&0&0
	\end{pmatrix}.
\end{align}
Thus the spinor representation of the $R$ transformation becomes 
\begin{align}
\begin{aligned}
\Lambda_{\frac12}=P_R^TUP_R
				=\begin{pmatrix}
					0&-\sin{\phi}_0&\cos{\phi}_0&0\\
					\sin{\phi}_0&0&0&\cos{\phi}_0\\
					-\cos{\phi}_0&0&0&\sin{\phi}_0\\
					0&-\cos{\phi}_0&-\sin{\phi}_0&0
				\end{pmatrix}.
\end{aligned}
\end{align}
The full boundary condition term is
\begin{align}
\begin{aligned}
	&S'V'=-\Lambda_{\frac12}SV\Lambda_{\frac12}\\
		&=V\begin{pmatrix}
			\alpha_1s_0^2+\alpha_2c_0^2&0&0&(\alpha_1-\alpha_2)s_0c_0\\
			0&s_0^2+\beta c_0^2&(\beta-1)s_0c_0&0\\
			0&(\beta-1)s_0c_0&c_0^2+\beta s_0^2&0\\
			(\alpha_1-\alpha_2)s_0c_0&0&0&\alpha_1c_0^2+\alpha_2s_0^2
		\end{pmatrix},
		\end{aligned}\label{spinor'}
\end{align}
where $\alpha_1=e^{2\pi ik_1/N}$, $\alpha_2=e^{2\pi ik_2/N}$, $\beta=e^{2\pi i(k_1+k_2)/N}$, $s_0=\sin{\phi}_0$, $c_0=\cos{\phi}_0$. These correspond to the transformation of coordinates under twists
\begin{align}
	\begin{pmatrix}
		z'_1\\z'_2
	\end{pmatrix}=
	\begin{pmatrix}
			\alpha_1z_1\\\alpha_2z_2
		\end{pmatrix}.
\end{align}
Due to positive flux eigenvalues, the Dirac spinor in $z'$ coordinates is
\begin{align}
	\Psi'&=\begin{pmatrix}
		\psi'_{++}\\
		0\\0\\0
	\end{pmatrix}.
\end{align}
This makes the boundary conditions
\begin{align}
	\Psi'&\rightarrow V\begin{pmatrix}		(\alpha_1\sin^2{\phi}_0+\alpha_2\cos^2{\phi}_0)\psi'_{++}\\
		0\\
		0\\
		(\alpha_1-\alpha_2)\sin{\phi}_0\cos{\phi}_0\psi'_{++}
	\end{pmatrix}.
\end{align}

\end{document}